\documentclass[]{aastex63}
\usepackage{lineno}
\usepackage{subfigure}
\usepackage{longtable}
\usepackage{graphicx}
\usepackage{appendix}
\usepackage{natbib}
\usepackage{amssymb,amsmath}

\shortauthors{Zhang et al.}

\begin{document}

\title{A Comprehensive Analysis of Text-Book-Version Afterglow Light curves of Gamma-Ray Bursts and Implication for Universal Radiation Physics of Baryonic Jets}

\correspondingauthor{En-Wei Liang}
\email{lew@gxu.edu.cn}
\author[0000-0003-0726-7579]{Lu-Lu Zhang}
\affil{Guangxi Key Laboratory for Relativistic Astrophysics, School of Physical Science and Technology, Guangxi University, Nanning 530004, People’s Republic of China}
\author[0000-0002-1766-6947]{Shu-Qing Zhong}
\affil{School of Science, Guangxi University of Science and Technology, Liuzhou 545006, People’s Republic of China}
\author[0000-0002-6162-9115]{Li-Ping Xin}
\affil{Key Laboratory of Space Astronomy and Technology, National Astronomical Observatories, Chinese Academy of Sciences, Beijing 100101, People’s Republic of China}
\author[0000-0002-7044-733X]{En-Wei Liang*}
\affil{Guangxi Key Laboratory for Relativistic Astrophysics, School of Physical Science and Technology, Guangxi University, Nanning 530004, People’s Republic of China}

\begin{abstract}
The standard external shock model in the thin-shell scenario predicts an onset bump in the early optical afterglow light curves of gamma-ray bursts (GRBs). We collect such a textbook-version light curve sample of $30$ GRBs, and derive the jet properties from our joint fit to their X-ray and optical afterglow light curves. It is found that the distributions of the isotropic initial Lorentz factors ($\Gamma_0$), the deceleration radii ($R_{\rm dec}$), and the magnetic field strength ($B_0$) are log-normal, but the distributions of the isotropic kinetic energy ($E_{\rm k, iso}$), medium density ($n_{0}$), and the magnetization parameter ($\sigma_{B}\equiv\epsilon_B/\epsilon_e$) are tentatively bimodal. A tight $R_{\rm dec}\mbox{-}B_{0}\mbox{-}\sigma_{B}$ relation is found. It infers a universal $\epsilon_e E_{\rm k,iso}$ among bursts, plausibly supporting the previous argument of a universal GRB radiation energy among GRBs. A jet break is required for modeling the light curves of $26$ GRBs. The distributions of the jet opening angles and the jet-corrected kinetic energies log-normally center at $\log \theta_{\rm j,c}/{\rm rad}=-1.51$ (standard deviation $\sigma=0.27$) and $\log (E_{\rm k, j,c}/{\rm erg})=51.78$ ($\sigma=0.54$), respectively. Those GRBs ($19$ GRBs), whose prompt gamma-ray emission is well estimated with broad energy-band observations, satisfy the previously discovered $L_{\rm \gamma, p, iso}-E_{\rm p,z}-\Gamma_{0}$ relation, and their gamma-ray radiation efficiencies 
log-normally distribute in the range from $0.04\%$ to $10\%$ with a central value of $0.42\%$. 
Such a low efficiency favors the baryonic fireball model, 
and the distribution of their baryon mass loading in the GRB ejecta log-normally centers at
$\log (M_{\rm fb,c}/M_{\sun})=-5$ ($\sigma=0.75$).   
\end{abstract}


\section{Introduction}
Gamma-ray bursts (GRBs) and their multiwavelength afterglows as 
amazing energetic transients are radiated from relativistic jets 
powered by newborn black holes or millisecond magnetars 
during the core collapse of rapidly spinning massive stars or 
the compact star mergers\citep{Usov_1992_Natur.357..472U, Woosley_1993_ApJ...405..273W, Fishman_1995_Meegan_ARAA..33..415F, Kluzniak_1998_Ruderman_ApJ...505L.113K, Piran_1999_PhR...314..575P}. 
The GRB survey during $1991-2000$ by the Burst and Transient Experiment (BATSE) 
on board the Compton Gamma-Ray Observatory (CGRO) collected a 
large sample (about $2068$ online triggered GRBs and $1838$ offline scanned GRBs) 
and discovered two types of GRBs, i.e., the long-duration GRBs and short-duration GRBs \citep{Kouveliotou_1993_Meegan_ApJ...413L.101K, Fishman_1995_Meegan_ARAA..33..415F, Paciesas_1999_Meegan_ApJS..122..465P, Stern_2001_Tikhomirova_ApJ...563...80S}. 
The precise localization and discovery of multiple wavelength afterglows 
of GRBs since $1997$ with BeppoSAX and ground-based telescopes, 
especially the very early afterglow observations with the X-Ray Telescope (XRT) on board the Swift mission since $2004$, revolutionized our understanding of the nature of this phenomenon~\citep{Costa_1997_Frontera_Natur.387..783C, Tavani_1997_ApJ...483L..87T, Piran_1999_PhR...314..575P, Zhang_2004_Meszaros_IJMPA..19.2385Z,Meszaros_2006_RPPh...69.2259M, Zhang_2006_Fan_ApJ...642..354Z, Gehrels_2009_Ramirez-Ruiz_ARAA..47..567G}. 
The GeV$\mbox{-}$TeV gamma-ray observations with the Large Area Telescope (LAT) on board the Fermi mission since $2008$ and the ground-based telescopes for very high energy (VHE) since $2018$, such as High Energy Stereoscopic System (H.E.S.S.), Major Atmospheric Gamma Imaging Cerenkov (MAGIC), and Large High Altitude Air Shower Observatory, 
further deepened our knowledge on the radiation physics of the GRB jets~\citep[e.g.,][]
{MAGIC_Collaboration_2019_Acciari_Natur.575..455M,
MAGIC_Collaboration_2019_Acciari_Natur.575..459M,
HESS_Collaboration_2021_Abdalla_Sci...372.1081H,
Zhang_2021_Ren_ApJ...917...95Z,
Zhang_2023_Ren_ApJ...952..127Z,
Cao_2023_Aharonian_SciA....9J2778C,
Ren_2023_Wang_arXiv231015886R}. 
More importantly, the discovery of a connection among GRBs, kilonova, 
and gravitational wave (GW) events since 2017, i.e., 
GRB~170817A/AT~2017gfo/GW170817~\citep[e.g.,][]{Abbott_2017_Abbott_ApJ...850L..39A,
Abbott_2017_Abbott_ApJ...850L..40A,
Abbott_2017_Abbott_ApJ...848L..13A,
Abbott_2017_Abbott_ApJ...841...89A,
Abbott_2017_Abbott_PhRvD..96b2001A,
Coulter_2017_Foley_Sci...358.1556C},
not only confirms that the short-duration GRBs are from the compact star mergers but also opens a new era for GRB study with multimessenger observations~\citep[e.g.,][]{Poggiani_2019f_LIGO_Scientific_Collaboration_rap.confE..13P, Poggiani_2020_LIGO_Scientific_Collaboration_mbhe.confE..19P, KANG_2023_LIU_SSPMA..53j0014K}. 

Although extensive investigations have been presented, 
the jet composition and radiation physics still remain open questions
\citep[e.g.,][]{Zhang_2011_CRPhy..12..206Z}. 
The proposed models are classified into two groups, 
i.e., the matter-dominated hot fireball model and the 
Poynting-flux-dominated cold fireball model~\citep[e.g.,][]{Sari_1995_Piran_ApJ...455L.143S, Zhang_2002_Meszaros_ApJ...566..712Z, Zhang_2004_Meszaros_IJMPA..19.2385Z, Zhang_2009_Peer_ApJ...700L..65Z}. 
The matter-dominated fireball model assumes that the 
prompt gamma rays are emitted by the synchrotron and/or 
inverse Compton scattering process of the 
relativistic electrons that accelerated via internal shocks 
of the GRB fireball shells
and the afterglows are emitted by the relativistic electrons accelerated 
via external shocks when the fireball propagates into the surrounding medium~\citep[e.g.,][]{Rees_1992_Meszaros_MNRAS.258P..41R, 
Meszaros_1993_Rees_ApJ...418L..59M, 
Meszaros_1997_Rees_ApJ...476..232M, 
Sari_1998_Piran_ApJ...497L..17S,
Sari_1999_Piran_ApJ...517L.109S, 
Sari_1999_Piran_ApJ...520..641S, 
Meszaros_1999_Rees_MNRAS.306L..39M, 
Hu_2019_Oates_AA...632A.100H}. 
The high radiation efficiency ($\eta_\gamma$) inferred from 
the observational data with the X-ray and GeV afterglows 
challenge the standard hot fireball model ~\citep[e.g.,][]{Ioka_2006_Toma_AA...458....7I, Zhang_2007_Liang_ApJ...655..989Z}). 
The cold fireball model can provide a natural 
explanation for the high efficiency observed in Fermi data~\citep[e.g.,][]{Beniamini_2015_Nava_MNRAS.454.1073B}. 
In addition, the nondetection of photosphere thermal emission 
in broad energy-band observations for some GRBs also disfavors 
the hot fireball model~\citep[e.g.,][]{Zhang_2009_Peer_ApJ...700L..65Z, Abdo_2009_Ackermann_Sci...323.1688A, Gehrels_2013_Razzaque_FrPhy...8..661G}. 
\cite{Zhang_2011_Yan_ApJ...726...90Z}
proposed an internal collision-induced magnetic reconnection and turbulence (ICMART) model to explain the data of some GRBs,
which can typically give a radiative efficiency of tens of percent.

The rich data of the prompt and early afterglow observations in the era of the Swift and Fermi missions since $2004$ reveal a great diversity among GRBs. 
Thus, the issues on the jet properties and radiation physics of GRBs are still under debated. 
In this paper, we focus on presenting a comprehensive analysis of the
GRB jet properties with a sample of ``textbook" version optical light curves. 
These light curves apparently show a smooth afterglow onset bump, which agrees with the hot fireball model prediction in the thin-shell scenario~\citep[e.g.,][]{Rees_1992_Meszaros_MNRAS.258P..41R, Sari_1995_Piran_ApJ...455L.143S, Sari_1999_Piran_ApJ...520..641S,
Liang_2010_Yi_ApJ...725.2209L, Liang_2013_Gao_ApJ...774...13L,
Ghirlanda_2018_Nappo_A_A...609A.112G,
Yi_2020_Wu_ApJ...895...94Y}. 
In the framework of the standard external shock model, 
we present a comprehensive analysis of the jet properties 
of this kind of GRBs by jointly fitting the X-ray and optical 
afterglow light curves.

The structure of this paper is as follows. 
The sample selection constraints and criteria are described in Section $\ref{section 2}$. 
We applied the standard external forward shock (FS) model to fit the optical and X-ray afterglow data and constrained the model parameters based on the nested sampling algorithm in Section $\ref{section 3}$. 
The modeling results with afterglow jet properties and radiation efficiency are presented in Section $\ref{section 4}$. 
The correlations about the fireball deceleration and magnetization are presented in Section $\ref{section 5}$. 
Finally, a summary is provided in Section $\ref{section 6}$. 
Throughout this paper, the notations $Q_n=Q/10^{n}$ in cgs and log$x$=log$_{10}x$ are adopted. 
We take the cosmology parameters as $H_0 = 67.8~\rm km~s^{-1}~Mpc^{-1}$,
and $\Omega_M = 0.308$ \citep{Planck_2016_Perovich_AGUFM.C21A0664P}.

\section{Sample Selection and Data}
\label{section 2}
We select only the ``textbook" version of optical light curves as predicted by the standard external shock model in the thin-shell scenario. The medium surrounding the jet is assumed to be the interstellar medium (ISM). The selection criteria are as follows. 
\begin{itemize}
    \item A clear onset bump is detected in the early optical light curve, and the decay slope post the peak time is consistent with that of the normal decay segment as predicted by the standard external model in the ISM without considering the late energy injection scenario, i.e. $0.75<\alpha<1.5$, 
    where $F\propto t^{-\alpha}$ is adopted.
    \item X-ray afterglow observation is available. 
    \item Redshift measurement is available. 
\end{itemize}

We make an extensive search for such optical light curves from the literature. All of optical data have been corrected for Galactic extinction \citep{Schlegel_1998_Finkbeiner_ApJ...500..525S}. 
The unabsorbed X-ray afterglow data are taken from the UK Swift Science Data Centre at the University of Leicester \citep{Evans_2009_Beardmore_MNRAS.397.1177E}. 
We finally obtain a sample of $30$~GRBs. 
They are listed in Table~\ref{Tab1}. Their light curves are illustrated in Figure~\ref{modelfit}. 

\section{Modelling Optical and X-ray Afterglow Light Curves }
\label{section 3}
\subsection{GRB Afterglow Model}

The dynamics of the standard GRB external FS model adopted for our analysis are oulined as 
(e.g., \citealp{Sari_1998_Piran_ApJ...497L..17S, Huang_2000_Gou_ApJ...543...90H,Nava_2013_Sironi_MNRAS.433.2107N, Zhang_2018_pgrb.book.....Z})
\begin{equation}
\frac{d \Gamma}{d r}=-\frac{\Gamma\left(\Gamma^{2}-1\right)(\hat{\gamma} \Gamma-\hat{\gamma}+1) \frac{d m}{d r} c^{2}-(\hat{\gamma}-1) \Gamma\left(\hat{\gamma} \Gamma^{2}-\hat{\gamma}+1\right)(3 U / r)}{\Gamma^{2}\left[m_{0}+m\right] c^{2}+\left(\hat{\gamma}^{2} \Gamma^{2}-\hat{\gamma}^{2}+3 \hat{\gamma}-2\right) U} ,
\end{equation}
\begin{equation}
 \frac{d U}{d r}=(1-\epsilon)(\Gamma-1) c^{2} \frac{d m}{d r}-(\hat{\gamma}-1)\left(\frac{3}{r}-\frac{1}{\Gamma} \frac{d \Gamma}{d r}\right) U ,  
\end{equation}
where $\Gamma$ is the bulk Lorentz factor of the fireball, 
$\hat{\gamma}=4/3$, $m_p$ is the proton mass, 
$c$ is the speed of light, $m$ is the swept-up mass, 
$U$ is the internal energy, and $\epsilon$ is the radiation efficiency of electrons. 
We have $dm/dr=4 \pi r^{2} n(r) m_{p}$, where $n(r)$ is the medium density. 
The continuity equation for electron distribution is \citep{Fan_2008_Piran_MNRAS.384.1483F}:
\begin{equation}
    \frac{\partial}{\partial r}\left(\frac{d N_{e}}{d \gamma_{e}^{\prime}}\right)+\frac{\partial}{\partial \gamma_{e}^{\prime}}\left[\frac{d \gamma_{e}^{\prime}}{d r}\left(\frac{d N_{e}}{d \gamma_{e}^{\prime}}\right)\right]=Q\left(\gamma_{e}^{\prime}, r\right) ,
\end{equation}
where the symbol `$\prime$' marks the comoving frame of shock front, and 
\begin{equation}
    \frac{d \gamma_{e}^{\prime}}{d r}=-\frac{\sigma_{\mathrm{T}}}{6 \pi m_{e} c^{2}} \frac{B^{\prime 2}}{\beta \Gamma}\left[1+Y\left(\gamma_{e}^{\prime}\right)\right] \gamma_{e}^{\prime 2}-\frac{\gamma_{e}^{\prime}}{r}
\end{equation}
is the cooling term of electrons, in which $B^{\prime}=\sqrt{32 \pi \epsilon_{B} n} \Gamma c$ is the internal magnetic field strength, $\epsilon_{B}$ denotes the fraction of the fireball internal energy allocated to the magnetic field, $Y$ is the inverse Compton parameter, given by $Y=\left(-1+\sqrt{1+4 \epsilon_{\rm{rad}} \eta_{\rm KN} \epsilon_{e} / \epsilon_{B}}\right) / 2$ \citep{Sari_2001_Esin_ApJ...548..787S, Fan_2008_Piran_MNRAS.384.1483F}. 
The cooling effects of synchrotron radiation, 
synchrotron self-Compton (SSC) processes, 
and adiabatic expansion are considered in our analysis. 
The shocked electrons from the medium is accelerated to 
a single power-law distribution in the energy space, 
which can be expressed as:
\begin{equation}
    Q\left(\gamma_{e}^{\prime},r\right) \propto \gamma_{e}^{\prime-p}, 
    \quad\quad\gamma_{e, \min }^{\prime} \leqslant \gamma_{e}^{\prime} \leqslant \gamma_{e, \max }^{\prime}. 
\end{equation}
We consider the scenario of $p>2$ for the injected electrons, and the values of $\gamma_{e,\min}^{\prime}$ and $\gamma_{e,\max}^{\prime}$ are determined by 
\begin{align}
&\gamma_{e, \min }^{\prime}=\epsilon_{e}\frac{(p-2)}{(p-1)} \frac{m_p}{ m_{e}} \Gamma ,\\
&\gamma_{e, \max }^{\prime}=\sqrt{\frac{9 m_{e}^{2} c^{4}}{8 B^{\prime} q_{e}^{3}}}
\end{align}
where $m_{e}$ and $q_{e}$ are the mass and charge of the electron, respectively~\citep[e.g.,][]{Sari_1998_Piran_ApJ...497L..17S, Kumar_2012_Hernandez_MNRAS.427L..40K}. 

We assume that the X-ray/optical afterglows are attributed to the synchrotron radiations of the electrons~\citep[e.g.,][]{MAGIC_Collaboration_2019_Acciari_Natur.575..455M, Zhang_2021_Ren_ApJ...917...95Z, Fukami_2022_Berti_icrc.confE.788F, Suda_2022_Artero_icrc.confE.797S, Giarratana_2022_Rhodes_AA...664A..36G, Zhang_2023_Ren_ApJ...952..127Z}. 
Assuming that the direction of the magnetic field 
is perpendicular to the electron velocity, the emitted spectral power of synchrotron radiation
at a given frequency $\nu'$ of a single electron is given by 
\begin{equation}
P^{\prime}_e(\nu^{\prime}, \gamma^{\prime}_e)
=\frac{\sqrt{3} q_e^3 B^{\prime}}{m_e c^2}
F\left(\frac{\nu'}{\nu'_c}\right),
\end{equation}
where $F(x)=x\int_{x}^{+\infty}K_{5/3}(k)dk$,
$K_{5/3}(k)$ is the modified Bessel function of $5/3$ order,
and $\nu'_{\rm c}=3q_e B^{\prime}{\gamma'_e}^2/(4\pi m_e c)$.
Thus, the spectral power of synchrotron radiation of electrons
${dN_e}/d\gamma_e^{\prime}$ at a given frequency $\nu'$ is
\begin{equation}
P'_{\rm syn}(\nu')=
\int_{\gamma^{\prime}_{e,\min}}^{\gamma^{\prime}_{e,\max}}
P'_e(\nu')\frac{dN_e}{d\gamma_e^{\prime}}d\gamma^{\prime}_e.
\end{equation}
In addition, the equal arrival time effect is taken into account 
by numerically computing the arrival times of photons \citep{Waxman_1997ApJ...485L...5W},
\begin{equation}
t_{\mathrm{obs}}=
(1+z) \int \frac{1-\beta\cos\Theta}{\beta c} dr \equiv \rm{const},
\end{equation}
where $\Theta$ represents the angle between the direction of motion
of a fluid element in the jet and the line of sight. We set the GRB jet as an on-axis-observed jet, i.e., $\theta_v=0$.
Assuming that the observed frequency is $\nu_{\rm obs}$, the frequency transformed to the comoving frame is denoted as
$\nu^{\prime}(\nu_{\rm obs})=(1+z)\nu_{\rm obs}/\mathcal{D}$, where $z$ is the redshift,
$\mathcal{D}=1/{\Gamma(1-\beta\cos\Theta)}$
is the Doppler factor.
For a detailed description of our model please refer to \cite{Zhang_2023_Ren_ApJ...952..127Z} and \cite{Ren_2023_Wang_arXiv231015886R}.

\subsection{Fitting Strategy and Results}

\begin{figure*}[htbp]
\centering
\includegraphics[angle=0,width=0.30\textwidth]{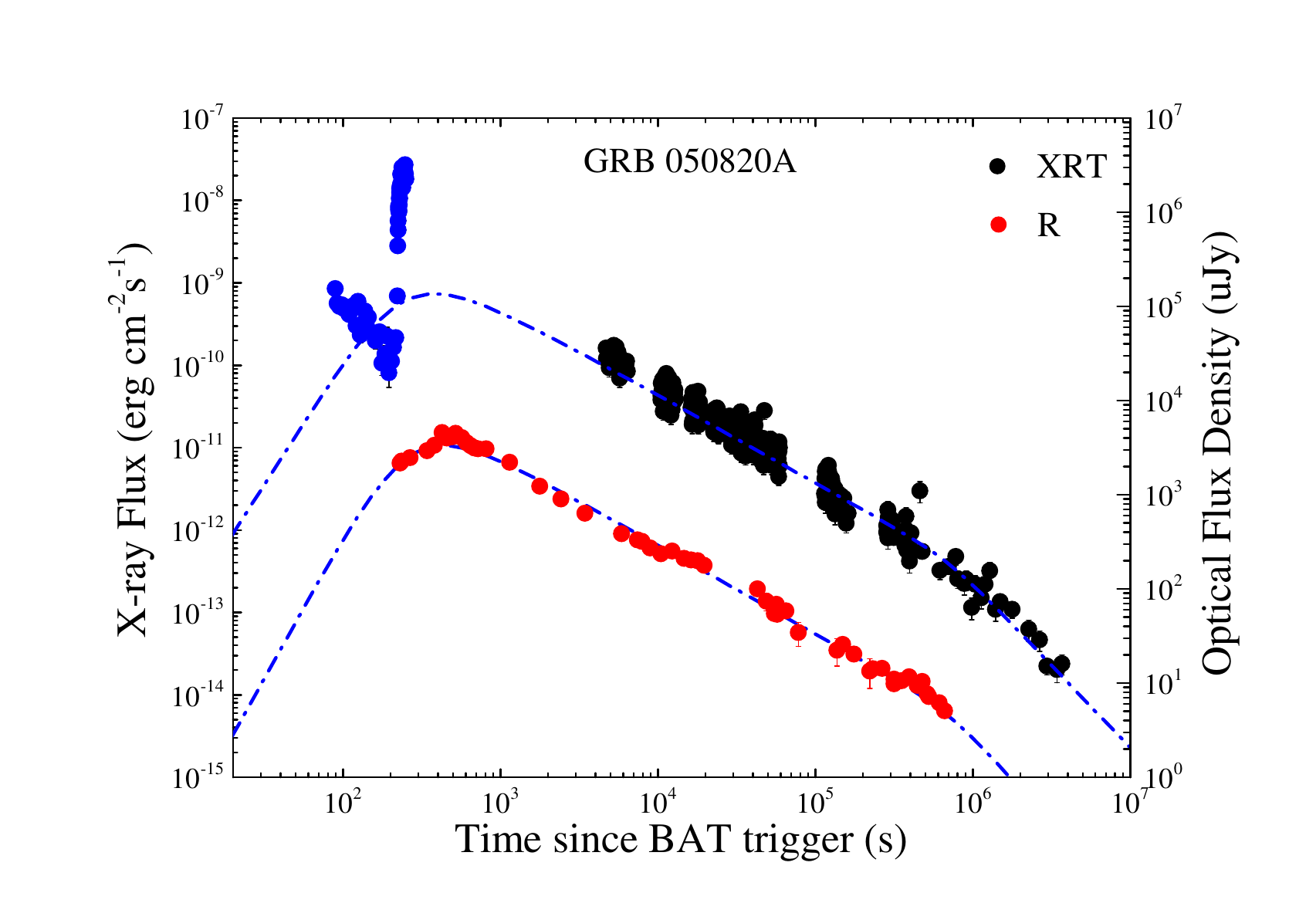}
\includegraphics[angle=0,width=0.30\textwidth]{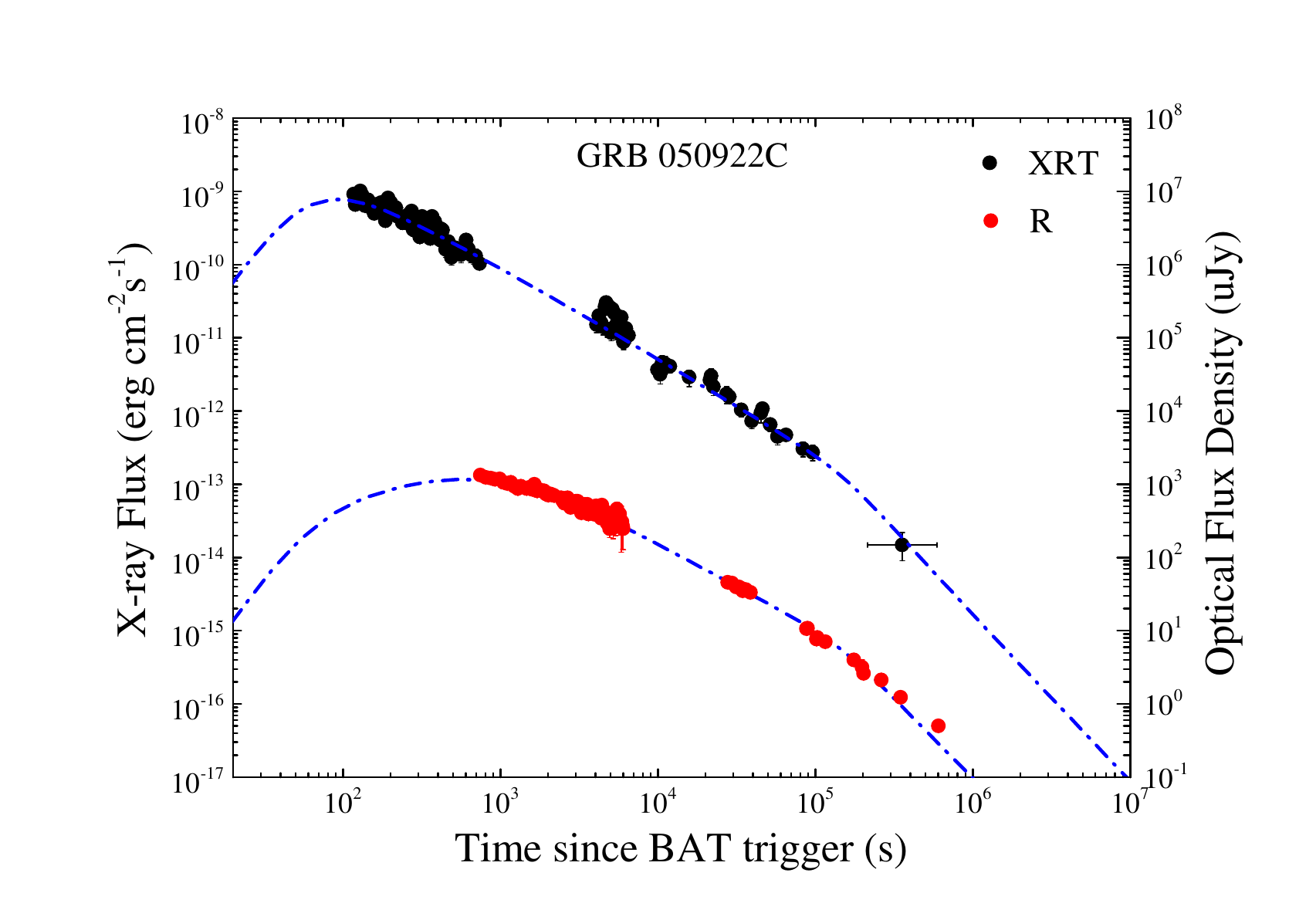}
\includegraphics[angle=0,width=0.30\textwidth]{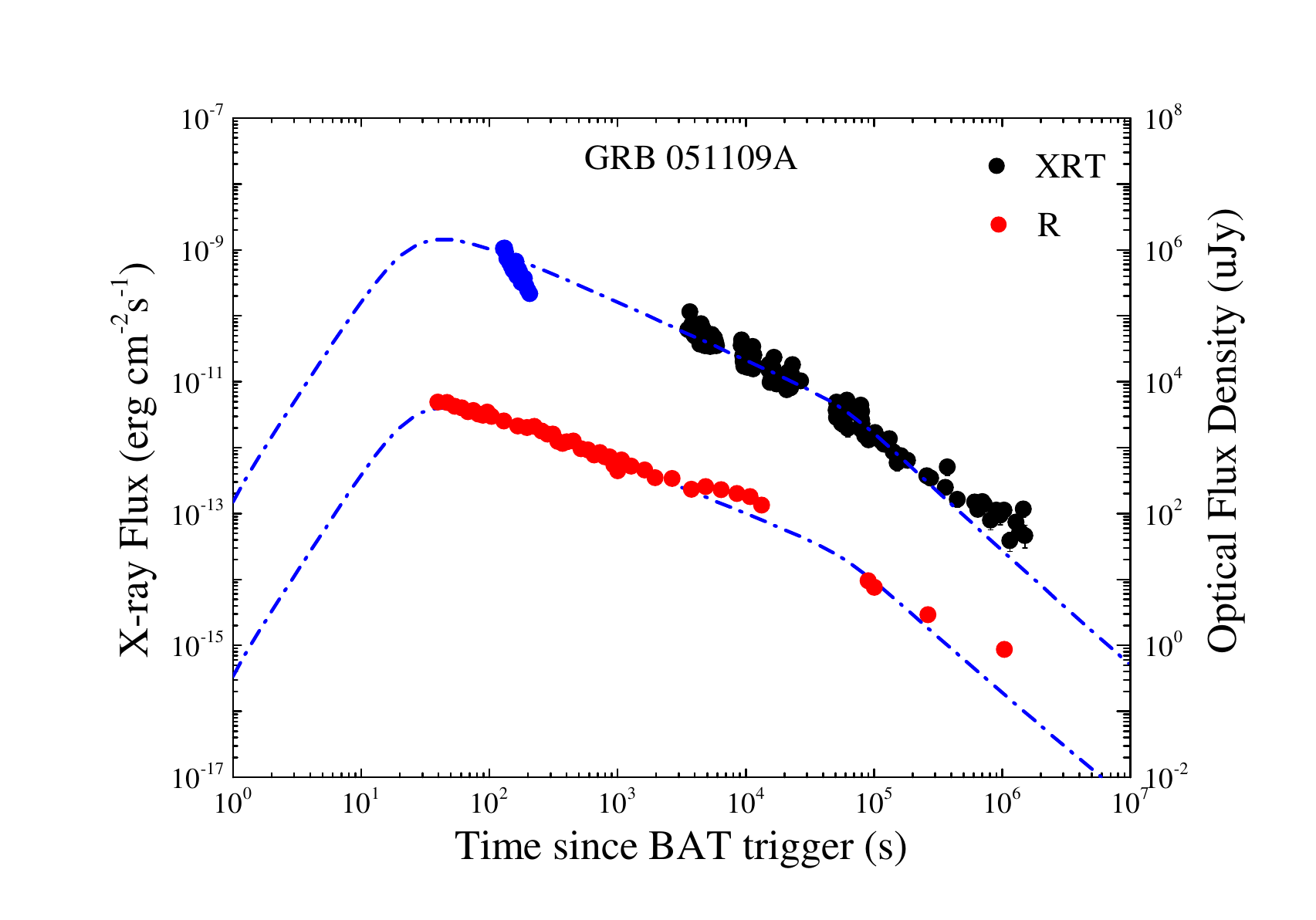}
\includegraphics[angle=0,width=0.30\textwidth]{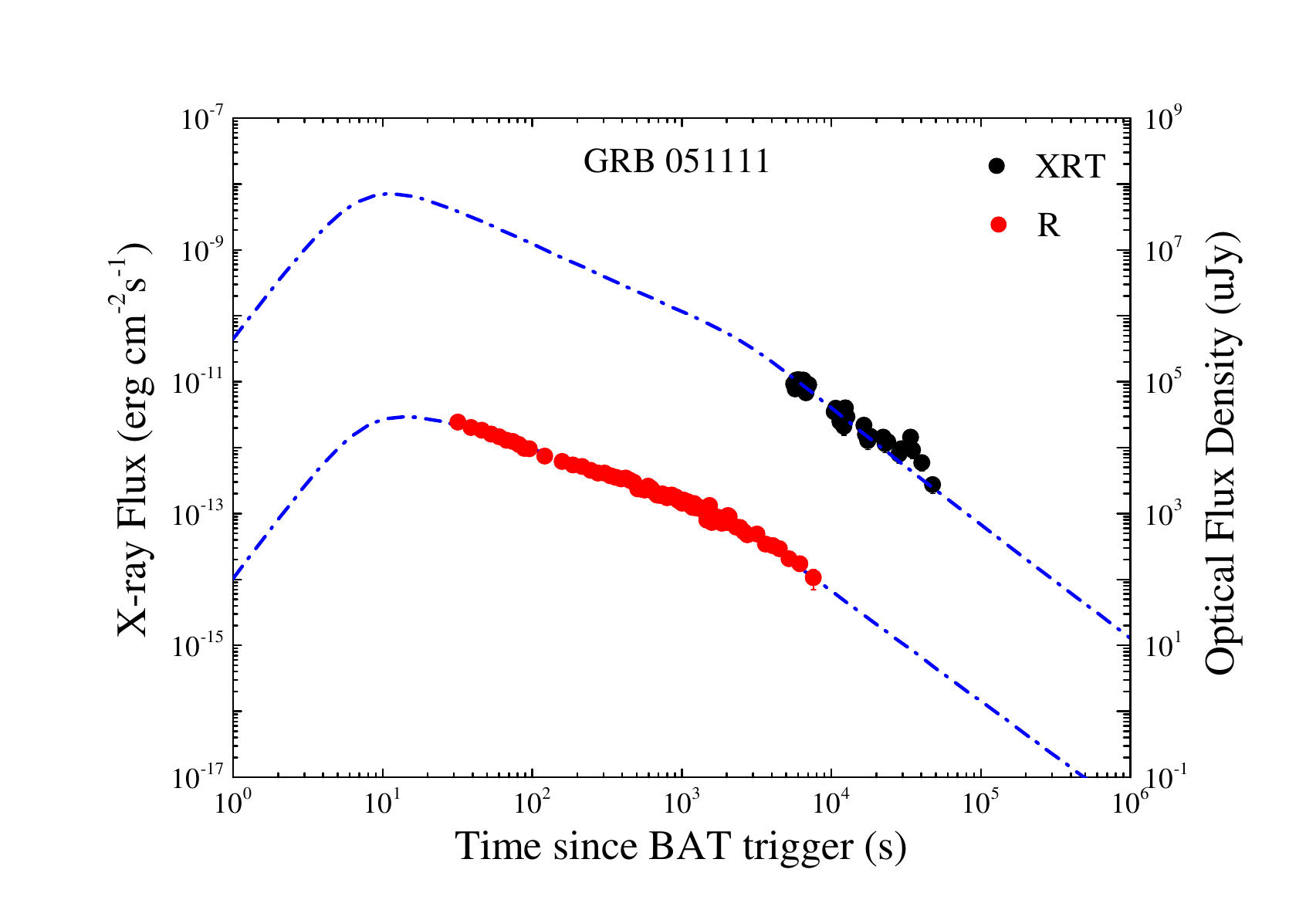}
\includegraphics[angle=0,width=0.30\textwidth]{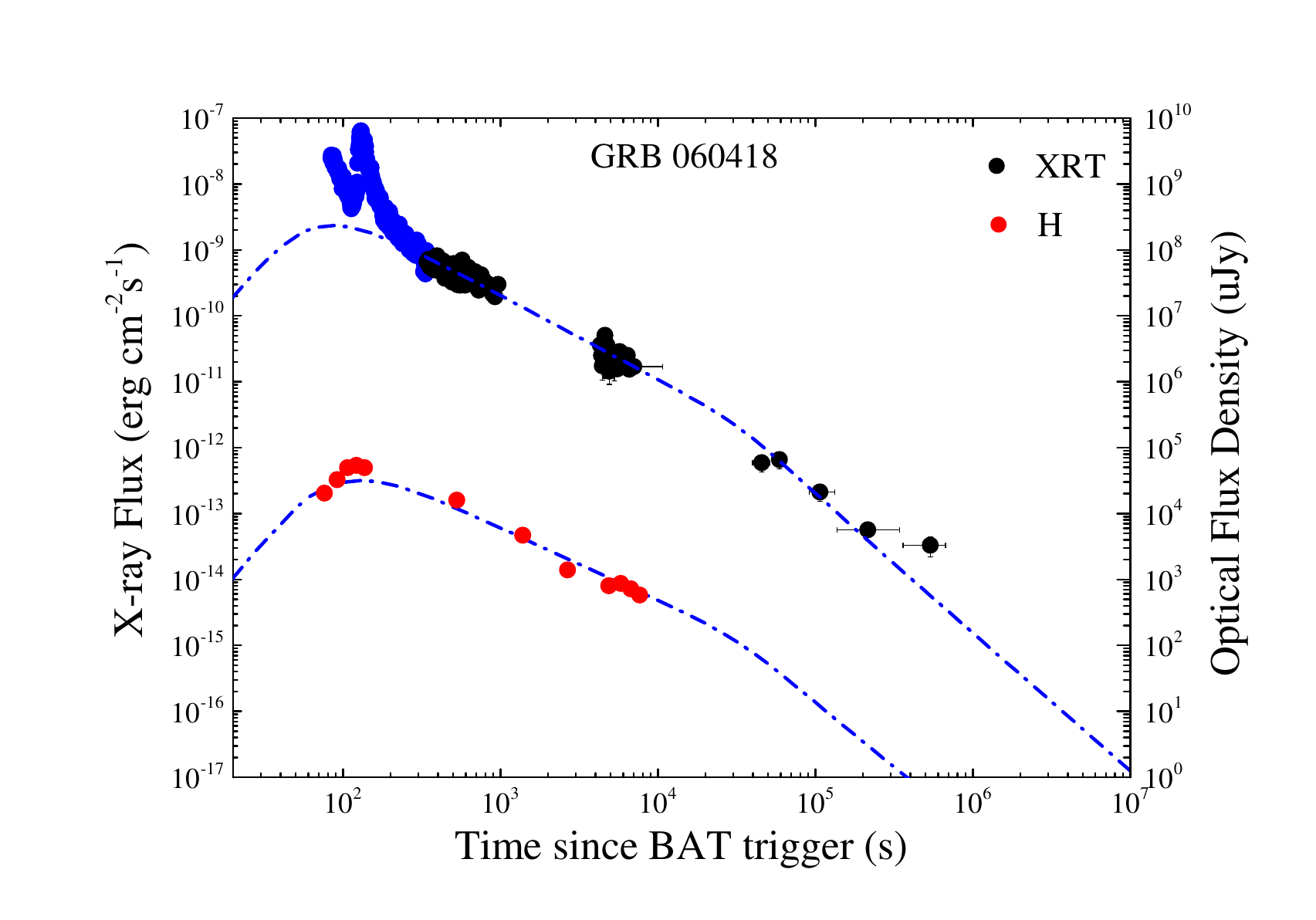}
\includegraphics[angle=0,width=0.30\textwidth]{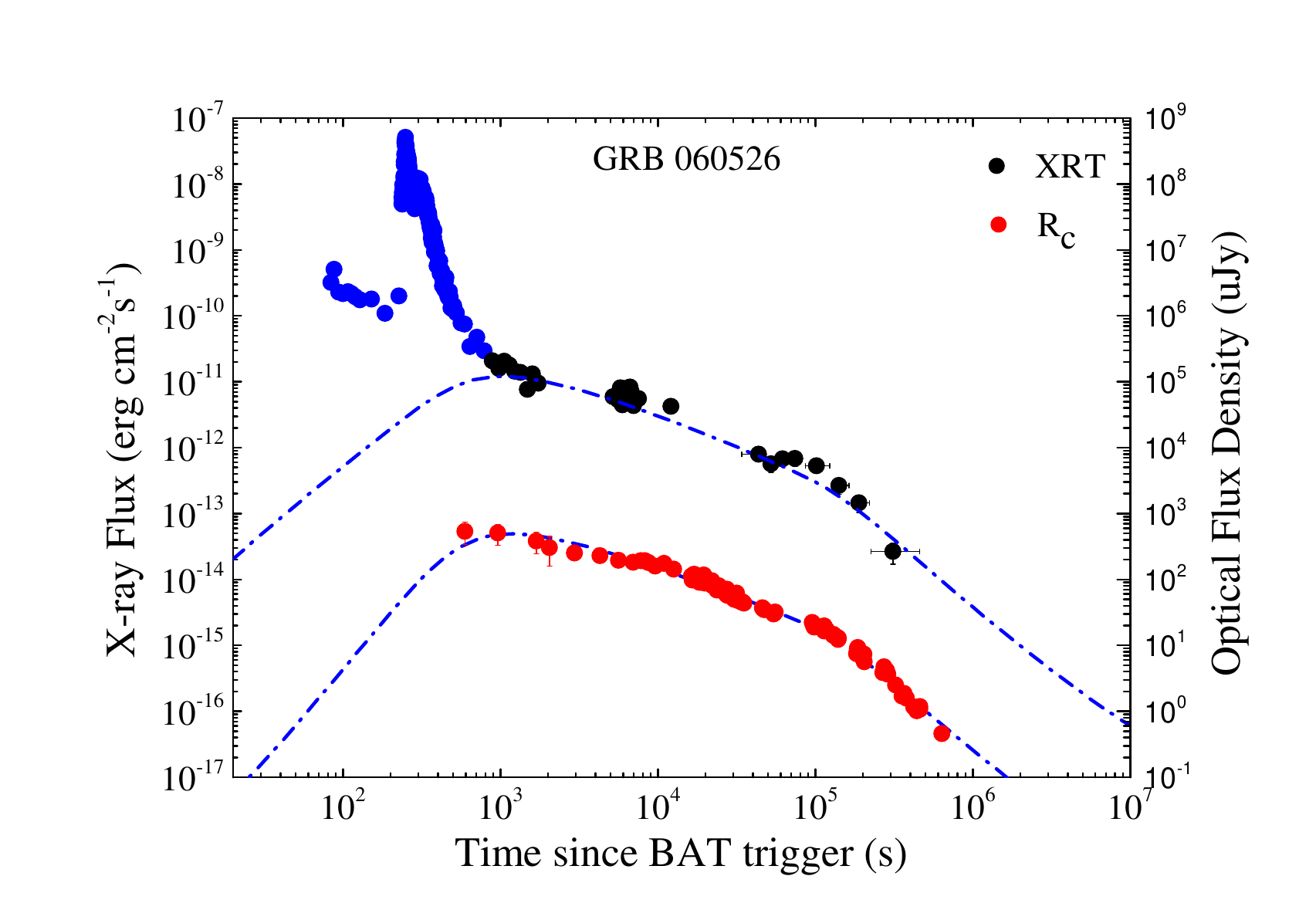}
\includegraphics[angle=0,width=0.30\textwidth]{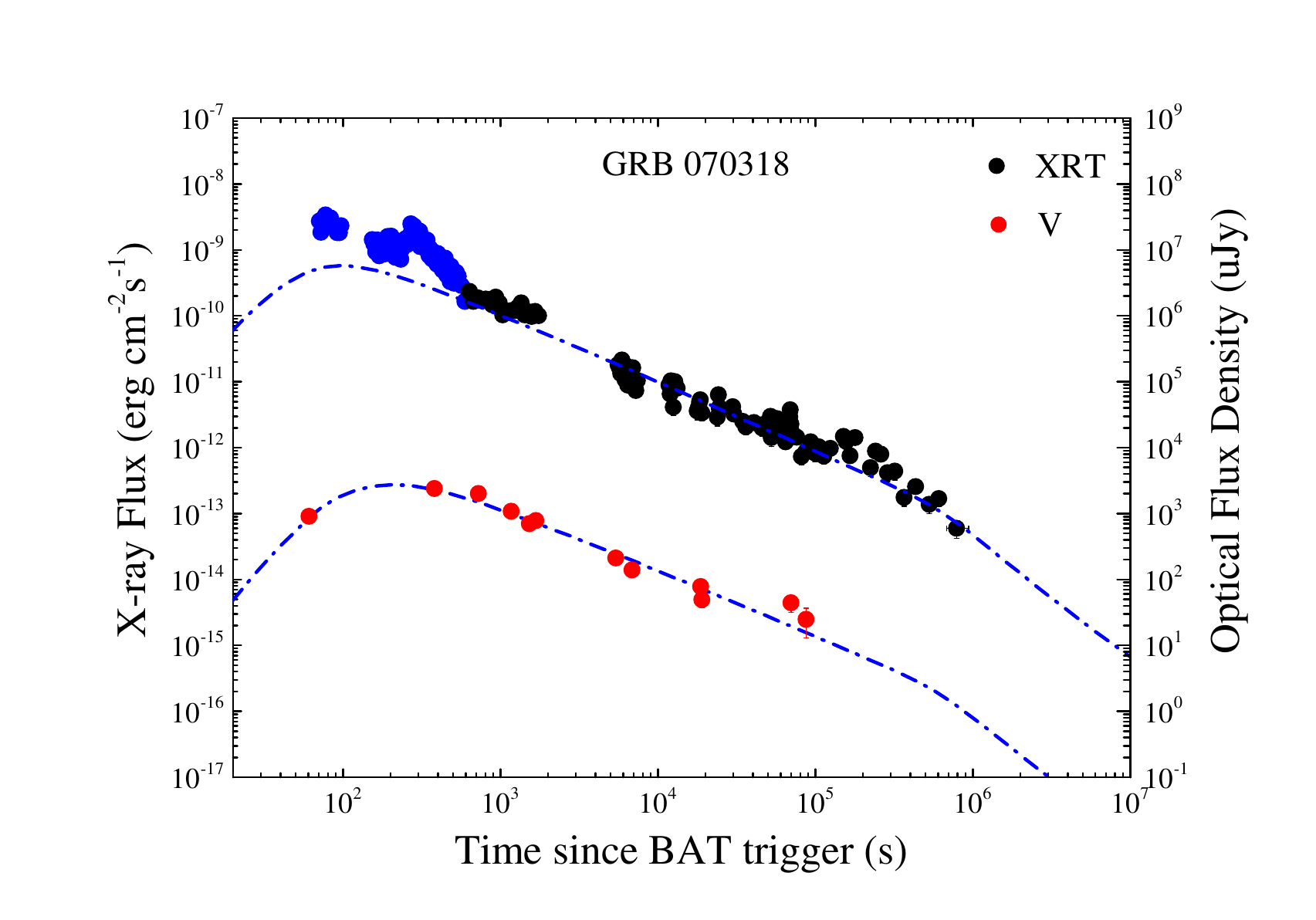}
\includegraphics[angle=0,width=0.30\textwidth]{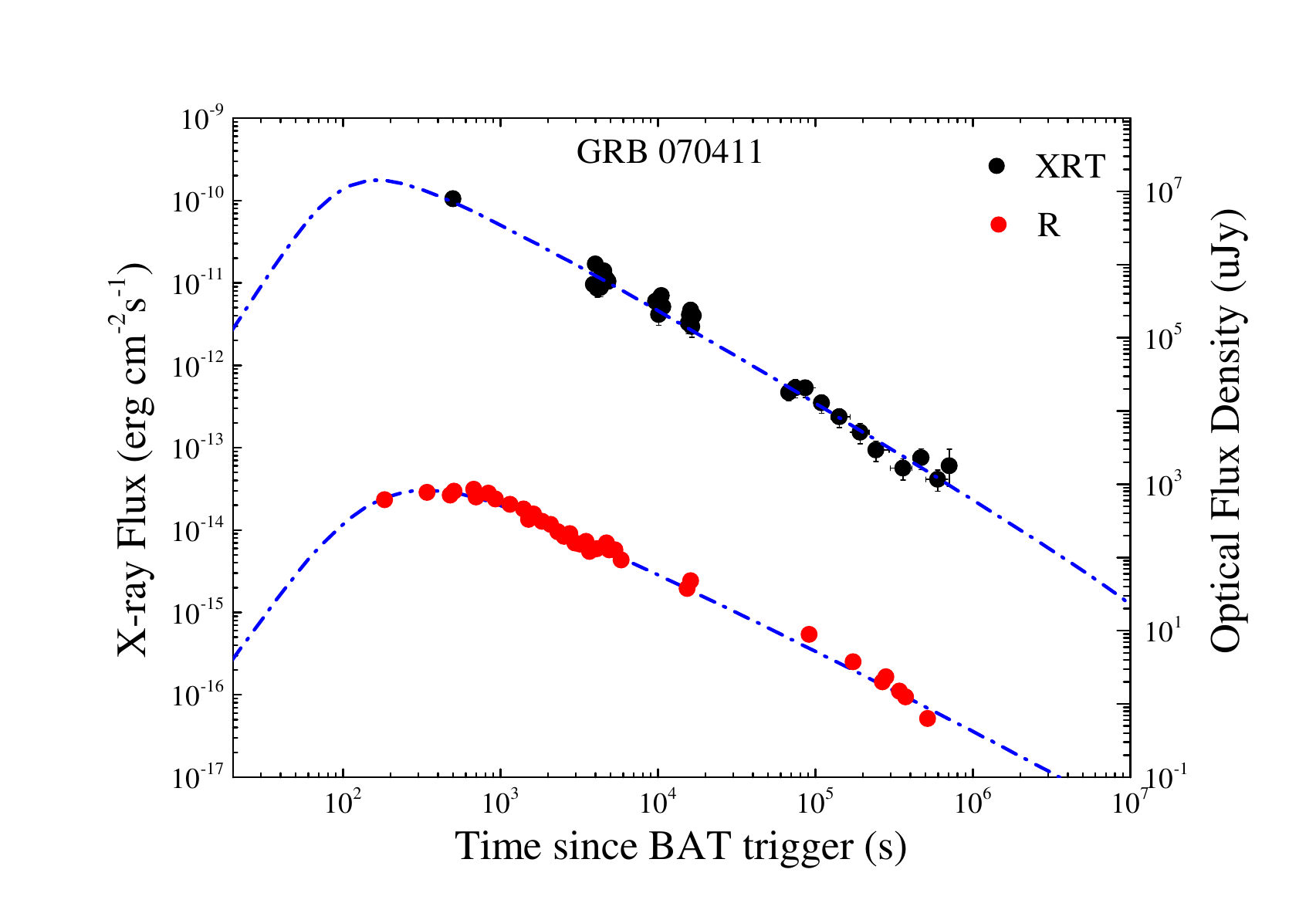}
\includegraphics[angle=0,width=0.30\textwidth]{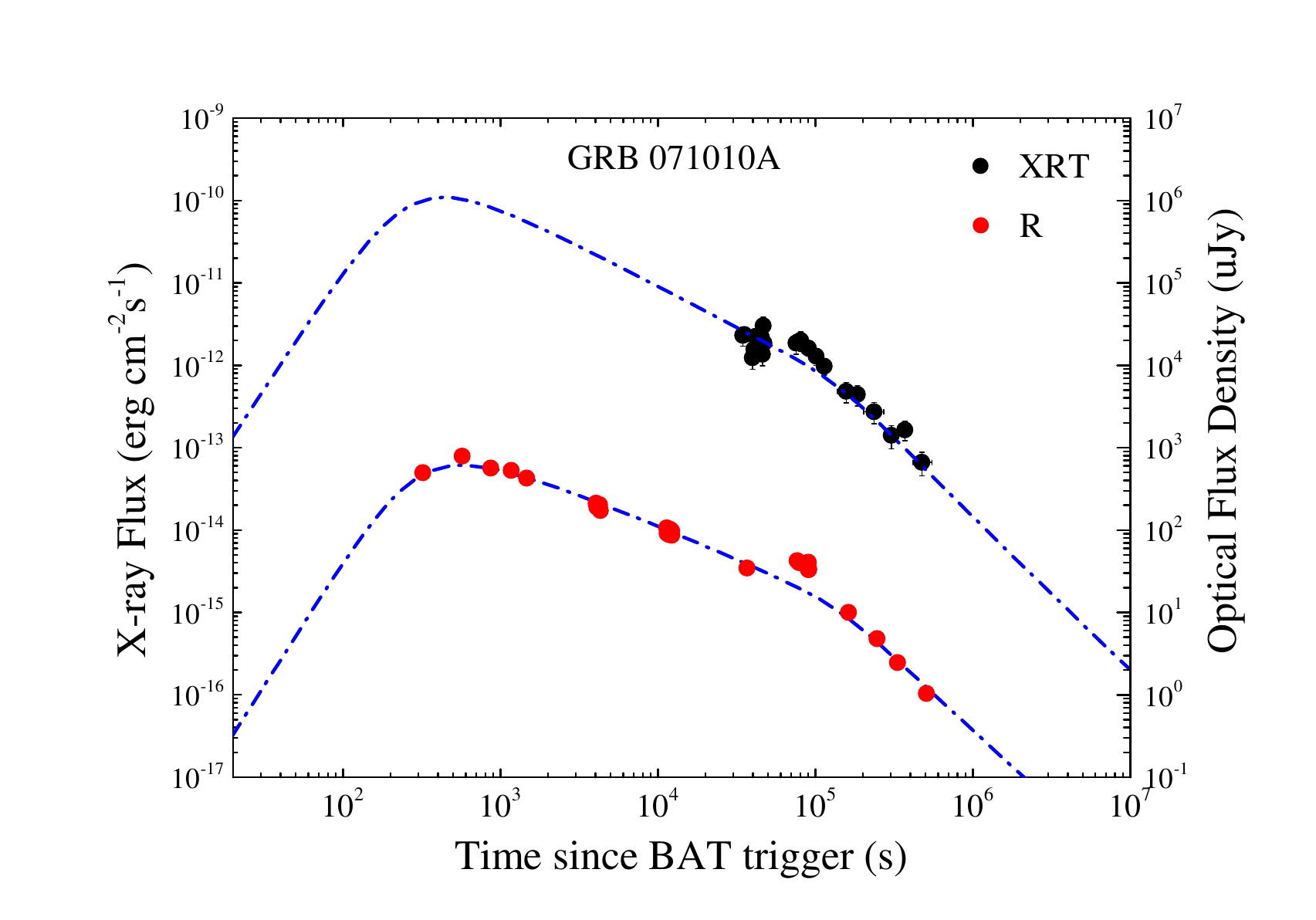}
\includegraphics[angle=0,width=0.30\textwidth]{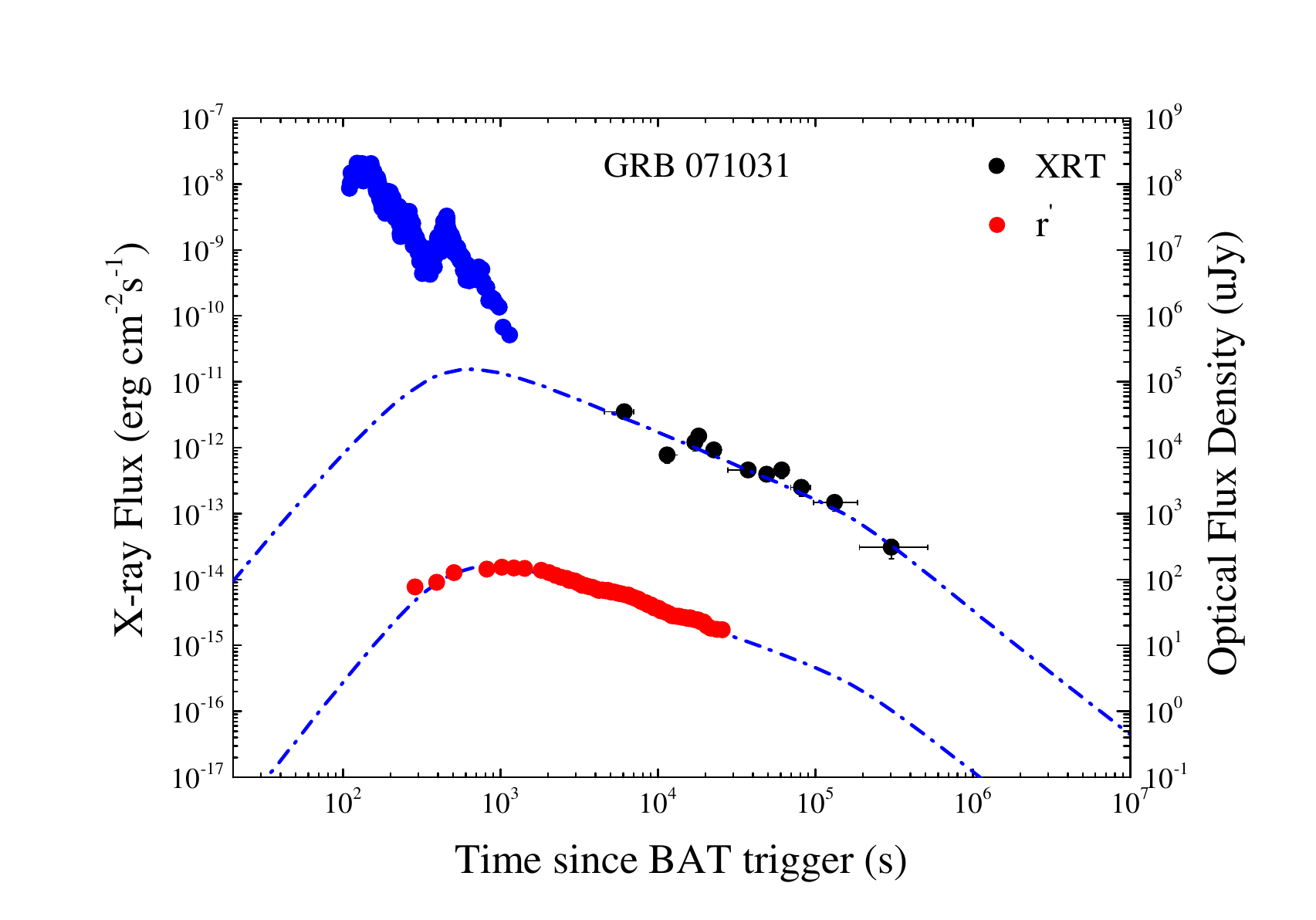}
\includegraphics[angle=0,width=0.30\textwidth]{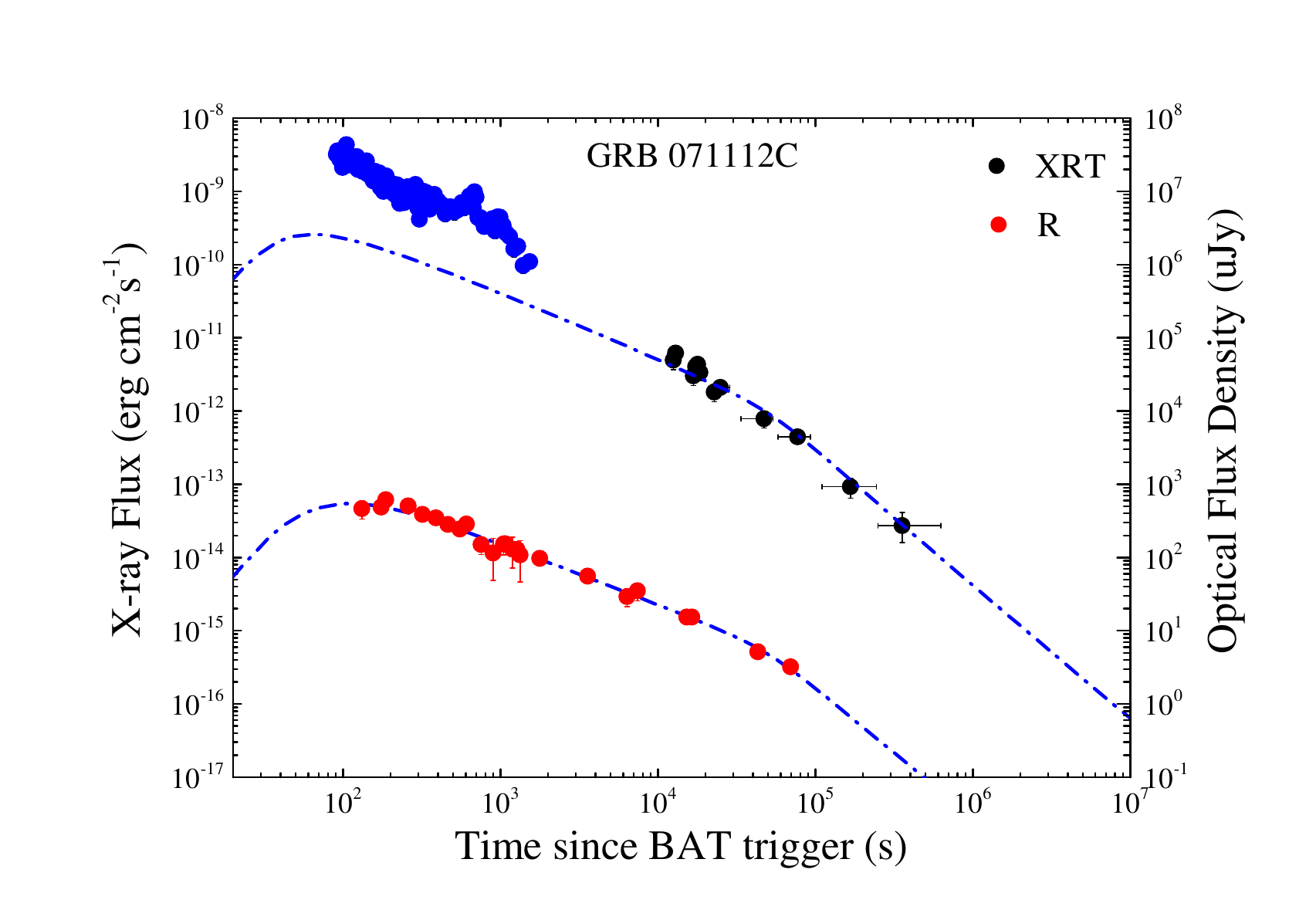}
\includegraphics[angle=0,width=0.30\textwidth]{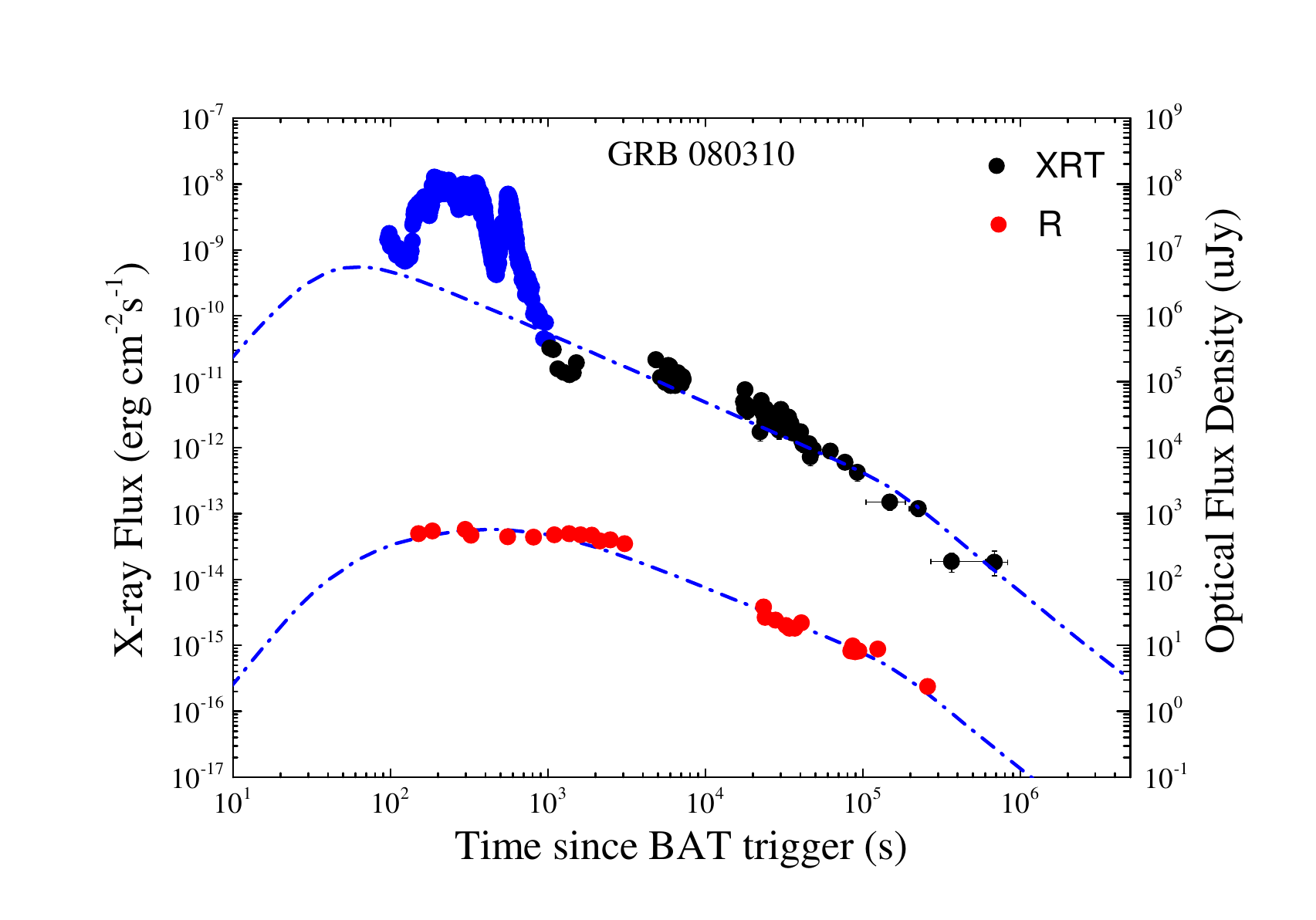}
\includegraphics[angle=0,width=0.30\textwidth]{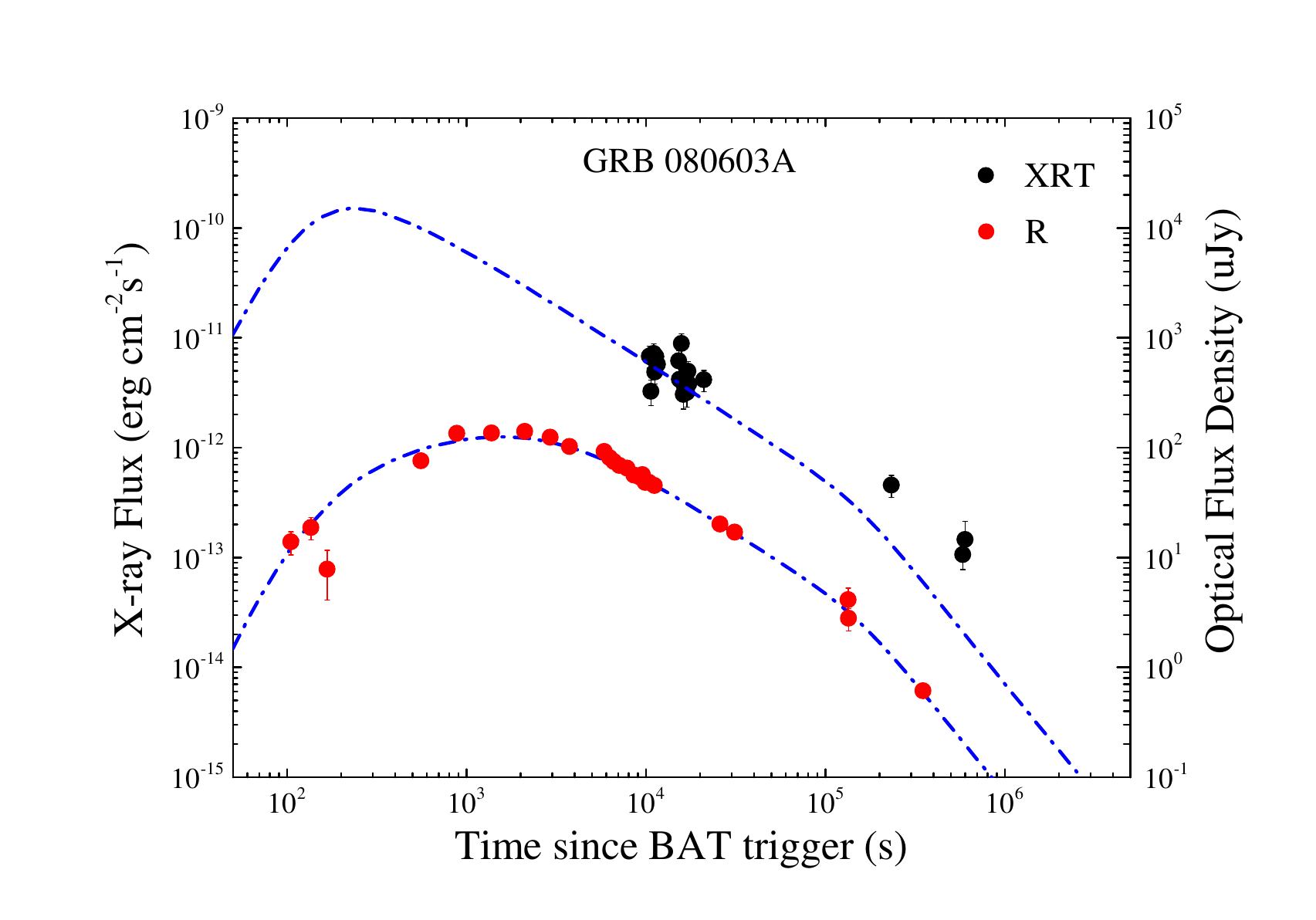}
\includegraphics[angle=0,width=0.30\textwidth]{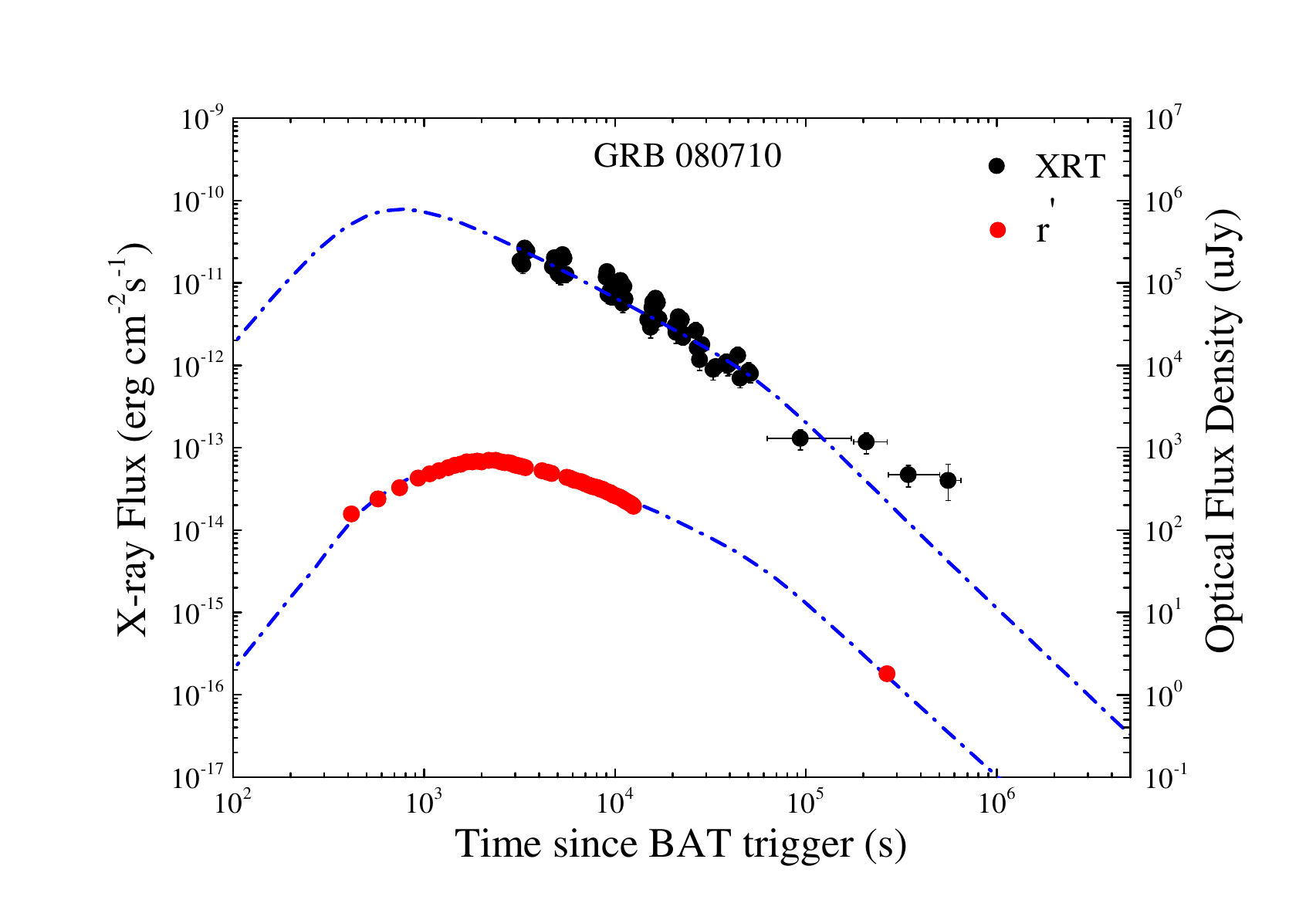}
\includegraphics[angle=0,width=0.30\textwidth]{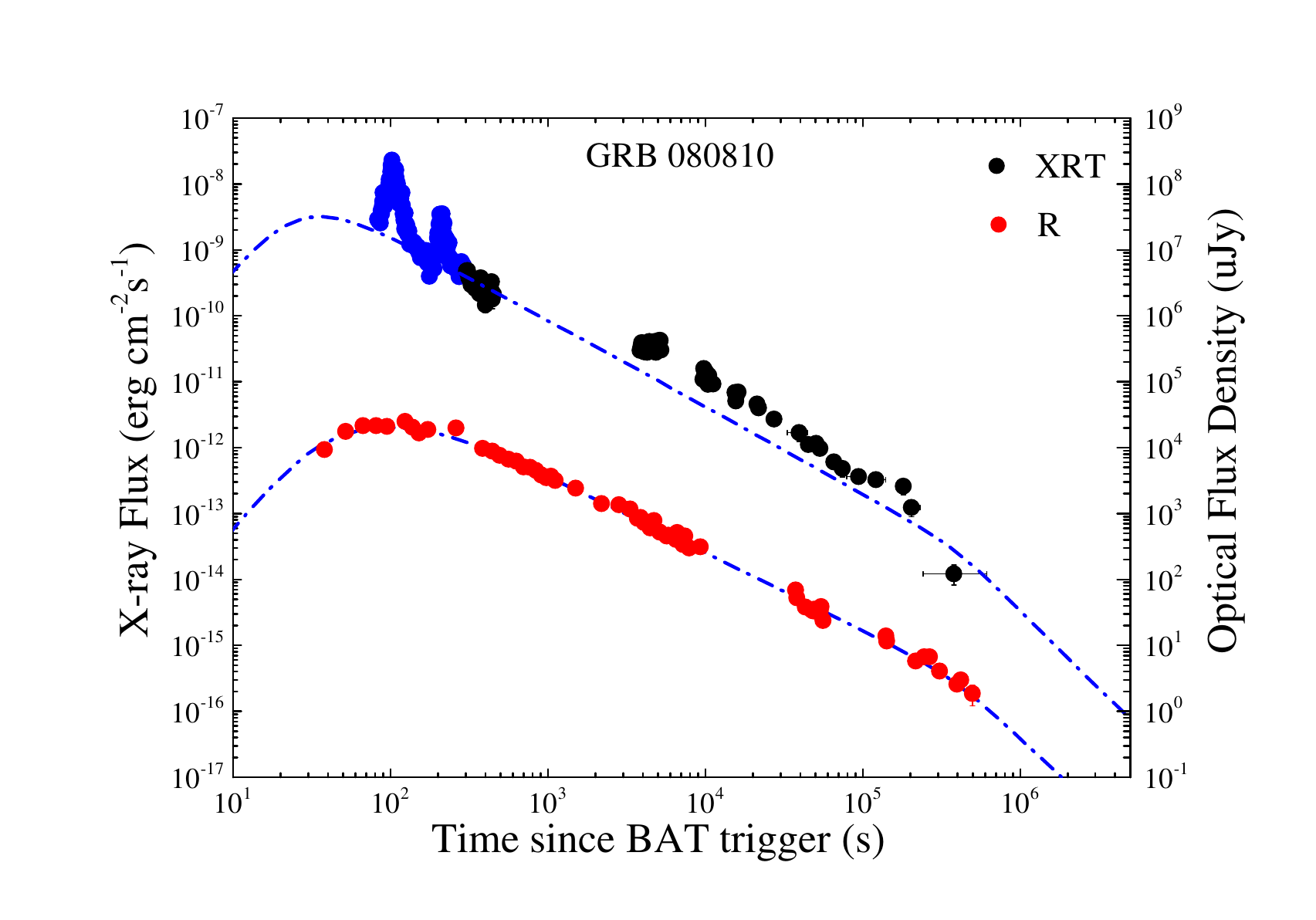}
\caption{Optical and X-ray afterglow light curves (data points) of the bursts in our sample and our theoretical fits (dashed-dotted lines) with the standard FS model by utilizing the {\tt pymultinest} package.}
\label{modelfit}
\end{figure*}

We jointly fit the optical and X-ray light curves with the standard FS model. 
The early X-ray light curves are usually superimposed with bright X-ray flares from late activity of the central engine, an X-ray plateau (X-ray shallow-decayed segment) from the late energy injection of the remnant, and/or a very steep decaying segment being due to the curvature effect of the high-latitude prompt gamma rays from the emitting fireball \citep{Zhang_2006_Fan_ApJ...642..354Z,Nousek_2006_Kouveliotou_ApJ...642..389N}. We identify a flare or a steep decay segment with a criterion of rising and/or decaying slopes being greater than $2$. Inspecting the light curves shown in Figure \ref{modelfit}, one can find that no X-ray plateau is found in these X-ray light curves but they usually start with irregular flares or a steep decaying segment, except for those that began to be observed at a late epoch ($t>10^3$ s post the GRB trigger). Note that the XRT has several operating modes, depending on the brightness of the source being observed. Bright flares and steep decaying segments are usually observed with XRT in the windowed timing (WT) mode. Therefore, We simply exclude these WT data during our fitting. Some flares (steep decaying segments) or partial of them are observed in the photon counting (PC) mode. We also excluded these PC mode data. We marked the data that are excluded in our fit with blue color in Figure \ref{modelfit}. The selected temporal intervals of the X-ray and optical data for our analysis are tabulated in Table~\ref{Tab1}. 

The model has seven free parameters. The {\tt PyMultiNest} Python package~\citep{Buchner_2014_Georgakakis_aap...564A.125B} is utilized for searching the probability distributions of the model parameters in our fits, which is based on a Nested Sampling Monte Carlo algorithm for enhancing computational efficiency in a large parameter space\footnote{\url{https://johannesbuchner.github.io/PyMultiNest/}}. The ranges of prior parameters are listed in Table~\ref{Tab2}.

The log-likelihood functional form for evaluating the goodness of the fit is adopted as  
\begin{equation}
\ln \mathcal{L}=-\frac{1}{2} \sum_i 
\left\{
\left(\frac{X_{i}-\hat{X}_{i}}{\sigma_{X,i}}\right)^2+ 
\ln(2\pi\sigma_{X,i}^2)+ 
w \left[
\left(\frac{O_{i}-\hat{O}_{i}}{\sigma_{O,i}}\right)^2+
\ln(2\pi\sigma_{O,i}^2)
\right]
\right\},
\end{equation}
where $X_i$ ($O_{i}$) and $\sigma_{X,i}$ ($\sigma_{O,i}$) are the $i$th X-ray (optical) data point and its error, $\hat{X_i}$($\hat{O_i}$) is the corresponding model result, and $w$ is a weight factor. 
We increase the weight of the optical data by normally setting $w=10$, because the optical data are better sampled and less contaminated by the emission from late central engine activities (flares and/or shallow decay phases/plateaus) as well as the tail emission of the prompt emission than the X-ray data.

Taking GRB~080310 as an example, Figure~\ref{fig1} shows the derived posterior distributions of the model parameters. The fitting curves are shown in Figure~\ref{modelfit} and the derived model parameters are reported in Table~\ref{Tab3}. One can find that the optical and X-ray afterglow light curves can be fitted by the FS model. The derived jet properties, including the direct model parameters and the inferred quantities, 
are summarized in Table~\ref{Tab4}. 

We discuss the derived jet properties in the next two sections. Note that the distributions of some parameters show a bimodal feature. Since the illustrated bimodal distribution feature depends on the selection of bin size, we test the bimodality with the kernel mixture modelling (KMM) algorithm in our analysis \citep{Ashman_1994_Bird_AJ....108.2348A}. To search for possible correlations among the jet properties, we make a Pearson pair-correlation analysis and a stepwise multiple observable regression analysis. We adopt the ordinary least squares (OLS) bisector algorithm for achieving the best linear fit to derive a possible linear correlation of the data set $\{x,y\}$,
$Y=\beta x+\alpha$,
where $\beta$ and $\alpha$ represent the slope and the intercept, respectively\footnote{The slope and intercept is given by 
$\beta=\left(\beta_{1}+\beta_{2}\right)^{-1}\left[\beta_{1} \beta_{2}-1+\sqrt{\left(1+\beta_{1}^{2}\right)\left(1+\beta_{2}^{2}\right)}\right]$ 
and $\alpha=\bar y -\beta \bar x$,
where $i$ denotes the $i$th data point $(x_i,y_i)$,  $\bar{x} =\frac{1}{n} \sum_{i=1}^{n} x_{i}$,
$\bar{y} =\frac{1}{n} \sum_{i=1}^{n} y_{i}$, $\beta_{1}$ and $\beta_2$ are the $\rm OLS(X|Y)$ and $\rm OLS(Y|X)$, respectively. $\beta_{1}=S_{xy}/S_{xx}$,  $\beta_{2}=S_{yy}/S_{xy}$, where $S_{x x} =\sum_{i=1}^{n}\left(x_{i}-\bar{x}\right)^{2}$, $S_{y y} =\sum_{i=1}^{n}\left(y_{i}-\bar{y}\right)^{2}$,
and $S_{x y} =\sum_{i=1}^{n}\left(x_{i}-\bar{x}\right)\left(y_{i}-\bar{y}\right)$.}. 
For details on the OLS bisector algorithm please refer to \cite{Isobe_1990_Feigelson_apj_v364.p104..104}.

\section{Properties of the Afterglow Jets}
\label{section 4}
\subsection{Jet Energetic, Opening Angle, and Radiation Efficiency}

The distribution of the derived $E_{\rm k, iso}$ values from our analysis 
is illustrated in Figure~\ref{Fig3a}.
The $\log(E_{\rm k,iso})$ distribution displays a plausible bimodal distribution. Our fit with a model of two Gaussian functions yields the two peaks as 
$\log(E_{\rm k,iso, c_1}/{\rm erg})=53.81$ 
and $\log(E_{\rm k,iso, c_2}/{\rm erg})=55.34$. Most of the bursts ($22/30$) are of the high-$E_{\rm k,iso}$ group. We compare the distribution of $\log (E_{\rm k, iso})$ with a sample of GRBs whose optical and X-ray afterglows can be explained with the FS model in the ISM scenario reported by \cite{Wang_2015_Zhang_ApJS..219....9W}. As shown in Figure~\ref{Fig3a}, the $\log (E_{\rm k, iso})$ distributions of the two samples cover almost the same range, but no clear bimodal distribution is found in the sample from \cite{Wang_2015_Zhang_ApJS..219....9W}. Since the apparent bimodal distribution depends on the selection of the bin size, we check the bimodality of the $\log(E_{\rm k,iso})$ distribution derived from our analysis with the KMM test, which yields a {\em p}-value of $0.005$. Hence the bimodality is only marginally claimed.

The jet opening angles are obtained from our fits for $26$ GRBs among the $30$ bursts\footnote{The fit to the lighcurves of GRBs 070411, 120815A, 141221A and 181110A are not required to assign a jet opening angle.}. 
The distribution of the derived $\theta_j$ values is shown in Figure~\ref{Fig3c}. It peaks at $\log (\theta_{\rm j}/\rm rad)=-1.4$ ($\theta_{\rm j}=2.3^{\circ}$) and the central value $\log(\theta_{\rm j, c}/\rm rad)=-1.51$ ($\theta_{\rm j, c}=1.8^{\circ}$). 
The peak value is almost consistent with that reported by \cite{Wang_2015_Zhang_ApJS..219....9W}. 
We calculate the jet-corrected kinetic energy as $E_{\rm k,j}=E_{\rm k,iso}(1-\cos \theta_{\rm j})$, 
and show its distribution in Figure~\ref{Fig3d}. 
It is found that $E_{\rm k,j}\in\{10^{50}\sim 10^{53}\}$~erg and the $\log (E_{\rm k,j})$ distribution can be fit with a Gaussian function, 
which yields a central value of $\log(E_{\rm k, j,c}/{\rm erg})=51.78$ and a standard deviation ($\sigma$) as $0.54$. 
The $R^2$ of the fit is $0.98$.

Broadband observations with Fermi Gamma-ray Burst Monitor (GBM, 8~\rm keV-40~\rm MeV), Konus-Wind ($\sim 20~\rm keV-20~\rm MeV$ for the triggered mode and $\sim 20~\rm keV-1.5~\rm MeV$ for the waiting mode;  \citealp{Tsvetkova_2021_Frederiks_ApJ...908...83T}), 
or Suzaku Wide-band All-sky Monitor (WAM, $50~\rm keV-5~MeV$, \citealp{Krimm_2009_Yamaoka_ApJ...704.1405K}) telescopes are available for $19$ out of the $30$ bursts\footnote{GRBs that only detected by 
{\em Swift}/BAT instrument are excluded. They are GRBs 060526, 070411, 071010A, 071031, 071112C, 080310, 080603, 080710, 090313, 120815A, and 170519A.}. 
Their isotropic prompt gamma-ray energy ($E_{\rm \gamma, iso}$) can be robustly 
estimated with the observations of these telescopes. 
We calculate their $E_{\rm \gamma, iso}$ in the $1-10^{4}$~keV band and show $\log E_{\rm \gamma, iso}$ as a function of $\log E_{\rm k, iso}$ in Figure~\ref{Fig3f}. 
It is found that $E_{\rm \gamma, iso}$ is not correlated with $E_{\rm k, iso}$ and the $\log E_{\rm \gamma, iso}$ distribution does not show a bimodal feature as that seen in the $\log E_{\rm k, iso}$ distribution. 
The $\log E_{\rm \gamma, iso}$ distribution is clustered around $\log(E_{\rm \gamma,iso}/{\rm erg})\sim 53$. Our fit with a Gaussian function yields $\log(E_{\rm \gamma,iso, c}/{\rm erg})=52.96$ ($\sigma=0.37$). Deriving the jet gamma-ray energy ($E_{\rm \gamma, j}$) by making geometrical correction with the derived jet opening angle, it is found that $E_{\rm \gamma, j}\in \{10^{48}, 7.2\times 10^{50}\}$~erg. 
The $\log E_{\rm \gamma, j}$ distribution can be fitted with a Gaussian function, yielding $\log E_{\rm \gamma, j, c}/{\rm erg}=49.47$ ($\sigma$=0.87). The $R^2$ of the fit is $0.99$. The normal distributions of $\log(E_{\rm \gamma,j})$ and $\log(E_{\rm k,j})$ are shown in Figure~\ref{Fig3d}. We should note that the distributions are similar to those reported by~\cite{Frail_2001_Kulkarni_ApJ...562L..55F} and \cite{Berger_2003_Kulkarni_ApJ...590..379B}, 
who proposed a standard energy reservoir among GRB jets. 

The radiation efficiency gives insights into the radiation physics and composition of the GRB fireball~\citep[e.g.,][]{Piran_1999_PhR...314..575P}. 
The hot fireball model predicts that the radiation efficiency is lower than $10\%$~\citep[e.g.,][]{Kumar_1999_ApJ...523L.113K, Panaitescu_1999_Spada_ApJ...522L.105P, Fan_2006_Piran_MNRAS.369..197F}. 
The inferred high $\eta_\gamma$ ($>10\%$ even up to $90\%$;~\citealp[e.g.,][]{Zhang_2007_Liang_ApJ...655..989Z, Wang_2015_Zhang_ApJS..219....9W}) using the X-ray and GeV gamma-ray afterglow data is a great challenge to the standard external shock model~\citep[e.g.,][]{Ioka_2006_Toma_AA...458....7I, Zhang_2007_Liang_ApJ...655..989Z}. 
We calculate the GRB radiative efficiency by $\eta_{\gamma}$= $E_{\rm \gamma,iso}$/$(E_{\rm \gamma, iso}$+$E_{\rm k, iso})$.
Figure \ref{Fig3e} shows the distribution of $\log (\eta_{\gamma}/\%)$, which can be fitted with a Gaussian function. 
One can find that the $\log (\eta_{\gamma}/\%)$ values range from $-1.4$ to $1$ ($\eta_\gamma=0.04\%-6.73\%$).
The median and mode of the $\log (\eta_{\gamma}/\%)$ distribution are $-0.27$ and $-0.35$ (corresponding to $\eta_\gamma=0.54\%$ and $0.42\%$). 
The radiative efficiencies of GRBs~051109, 051111, 120729A and 140629A are lower than or equal to $0.1\%$. We also mark the lines for $\eta_\gamma=0.04\%$, $1\%$, and $10\%$ in Figure~\ref{Fig3f}. 
The derived low radiative efficiency for the bursts in our sample favors the viewpoint that the jets of these GRBs are matter-dominated.

\subsection{The Fireball Initial Lorentz Factor, Medium Density, and Deceleration Radius}
In the initial stage, the fireball evolves through a radiation-dominated acceleration phase to a matter-dominated ``coasting" phase. During the coasting phase, the initial Lorentz factor ($\Gamma_0$) remains approximately constant until the fireball accumulates a significant amount of mass from the surrounding medium and begins to decelerate. The observed afterglow onset peak marks the deceleration of the GRB fireball in the thin-shell scenario of the standard fireball model~\citep[e.g.,][]{Sari_1999_Piran_ApJ...520..641S, Liang_2010_Yi_ApJ...725.2209L, Liang_2013_Gao_ApJ...774...13L}. Hence, $\Gamma_0$ is sensitive to the peak time of the afterglow light curves. The derived $\Gamma_0$ values of most bursts ($26/30$) are greater than $100$, being consistent with the expectation of the GRB fireball model for avoiding the compactness problem~\citep[e.g.,][]{Lithwick_2001_Sari_ApJ...555..540L, Abdo_2009_Ackermann_Sci...323.1688A, Zhang_2022_Xin_ApJ...941...63Z}. 
The $\log(\Gamma_0)$ distribution derived from our fits is shown in Figure~\ref{Fig4a}. Although the log($\Gamma_0$) distribution can be roughly fitted with a Gaussian function, i.e., $\log \Gamma_{0,c}=2.63$ ($\sigma=0.27$), it shows a sharp cut at the right side. 
This would be due to our sample selection effect. The GRBs included in our analysis are selected with the detection of the optical afterglow onset peak. The deceleration time of extremely bright GRBs occurs very early, which can often be missed by observation or affected by the prompt optical emission and the optical emission of the reverse shock~\citep[e.g.,][]{Racusin_2008_Karpov_Natur.455..183R, Xin_2023_Han_NatAs...7..724X}. 
In fact, the Lorentz factors of some GRBs estimated with the Fermi/LAT observations are greater than $1000$, which is essential to prevent high-energy photons from undergoing annihilation through the creation of $e^{\pm}$ pairs~\citep[e.g.,][]{Abdo_2009_Ackermann_Sci...323.1688A, Abdo_2009_Ackermann_Natur.462..331A, Abdo_2009_Ackermann_ApJ...706L.138A, Ioka_2010_PThPh.124..667I}. 

\cite{Ghirlanda_2018_Nappo_A_A...609A.112G} calculated the $\Gamma_0$ values of the bursts with the detection of the optical onset bump in both the ISM and wind medium scenarios. In the ISM scenario, they adopted $n_{0}=1~\rm cm^{-3}$ and estimated the $E_{\rm k,iso}$ value with $E_{\rm k,iso}=E_{\rm \gamma, iso}/\eta_\gamma$, where $\eta_\gamma$ is assumed to be a universal value of $20\%$ for all bursts. We compare the $\log(\Gamma_0)$ distributions between ours and that calculated with the data reported by \cite{Ghirlanda_2018_Nappo_A_A...609A.112G} in Figure~\ref{Fig4a}. It is found that the central value $\Gamma_0$ derived from our analysis is larger than that from \cite{Ghirlanda_2018_Nappo_A_A...609A.112G} by a factor of $\sim 2$, i.e.,    
$\Gamma_0=426^{+794}_{-229}$ ($1\sigma$ range in log-scale) vs. $\Gamma_0=204^{+389}_{-107}$. The discrepancy is mostly due to the difference of $n_0$ and $\eta_\gamma$ values since $\Gamma_0\propto E_{\rm k, iso}^{1/8} n^{1/8} t^{-3/8}_{\rm peak,z}$. As shown in Figure \ref{Fig3}, the $\eta_\gamma$ values derived from our analysis are mostly in the range of $0.04\% \sim 10\%$ with a typical value of $\sim 0.4\%$. It is smaller than that adopted by \cite{Ghirlanda_2018_Nappo_A_A...609A.112G} by a factor of 50. This leads to an underestimation of the $\Gamma_0$ value by a factor of $50^{1/8}\sim 1.6$. This can reasonably explain the discrepancy between our results and those reported by \cite{Ghirlanda_2018_Nappo_A_A...609A.112G}.

Furthermore, we calculate the baryon mass loading in the GRB ejecta by 
$M_{\rm fb}=E_{\rm k,j}/(\Gamma_{0}c^{2})$. 
Figure~\ref{Fig4b} shows the $\log (M_{\rm fb}/M_{\odot})$ distribution, which covers from $6.54\times10^{-7}M_{\odot}$ to $2.02\times10^{-5}M_{\odot}$ with a mode of $M_{\rm fb,p}=1.48\times10^{-5} M_{\odot}$. \cite{Ghirlanda_2018_Nappo_A_A...609A.112G} derived the $\log (M_{\rm fb}/M_{\odot})$ distribution for a sample of GRBs with the detection of the optical afterglow onset bump. The $\log (M_{\rm fb}/M_{\odot})$ distribution derived from their analysis mainly covers the range of $10^{-7}\sim 10^{-4} M_{\odot}$, being averagely smaller than that derived from our analysis by a factor of 4, as shown in Figure \ref{Fig4b}. We should note that \cite{Ghirlanda_2018_Nappo_A_A...609A.112G} estimated the $E_{\rm k,j}$ values with $E_{\rm k,j}=E_{\rm \gamma, iso}(1-\rm cos\theta_j)/\eta_\gamma$ by taking $\theta_j=5^{\circ}$ and $\eta_\gamma=20\%$ for all bursts. The $E_{\rm k,j}$ and $\eta_{\gamma}$ values in our analysis are derived from theoretical modeling. This would result in the difference between ours and that reported by \cite{Ghirlanda_2018_Nappo_A_A...609A.112G}.

The derived medium density among the bursts ranges from $10^{-3}$ to $10^{3}$ cm$^{-3}$, as shown in Figure \ref{Fig4c}. 
Its distribution shows a clear bimodal feature, with a separation at $\log (n_{0}/\rm cm^{-3})\sim -0.1$. 
The low-density group ($12/30$ of bursts) narrowly cluster at $\log (n_{0}/\rm cm^{-3})=-2.5$, and the high-density group ($18/30$ of bursts) normally peaks at around $\log (n_{0}/\rm cm^{-3})=1$. 
Our fit with a bimodal model yields $\log (n_{0,c_1}/\rm cm^{-3})=-2.5$ ($\sigma=0.5$) and $\log (n_{0,c_2}/\rm cm^{-3})=1.0$ ($\sigma=1.3$). The bimodality test with the KMM algorithm gives a {\em p}-value of 0.037.
The bimodality $\log n_0$ distribution suggests that two groups of GRBs should be in difference environments (e.g., \citealp{Fruchter_2006_Levan_Natur.441..463F}). The GRBs of the high-density (or low-density) group may be close to (or far away from) the star formation regions of their host galaxies. 

The fireball is decelerated by the surrounding medium. The deceleration radius is defined as
\begin{equation}
R_{\rm dec}=\left(\frac{3E_{\rm k,iso}}
{2\pi{\hat\gamma}\Gamma_{0}^{2}n_{0} m_{p}c^{2}}\right)^{1/3}
\approx\left(6.2\times10^{16}\rm cm\right)
E_{\rm k,iso,52}^{1/3}\Gamma_{0,2}^{-2/3}n_{0}^{-1/3}.
\label{12}
\end{equation}
The derived $R_{\rm dec}$ distribution is shown in Figure~\ref{Fig4d}. It is found that the $R_{\rm dec}$ distribution covers the range of $2.22\times10^{16} - 3.57\times10^{18}~\rm cm$ and it is well fitted with a log-normal function, i.e., $\log (R_{\rm dec, 17, c}/{\rm cm})=0.59$ ($\sigma=0.56$).

It is interesting that the derived $R_{\rm dec}$ distribution is log-normal, although the $n_{0}$ distributions are tentatively bimodal. We suspect that this results from the degeneration of $E_{\rm k, iso}$ and $n_0$ in our light curve fitting. We first check whether $E_{\rm k, iso}$ and $n_{0}$ have degenerated in the MCMC fit for individual GRBs, but do not find clear degeneration feature that shows as a narrow band in the probability contour in the $\log E_{\rm k, iso}-\log n_{0}$ plane. Second, we examine the correlations of the two parameters among the bursts, but we do not find any correlation, as shown in Figure \ref{Fig7a}. Therefore, the bimodal distribution of $n_{0}$ should be not due to the degeneration of $E_{\rm k, iso}$ and $n_{0}$. 
Third, the log-normal distribution of $R_{\rm dec}$ suggests a correlation among  $E_{\rm k, iso}$, $n_0$, and $\Gamma_0$ based on Eq. \ref{12}. We further test the correlation between the ratio of $\varepsilon\equiv E_{\rm k,iso}/n_{0}$ and $\Gamma^{2}_{0,2}$, where $\varepsilon$ is in sense of the isotropic kinetic energy per particle density. Figure \ref{Fig7b} shows $\varepsilon$ as a function of $\Gamma^{2}_{0,2}$. It is found that the two quantities are tightly correlated. The multiple linear regression analysis gives a correlation coefficient of $0.88$ and a chance probability of $p<10^{-4}$. The best fit with the OLS bisector algorithm yields a relation of $\varepsilon= (3.08\pm 0.24)\log(\Gamma_{0,2}^{2})\pm(0.31\pm 0.31)$. The dispersion of the relation is $\delta=1.19$. This correlation infers a quasi-universal $R_{\rm dec}$ (a log-normal distribution of $R_{\rm dec}$) since $R_{\rm dec}\propto \varepsilon{1/3}\Gamma^{-2/3}_{0,2}$.

\subsection{The jet energy equipartition and composition}
The internal energy of the fireball is assumed to be carried by the electrons, protons, and magnetic field with the electron equipartition parameter $\epsilon_e$ and the magnetic equipartition parameter $\epsilon_B$~\citep[e.g.,][]{Zhang_2011_CRPhy..12..206Z}.
The distribution of log($\epsilon_e$) for the bursts in our sample is shown in Figure~\ref{Fig5a}. 
One can observe that the log($\epsilon_e$) distribution illustrates two peaks at $-2.39$ ($\sigma=0.36$) and $-0.95$ ($\sigma=0.39$), which are derived from our bimodal function fit ($R^{2}=0.88$). 
The {\em p}-value of the KMM test is $0.014$, and the inferred two peaks are at $-2.40$ and $-0.94$.   
The broad log($\epsilon_e$) distribution might suggest the efficiency of the electron acceleration among bursts is dramatically different. 
The low-$\epsilon_e$ group likely suggests the existence of a scenario in which the electrons are unable to be efficiently accelerated by the GRB blast wave or completely radiative even when electrons are in the fast-cooling regime~\citep[e.g.,][]{Gao_2013_Lei_NewAR..57..141G}. 
This may lead to a low-value $\eta_{\gamma}$~\citep[e.g.,][]{Wang_2015_Zhang_ApJS..219....9W}. We examine the correlation between $\epsilon_{e}$ and $\eta_{\gamma}$, as shown in Figure \ref{Fig5_6}. Our analysis with the OLS bisector algorithm gives a relation of $\log(\eta_{\gamma}) =(0.93\pm 0.15)\log(\epsilon_{e})+(1.66\pm 0.35)$, and the Pearson correlation coefficient is $r=0.64$. The dispersion of this relation is $\delta=0.50$. This relation suggests that fireballs
with a lower $\epsilon_{e}$ value tend to have a smaller gamma-ray radiation efficiency $\eta_{\gamma}$.

The $\log(\epsilon_B$) distribution spans a wide range from $-7$ to $-1.5$ with a mode of $-3.32$ and a median of $\log(\epsilon_{B,c})=-4.09\pm 0.14$, as shown in Figure~\ref{Fig5b}. This statistically agrees with that reported by the literature \citep{Santana_2014_Barniol_Duran_ApJ...785...29S,Japelj_2014_Kopac_ApJ...785...84J,
Santana_2014_Barniol_Duran_ApJ...785...29S,Gao_2015_Wang_ApJ...810..160G,
Wang_2015_Zhang_ApJS..219....9W}. 
Adopting a ratio $\sigma_{B}=\epsilon_{B}/\epsilon_{e}$ as a measurement of the fireball magnetization, we find that the $\log \sigma_{B}$ values range in $[-5.5,0.5]$ and its distribution illustrates as
a plausible bimodal distribution that peaks at $\log \sigma_{B,c_1}=-4.3\pm 0.3$ and $\log \sigma_{B,c_2}=-1.3\pm 0.1$, 
as shown in Figure \ref{Fig5c}.
The {\em p}-value of the KMM test is 0.09. 
We also calculate the initial magnetic field strength of the fireball with 
\begin{equation}
 B_{0}=(32\pi m_p\epsilon_B n_{0})^{1/2}\Gamma_0 c.
\label{13}
\end{equation}
The distribution of $\log (B_0/{\rm Gs})$ is present in Figure~\ref{Fig5e}. It spans from $-1.6$ to $1.2$ and can be fitted with a Gaussian function, i.e., $\log (B_{0, c}/{\rm Gs})=-0.17$ and $\sigma$ $=0.54$.    

The electrons is assumed to be accelerated by the shocks in the jets. The derived power-law indices $p$ of the electron spectrum cover the range of from $2.05$ to $2.6$, but a large fraction of them narrowly cluster at $[2.05, 2.1]$, as shown in Figure~\ref{Fig5d}.
The derived electron spectra of these bursts are systematically harder than that predicted by the particle acceleration via shocks~\citep[e.g.,][]{Achterberg_2001_Gallant_MNRAS.328..393A, Lemoine_2003_Pelletier_ApJ...589L..73L}.

\section{Correlation Analysis of the Jet Properties}
\label{section 5}
\subsection{$R_{\rm dec}-\sigma_{B}-B_0$ Relation}

The deceleration of the fireball is a dynamic feature. We first explore whether the deceleration radius $R_{\rm dec}$ is correlated with the magnetization of the jet. Interestingly, we find a tentative correlation between $R_{\rm dec}$ and $\sigma_{B}$,
as shown in Figure~\ref{Fig6a}. 
Our fit gives  
\begin{equation}
\log(R_{\rm dec, 17}) =(0.38\pm 0.05)\log(\sigma_{B})+(1.41\pm 0.11),
\label{R_dec-sigma}
\end{equation}
with a Pearson correlation coefficient of $r=0.65$ and a chance probability $p$-value $<10^{-4}$, the dispersion of the relation is $\delta=0.47$. This relation suggests that a fireball with a higher degree of magnetization tends to have a larger deceleration radius, hence a longer deceleration timescale.  
We also search the relation of $R_{\rm dec,17}$ to the initial magnetic field strength $B_0$. 
We find that $B_0$ is anticorrelated with $R_{\rm dec,17}$ in Figure~\ref{Fig6b}, e.g.,  
\begin{equation}
{\rm log}\left(B_0/{\rm Gs}\right) =
\left(-0.89\pm 0.12\right) {\rm log} \left(R_{\rm dec,17}\right) +\left(0.31\pm 0.09\right).
\end{equation}
Furthermore, we make stepwise multiple variable regression analysis among $R_{\rm dec,17}$, $B_{0}$, and $\sigma_{B}$. We obtain a $R^{r}_{\rm dec,17}(B_{0},\sigma_{B})$ relation as  
\begin{equation}
\log\left(R^{\rm r}_{\rm dec,17}\right)=(1.00\pm 0.05)-(0.83\pm 0.06)\log({B}_0)+(0.26\pm 0.02)\log(\sigma_{B}).
\label{14}
\end{equation}
The regression analysis yields an F-test probability is $204.25$. Figure~\ref{Fig6c} shows $R^{\rm r}_{\rm dec,17}$ as a function of $R_{\rm dec,17}$. The Pearson linear correlation coefficient of the two quantities is $r=0.97$ and the chance probability is $p=0$. 
Our best linear fit gives $R^{\rm r}_{\rm dec,17}=(0.97\pm 0.04)R_{\rm dec,17}\pm (0.02\pm 0.04)$, and the standard deviation is $\delta=0.16$. These results indicate that the $R^{r}_{\rm dec,17}(B_{0},\sigma_{B})$ relation is very tight. 
Note that, GRBs~060526 and 080603A, which are masked with red stars in Figure~\ref{Fig6a}, are far away from the $95\%$ confidence band of the $\log R_{\rm dec}-\log \sigma_B$ relation. GRB~060526 has the smallest $R_{\rm dec} $ and GRB 080603A has the smallest $\sigma_{B}$ value among the bursts in our sample. They do not display other unique characteristics of their prompt and afterglow emission. 
It is interesting that they are still within the $95\%$ confidence band of the $R^{r}_{\rm dec,17}(B_{0},\sigma_{B})$ relation.   

Apparently, the tight $R^{r}_{\rm dec,17}(B_{0},\sigma_{B})$ relation indicates that the deceleration radius depends on the magnetization of the fireball. The physics behind this relation is of our interest. 
Note that $R^{r}_{\rm dec,17}$, $B_{0}$, and $\sigma_{B}$ are not observational data, but are calculated with the model parameters derived from our fits. Based on Eqs.~(\ref{12}) and (\ref{13}),
we have $R_{\rm dec}\propto~(B_0^{-2}\sigma_{B})^{1/3}(\epsilon_e E_{\rm k,iso})^{1/3}$, i.e., $\log R_{\rm dec}\propto -\frac{2}{3}\log B_0+\frac{1}{3}\log \sigma_{B}+\frac{1}{3}\log (\epsilon_e E_{\rm k,iso})$. This is roughly consistent with Eq.~\ref{14} if $\epsilon_e E_{\rm k,iso}$ is a constant among bursts.
Thus, this relation hints that the a certain fraction of the kinetic energy carried by the electrons ($\epsilon_e E_{\rm k, iso}$) is almost constant. 

\subsection{$L_{\gamma, p,\rm iso}-E_{p,z}-\Gamma_0$ Relation}
\cite{Liang_2015_Lin_ApJ...813..116L} discovered a tight $L^{\rm r}_{\gamma,\rm iso, 52}(E_{p,z},\Gamma_{0})$ relation, where $E_{p,z}=E_p(1+z)$.
This relation is incorporated in both the $L_{\rm \gamma,p,iso}-E_{p,z}$~\citep
{Amati_2002_Frontera_AA...390...81A,Liang_2004_Dai_ApJ...606L..29L} and $L_{\rm \gamma,p,iso}-\Gamma_0$ relations~\citep{Liang_2010_Yi_ApJ...725.2209L}, and significantly reduces the intrinsic scatters of the $L_{\gamma,\rm iso}-E_{p,z}$ 
and $L_{\gamma,\rm iso}-\Gamma_0$ relations. Extremely bright GRBs~\citep[e.g., GRB~221009A,][]{Lan_2023_Gao_ApJ...949L...4L} and other bright GRBs with detection 
of TeV/sub-TeV gamma-ray afterglows (GRBs~180720B, 190114C, 190829A, 201015A, 201216C; \citealp{Huang_2020_Liang_ApJ...903L..26H, Zhang_2023_Ren_ApJ...952..127Z}) well satisfy this relation.
Note that~\cite{Liang_2015_Lin_ApJ...813..116L} calculated the $\Gamma_0$ value with $\Gamma_0\propto (n_{0}\eta_\gamma)^{-1/8}(E_{\rm \gamma, iso, 52} t^{-3}_{p, z})^{1/8}$ by adopting universal values of $n_{0}$ and $\eta_\gamma$ among bursts, i.e., $n_{0}=1~\rm cm^{-3}$ and $\eta_{\gamma}=0.2$. 
As shown in this analysis, these values are diverse among bursts. We examine whether the GRBs in our sample agree with this relation by using the $\Gamma_0$ values derived from our analysis. 
We use only those GRBs whose their $L_{\rm \gamma,p,iso}$ and $E_{p, z}$ values can be robustly derived from observations with broadband gamma-ray instruments as mentioned above\footnote{We excluded the following GRBs from the sample due to the lack of the value of $E_{p}$ or $L_{\rm p,iso}$: 
GRBs~071112C, 090313, 120729A, 120815A, 160203A and 170519A. 
At the same time, we calculated the isotropic peak luminosity $L_{p,\rm iso}$ using the data from Fermi/GBM, analysis data via \texttt{GBM Data Tools}~\citep{GbmDataTools},
which included GRBs~080810, 081008, 081109A, 130610A and 141221A.}. 
The $L^{\rm r}_{\rm \gamma, iso, 52}(E_{p,z},\Gamma_{0})$ relation reported by~\cite{Liang_2015_Lin_ApJ...813..116L} is written as\footnote{Adopting the multiply regression analysis, we also derive this relation from the data reported in this sample as 
$L^{\rm r}_{\gamma, p,\rm iso,52}=10^{-5.26\pm 1.11} {(E_{p,z}/{\rm keV})}^{0.94\pm 0.35}\Gamma_{0}^{1.16\pm 0.41}$. 
The derived parameters are slightly different from those reported by~\cite{Liang_2015_Lin_ApJ...813..116L}.} 
\begin{equation}
L^{\rm r}_{\rm \gamma, iso, 52}(E_{p,z},\Gamma_{0})=10^{-6.38\pm 0.35}{(E_{p,z}/{\rm keV})}^{1.34\pm 0.14}\Gamma_{0}^{1.32\pm 0.19} 
\label{Liang-relation}
\end{equation}
and its dispersion is illustrated in Figure~\ref{three parameters} by the correlation of $L^{\rm r}_{\gamma, p,\rm iso, 52}-L_{\gamma, p,\rm iso, 52}$. We calculate the $L^{\rm r}_{\gamma, p,\rm iso, 52}$ values 
with Eq.~\ref{Liang-relation} by using the $\Gamma_0$ values derived from our analysis. Our results are compared with those reported by~\cite{Liang_2015_Lin_ApJ...813..116L}
in the $L^{\rm r}_{\gamma, p,\rm iso, 52}-L_{\gamma, p,\rm iso, 52}$ plane. One can observe that the  $L^{\rm r}_{\rm \gamma, iso, 52}(E_{p,z},\Gamma_{0})$ relation is confirmed by our results, although the data points have large uncertainties that are attributed to the large uncertainties of $\Gamma_0$ from our fits to the light curves.

\section{Summary}
\label{section 6}

By jointly fitting the optical and X-ray afterglow light curves of a sample of $30$~GRBs
that have detected of an early onset bump in their optical afterglow with a standard external shock model, we derive the jet properties of these bursts. They are summarized below. 
\begin{itemize}
    \item The distribution of $\log (\theta_{j})$ for 26 out of the $30$~bursts normally peaks at 
    $\log (\theta_{\rm j,c}/\rm rad)=-1.4$ ( i.e., $\theta_{j,c}=2.3^{\circ}$), and the distributions of $\log E_{\rm k, j}$ and $\log {E}_{\rm \gamma, j}$ (for $19$ out of the $26$ bursts) normally peaks at $\log (E_{\rm k, j,c}/{\rm erg})=51.78$ and $\log ({E}_{\rm \gamma, j, c}/{\rm erg})=49.47$. These results are consistent with the idea of a standard energy reservoir among GRBs previously reported by \cite{Frail_2001_Kulkarni_ApJ...562L..55F} and \cite{Berger_2003_Kulkarni_ApJ...590..379B}, although the derived ${E}_{\rm \gamma, j, c}$ of our sample is lower than that reported in \cite{Frail_2001_Kulkarni_ApJ...562L..55F} by about $1$ order of magnitude.
    
    \item The inferred radiation efficiencies of the $30$~bursts are lower than $10\%$ and even down to $0.04\%$, being consistent with the model prediction for the material-dominated jet scenario. The initial magnetic field strength $B_{0}$ has a normal distribution after taking the logarithm, yielding the central value is $B_{0,c}=0.68$~Gs. 
    The normal distribution of $\log (M_{\rm fb}/M_{\odot})$ peaks concentrated at $M_{\rm fb,p}=1.48\times10^{-5} M_{\odot}$. 
    Measuring the fireball magnetization with $\sigma_{B}=\epsilon_{B}/\epsilon_{e}$, we find $\sigma_{B}<1$ for $90\%$ of the GRBs in our sample, suggesting a relatively weak magnetization of these GRBs. 
    Interestingly, we find a tight correlation of $R^{\rm r}_{\rm dec,17}=10^{0.10\pm 0.05}{B}_0^{-0.83\pm 0.06}\sigma_{B}^{0.26\pm 0.02}$. This relation infers a universal kinetic energy carried by the electrons ($\epsilon_e E_{\rm k,iso}$) among the GRBs.

    \item The distributions of the initial Lorentz factor and deceleration radius of the fireball are normally peak at $\log (\Gamma_{0,c})=2.63$ and $\log (R_{\rm dec,c})/{\rm cm}=17.59$, although the circumburst density shows a bimodal distribution with separation at $0.1~\rm cm^{-3}$.
    Adopting the $\Gamma_0$ values derived from analysis, we confirm the tight $L_{\rm p,iso}-E_{\rm p,z}-\Gamma_{0}$ relation previously reported by~\cite{Liang_2015_Lin_ApJ...813..116L}.
\end{itemize}

\acknowledgments
 We are deeply grateful for the insightful comments provided by the referee. This work used data and software provided by the Fermi Science Support Center and data supplied by the UK Swift Science Data Centre at the University of Leicester. This work is supported by the National Natural Science Foundation of China (Grant No.12133003). S.Q.Z. acknowledges support from the National Natural Science Foundation of China (grant No. 12247144), China Postdoctoral Science Foundation (grant Nos. 2021TQ0325 and 2022M723060) and the starting Foundation of Guangxi University of Science and Technology (grant No. 24Z17).

\software{\texttt{Matplotlib}
\citep{Hunter_2007_CSE.....9...90H},
\texttt{PyMultiNest} \citep{Buchner_2016_ascl.soft06005B},
\texttt{Numpy} \citep{Harris_2020_Millman_Natur.585..357H},
\texttt{GBM Data Tools} \citep{GbmDataTools}, 
\texttt{Astropy} \citep{Astropy_Collaboration_2013_Robitaille_AA...558A..33A}}

\bibliographystyle{aasjournal}
\bibliography{export-bibtex}

\clearpage

\begin{deluxetable*}{lll}
\label{Tab1}
\tabletypesize{\small}
\tablewidth{0pt}
\tablecaption{The selected time intervals of the optical and X-ray afterglow light curves for our fits with the FS model}
\tablehead{
\colhead{GRBs} & \colhead{X-ray afterglows} & \colhead{Optical afterglows} \\
\colhead{} & \colhead{(s)} & \colhead{(s)} 
} 
\startdata
\object{GRB 050820A} &	$4.68\times10^{3}-3.69\times 10^{6}$ &	$230-6.63\times10^{5}$  \\
\object{GRB 050922C} &	$1.17\times10^{2}-3.56\times10^{5}$	 &	$7.45\times10^{2}-6.06\times10^{5}$  \\
\object{GRB 051109A} &	$3.56\times10^{3}-1.50\times10^{6}$	 &	$39.7-1.04\times10^{6}$ \\
\object{GRB 051111}	 &	$5.63\times10^{3}-4.76\times10^{4}$	 &	$31.9-7587$           \\
\object{GRB 060418}	 &	$342.22-5.37\times10^{5}$	         &	$76-7659$	          \\
\object{GRB 060526}	 &	$885.34-3.11\times10^{5}$	         &	$593.10-6.38\times10^{5}$ \\
\object{GRB 070318}	 &	$637.99-7.88\times10^{5}$	         &	$60.72-8.74\times10^{4}$  \\
\object{GRB 070411}	 &	$499.28-7.10\times10^{5}$	         &	$183.47-5.17\times10^{5}$	\\
\object{GRB 071010A} &	$3.45\times10^{4}-4.73\times10^{5}$	 &	$320.90-5.07\times10^{5}$	\\
\object{GRB 071031}	 &	$6.13\times10^{3}-3.04\times10^{5}$	 &	$287.30-2.57\times10^{4}$	\\
\object{GRB 071112C} &	$1.25\times10^{4}-3.56\times10^{5}$	 &	$131.99-6.96\times10^{4}$	\\
\object{GRB 080310}	 &	$1.03\times10^{3}-6.86\times10^{5}$	 &	$298.3-1.24\times10^{5}$	\\
\object{GRB 080603A} &	$1.05\times10^{4}-6.02\times10^{5}$	 &	$105-3.50\times10^{5}$	\\
\object{GRB 080710}	 &	$3.19\times10^{3}-5.56\times10^{5}$	 &	$416.9-2.67\times10^{5}$  \\
\object{GRB 080810}	 &	$304.25-3.79\times10^{5}$	         &	$38-4.98\times10^{5}$	\\
\object{GRB 081008}	 &	$488.38-2.99\times10^{5}$	         &  $108.82-1.85\times10^{5}$ \\
\object{GRB 081109A} &	$338.56-4.76\times10^{5}$	         &	$168.9-6.66\times10^{4}$ \\
\object{GRB 081203A} &	$150.99-3.41\times10^{5}$	         &	$28.48-5.71\times10^{3}$ \\
\object{GRB 090313}	 &	$2.68\times10^{4}-7.17\times10^{5}$	 &	$204.6-3.75\times10^{5}$ \\
\object{GRB 090618}	 &	$311.90-2.90\times10^{6}$	         &	$76-7.26\times10^{4}$	\\
\object{GRB 100906A} &	$206.11-2.02\times10^{5}$	         &	$51.35-1.09\times10^{4}$ \\
\object{GRB 120729A} &	$59.87-4.25\times10^{4}$	         &	$30.4-5.44\times10^{4}$ \\
\object{GRB 120815A} &	$2.74\times10^{3}-4.88\times10^{4}$	 &	$169-11053$	\\
\object{GRB 130610A} &	$324.9-2.28\times10^{5}$	         &	$122-2.25\times10^{3}$	\\
\object{GRB 140629A} &	$82.67-9.16\times10^{4}$	         &	$71.80-35347$ \\
\object{GRB 141221A} &	$88.52-137.42; 3.90\times10^{3}-1.90\times10^{5}$  &  $68.7-1600$	\\
\object{GRB 160203A} &	$318.85-1.53\times10^{5}$	         &	$258-3680$	\\
\object{GRB 170202A} &	$433.86-3.42\times10^{5}$			 &  $68-9309$             \\
\object{GRB 170519A} &	$467.91-1.67\times10^{5}$	         &	$1119-2.99\times10^{5}$	\\
\object{GRB 181110A} &	$422.88-1.15\times10^{5}$	         &	$720-5.58\times10^{4}$ 
\enddata
\end{deluxetable*}

\begin{deluxetable}{lccccccc}
\label{Tab2}
\tabletypesize{\small}
\tablecaption{The prior range of model parameters of FS.}
\tablehead{
\colhead{Parameter} & \colhead{$\log E_{\rm k, iso}$/erg} & \colhead{$\log \Gamma_{0}$} & \colhead{$\log n/\rm cm^{-3}$} & \colhead{$p$} & \colhead{$\log \theta_{j}$} & \colhead{$\log \epsilon_{e}$} & \colhead{$\log \epsilon_{B}$}} 
\startdata
Range & $[52.5,56.5]$ & $[1,3.5]$ &  $[-3,3]$  & $[2.05,3]$ & $[-2,-0.5]$ & $[-3.3,-0.3]$ & $[-8,-2]$ \\
\enddata  
\end{deluxetable}

\clearpage

\begin{deluxetable*}{cccccccccccccccc}
\tablewidth{0pt}
\rotate
\tabletypesize{\tiny}
\tablecaption{The derived model parameters from our fits to the optical and X-ray afterglow light curves with the standard external shock model, along with several inferred quantities such as prompt and afterglow properties.}
\tablehead{
\colhead{GRB(Band)$^{\rm Ref}$} &
\colhead{$z$ } &
\colhead{$\rm log(E_{k, iso}$)} &
\colhead{$\rm log(n)$ } &
\colhead{$\rm log(\Gamma_0)$} &
\colhead{$p$} &
\colhead{$\log(\theta_j $/rad)} &
\colhead{$\rm log(\epsilon_{e}$)} &
\colhead{$\rm log(\epsilon_{B}$)} &
\colhead{$E_{\rm \gamma, iso}$} &
\colhead{$L_{\rm p,iso}$} &
\colhead{$E_{\rm p, z}$}  &
\colhead{$\eta_{\gamma}$} &
\colhead{$B_{0}$} &
\colhead{$R_{dec}$}\\
\colhead{ } &
\colhead{ } &
\colhead{(erg)} &
\colhead{ (cm$^{-3}$)} &
\colhead{ } &
\colhead{ } &
\colhead{ (rad)} &
\colhead{ } &
\colhead{ } &
\colhead{(erg)} &
\colhead{($10^{52}~{\rm erg/s}$)} &
\colhead{Kev}  &
\colhead{ ($\%$)} &
\colhead{ (Gs)} &
\colhead{ ($10^{17}\rm cm$)}
}

\startdata
\object{GRB 050820A$(R)^{(1)}$}  &  2.612  &  $55.97^{+0.02}_{-0.02}$  &  $-2.99^{+0.01}_{-0.01}$  & 
$2.84^{+0.01}_{-0.02}$   & $2.22^{+0.01}_{-0.01}$   &  $-1.59^{+0.01}_{-0.01}$  &  $-2.25^{+0.01}_{-0.01}$  &  $-3.28^{+0.02}_{-0.02}$  &  $8.30e53$  &   $5.78^{+3.05}_{-3.05}$   &  $978.85$  &  $1.03$  & $0.20$  &  $35.68$ \\
\object{GRB 050922C$(R)^{(2)}$}  &  2.198  &  $54.05^{+0.06}_{-0.06}$  &  $-0.76^{+0.18}_{-0.13}$  & 
$2.53^{+0.01}_{-0.02}$   & $2.58^{+0.02}_{-0.02}$   &  $-1.30^{+0.02}_{-0.02}$  &  $-1.47^{+0.05}_{-0.04}$  &  $-3.82^{+0.18}_{-0.22}$  &  $8.10e52$  &   $5.17^{+28.00}_{-0.01}$  &  $457.31$  &  $6.73$  & $0.68$  &  $2.38$ \\
\object{GRB 051109A$(R)^{(3)}$}   &  2.346   &  $55.99^{+0.01}_{-0.01}$  &  $1.74^{+0.03}_{-0.04}$   & 
$2.60^{+0.01}_{-0.01}$   & $2.05^{+0.01}_{-0.01}$   &  $-1.39^{+0.01}_{-0.01}$  &  $-1.87^{+0.02}_{-0.02}$  &  $-6.96^{+0.05}_{-0.03}$  &  $5.00e52$   &   $10.11^{+9.14}_{-9.14}$  &  $538.71$  &  $0.05$ & $0.38$  &  $1.39$ \\
\object{GRB 051111$(R)^{(4)}$}   &  1.55   &  $56.09^{+0.18}_{-0.24}$  &  $1.47^{+0.19}_{-0.24}$   & 
$2.80^{+0.01}_{-0.01}$   & $2.05^{+0.01}_{-0.01}$   &  $-1.91^{+0.01}_{-0.01}$  &  $-3.11^{+0.22}_{-0.14}$  &  $-5.37^{+0.31}_{-0.30}$  &  $6.30e52$  &   $1.55^{+0.61}_{-0.33}$   &  $1139.85$  &  $0.05$ & $2.75$  &  $2.38$ \\
\object{GRB 060418$(H)^{(5)}$}   &  1.49   &  $55.75^{+0.17}_{-0.25}$  &  $-2.72^{+0.22}_{-0.17}$  & 
$2.94^{+0.03}_{-0.03}$   & $2.41^{+0.06}_{-0.05}$   &  $-1.97^{+0.03}_{-0.02}$  &  $-2.92^{+0.14}_{-0.14}$  &  $-2.61^{+0.33}_{-0.39}$  &  $9.00e52$   &   $5.75^{+2.91}_{-3.18}$   &  $572.70$  &  $0.16$  &  $0.73$  &  $21.01$ \\
\object{GRB 060526$(Rc)^{(6)}$}  &  3.21   &$52.66^{+0.02}_{-0.02}$ &  $2.84^{+0.16}_{-0.14}$   & 
$1.58^{+0.02}_{-0.02}$   & $2.05^{+0.01}_{-0.01}$   &  $-0.76^{+0.02}_{-0.02}$  &  $-0.90^{+0.02}_{-0.03}$  &  $-3.10^{+0.10}_{-0.13}$  &  $--$   &   $5.04^{+4.05}_{-4.05}$   &  $164.19$  &  $--$  &  $10.97$  &  $0.22$ \\
\object{GRB 070318$(V)^{(7)}$}   &  0.84   &  $53.69^{+0.07}_{-0.08}$  &  $2.33^{+0.22}_{-0.26}$   &  
$1.98^{+0.03}_{-0.02}$   &  $2.07^{+0.01}_{-0.01}$   &  $-0.50^{+0.02}_{-0.02}$  &  $-0.53^{+0.02}_{-0.02}$  &  $-5.52^{+0.39}_{-0.35}$  &  $1.45e52$   &   $0.18^{+0.05}_{-0.05}$   &  $671.60$  &  $6.51$  &  $0.94$  &  $0.39$ \\
\object{GRB 070411$(R)^{(8)}$}   &  2.95   &  $54.75^{+0.05}_{-0.04}$  &  $-2.40^{+0.76}_{-0.43}$  &  
$2.72^{+0.05}_{-0.09}$   &  $2.26^{+0.02}_{-0.02}$   &  $--$  &  $-2.02^{+0.12}_{-0.06}$  &  $-2.74^{+0.37}_{-0.63}$  &  $--$   &   $1.77^{+0.51}_{-0.51}$   &  $474$  &  $--$  &  $0.55$  &  $10.69$ \\
\object{GRB 071010A$(R)^{(9)}$}  &  0.98   &  $54.44^{+0.04}_{-0.08}$  &  $-0.23^{+0.26}_{-0.26}$  &  
$2.18^{+0.03}_{-0.03}$	 &  $2.05^{+0.01}_{-0.01}$   &  $-1.23^{+0.03}_{-0.03}$  &  $-2.25^{+0.07}_{-0.04}$  &   $-3.18^{+0.20}_{-0.19}$  &  $--$  &   
$0.03^{+0.02}_{-0.02}$   &  $69.3$  &  $--$  &  $1.17$ &  $3.71$  \\
\object{GRB 071031$(r^{'})^{(10)}$} & 2.69  &  $54.46^{+0.06}_{-0.10}$  &  $0.30^{+0.26}_{-0.27}$   &  
$2.14^{+0.03}_{-0.03}$   &  $2.05^{+0.01}_{-0.01}$   &  $-1.22^{+0.03}_{-0.03}$  &  $-2.22^{+0.09}_{-0.05}$  &  $-3.08^{+0.20}_{-0.20}$  &  $--$   &  
$0.14^{+0.04}_{-0.04}$   &  $110.7$  &  $--$  &  $2.19$  &  $2.62$ \\
\object{GRB 071112C$(R)^{(11)}$}    &  0.82 &  $53.49^{+0.12}_{-0.14}$  &  $-0.85^{+0.25}_{-0.11}$  & 
$2.46^{+0.03}_{-0.03}$	 &  $2.12^{+0.02}_{-0.03}$   &  $-1.33^{+0.04}_{-0.02}$  &  $-1.37^{+0.12}_{-0.09}$  &   $-4.03^{+0.22}_{-0.26}$  &  $--$   &   
$8.71^{+6.62}_{-5.01}$   &  $--$  &  $--$  &  $0.41$  &  $1.84$ \\
\object{GRB 080310$(R)^{(12)}$}     &  2.42 &  $53.87^{+0.04}_{-0.04}$  &  $1.01^{+0.12}_{-0.15}$   & 
$2.34^{+0.03}_{-0.02}$   &  $2.10^{+0.01}_{-0.01}$   &  $-1.03^{+0.01}_{-0.02}$  &  $-0.63^{+0.03}_{-0.03}$  &   $-4.60^{+0.24}_{-0.20}$  &  $--$   &   
$0.31^{+0.09}_{-0.09}$   &  $257.87$  &  $--$  &  $1.36$  &  $0.71$ \\
\object{GRB 080603A$(R)^{(13)}$}     &  2.69 &  $54.87^{+0.08}_{-0.09}$  &  $-0.93^{+0.30}_{-0.31}$  &  
$2.53^{+0.05}_{-0.04}$   &  $2.41^{+0.01}_{-0.01}$   &  $-1.44^{+0.04}_{-0.05}$  &  $-0.58^{+0.05}_{-0.05}$  &   $-6.38^{+0.17}_{-0.15}$  &  $--$   &   
$--$   &  $160$  &  $--$  &  $0.03$  &  $5.08$ \\
\object{GRB 080710$(r^{'})^{(14)}$} & 0.85 &  $52.94^{+0.15}_{-0.07}$  &  $-1.25^{+0.16}_{-0.19}$  &       
$2.00^{+0.04}_{-0.03}$   &  $2.45^{+0.11}_{-0.03}$   &  $-1.29^{+0.02}_{-0.02}$  &  $-1.15^{+0.04}_{-0.04}$  &   $-2.04^{+0.03}_{-0.29}$  &  $--$   &   
$0.04_{-0.01}^{+0.01}$  &  $200$  &  $--$  &  $0.88$  &  $3.33$ \\
\object{GRB 080810$(R)^{(15)}$}     & 3.35   & $55.50^{+0.01}_{-0.01}$  &  $0.07^{+0.06}_{-0.06}$   &         $2.78^{+0.01}_{-0.01}$   &  $2.50^{+0.01}_{-0.01}$   &  $-1.27^{+0.01}_{-0.01}$  &  $-2.37^{+0.01}_{-0.01}$  &   $-3.98^{+0.05}_{-0.05}$  &  $3.00e53$   &   $7.84$   &   $2475.15$  &  $0.94$  &  $2.60$  &  $2.60$ \\
\object{GRB 081008$(r^{'})^{(16)}$} & 1.967  & $55.35^{+0.05}_{-0.05}$  &  $-2.94^{+0.08}_{-0.04}$  &         $2.86^{+0.01}_{-0.01}$   &  $2.21^{+0.01}_{-0.01}$   &  $-1.96^{+0.01}_{-0.01}$  &  $-2.20^{+0.01}_{-0.01}$  &   $-3.70^{+0.09}_{-0.11}$  &  $6.30e52$   &  $0.92$   &   $261.7$   &  $0.60$  &  $0.13$  &  $20.69$ \\
\object{GRB 081109A$(H)^{(17)}$	}   & 0.9787 & $55.38^{+0.09}_{-0.18}$  &  $-3.51^{+0.25}_{-0.16}$  &         $2.81^{+0.03}_{-0.04}$	 &  $2.06^{+0.01}_{-0.01}$   &  $-1.93^{+0.04}_{-0.03}$  &  $-3.08^{+0.17}_{-0.09}$  &  $-2.12^{+0.08}_{-0.17}$  &  $5.00e52$   &   
$0.43$   &  $474.89$  &  $0.21$  &  $0.38$  &  $35.40$ \\
\object{GRB 081203A$(U)^{(18)}$}    & 2.05   & $54.92^{+0.05}_{-0.07}$  &  $1.73^{+0.11}_{-0.14}$   &         $2.33^{+0.02}_{-0.01}$   &  $2.57^{+0.01}_{-0.01}$   &  $-1.39^{+0.01}_{-0.01}$  &  $-1.38^{+0.03}_{-0.03}$  &  $-6.10^{+0.23}_{-0.15}$  &  $3.50e53$   &   
$2.82_{-0.19}^{+0.19}$  &  $1762.9$  &  $4.04$  &  $0.54$  &  $0.94$ \\
\object{GRB 090313$(r^{'})^{(19)}$}  & 3.375 & $55.31^{+0.11}_{-0.28}$  &  $-2.19^{+0.24}_{-0.19}$  &         $2.59^{+0.03}_{-0.03}$	 &  $2.07^{+0.01}_{-0.01}$   &  $-1.73^{+0.03}_{-0.03}$  &  $-2.15^{+0.26}_{-0.11}$  &  $-2.24^{+0.15}_{-0.20}$  &  $--$   &   
$0.36_{-0.31}^{+0.31}$  &  $--$  &  $--$  &  $0.92$  &  $17.08$ \\
\object{GRB 090618$(R)^{(20)}$}     &  0.54  & $55.45^{+0.04}_{-0.07}$  &  $-2.05^{+0.15}_{-0.16}$  &         $2.74^{+0.02}_{-0.02}$   &  $2.05^{+0.01}_{-0.01}$   &  $-1.75^{+0.02}_{-0.02}$  &  $-2.63^{+0.07}_{-0.04}$  &  $-3.38^{+0.11}_{-0.11}$  &  $2.00e53$   &   
$1.91_{-0.22}^{+0.22}$  &  $338.8$  &  $0.71$  &  $0.41$  &  $13.56$ \\
\object{GRB 100906A$(V)^{(21)}$}    &  1.727 & $55.50^{+0.01}_{-0.01}$  &  $0.20^{+0.06}_{-0.06}$   &         $2.56^{+0.01}_{-0.01}$	 &  $2.51^{+0.01}_{-0.01}$   &  $-1.50^{+0.01}_{-0.01}$  &  $-1.97^{+0.01}_{-0.01}$  &   $-5.34^{+0.05}_{-0.05}$  &  $2.20e53$   &   
$1.51_{-0.07}^{+0.07}$  &  $387.23$  &  $0.69$  &  $0.38$  &  $3.30$ \\
\object{GRB 120729A$(R)^{(22)}$}     &  0.8 &  $55.38^{+0.38}_{-0.31}$  &  $0.59^{+0.37}_{-0.31}$   & 
$2.61^{+0.01}_{-0.01}$   &  $2.36^{+0.01}_{-0.01}$   &  $-1.78^{+0.01}_{-0.01}$  &  $-2.90^{+0.23}_{-0.27}$  &   $-4.89^{+0.31}_{-0.38}$  &  $2.30e52$   &   
$--$   & $--$ &  $0.10$  &  $1.12$  &  $2.07$ \\
\object{GRB 120815A$(r^{'})^{(23)}$} & 2.358 & $53.88^{+0.02}_{-0.02}$  &  $-2.61^{+0.27}_{-0.23}$  &         $2.99^{+0.01}_{-0.02}$ 	 &  $2.07^{+0.01}_{-0.01}$   &  $--$  &  $-0.96^{+0.0}_{-0.05}$   &   $-2.44^{+0.17}_{-0.22}$  &  $--$  &    
$0.76^{+0.38}_{-0.38}$  & $--$ &  $--$  &  $1.14$  &  $4.26$ \\
\object{GRB 130610A$(I)^{(24)}$}     & 2.092 & $55.28^{+0.16}_{-0.27}$  &  $-1.94^{+0.10}_{-0.05}$  &         $2.72^{+0.02}_{-0.03}$	 &  $2.34^{+0.01}_{-0.01}$   &  $-1.72^{+0.04}_{-0.03}$  &  $-2.07^{+0.10}_{-0.10}$  &   $-4.93^{+0.53}_{-0.40}$  &  $6.18e52$  &    
$1.08$   & $911.83$ &  $0.32$  &  $0.07$  &  $11.28$ \\
\object{GRB 140629A$(R)^{(25)}$}     & 2.275 & $56.00^{+0.01}_{-0.01}$  &  $0.67^{+0.07}_{-0.06}$   &          $2.55^{+0.01}_{-0.01}$   &  $2.59^{+0.01}_{-0.01}$   &  $-1.58^{+0.01}_{-0.01}$  &  $-2.25^{+0.04}_{-0.01}$  &   $-5.86^{+0.05}_{-0.10}$  &  $4.40e52$  &    
$1.86^{+0.69}_{-0.73}$  & $281.65$  &  $0.04$  &  $0.35$  &  $3.43$ \\
\object{GRB 141221A$(I)^{(26)}$}     & 1.45  & $55.48^{+0.01}_{-0.03}$  &  $-2.85^{+0.19}_{-0.11}$  &          $2.89^{+0.01}_{-0.02}$ 	 &  $2.37^{+0.01}_{-0.01}$   &  $--$  &  $-2.43^{+0.03}_{-0.02}$  &   $-3.83^{+0.09}_{-0.14}$  &  $6.99e52$  &    
$0.77$  & $328.13$ &  $0.23$  &  $0.14$  &  $20.37$ \\
\object{GRB 160203A$(I)^{(27)}$}     & 3.52  & $55.16^{+0.23}_{-0.30}$  &  $1.35^{+0.49}_{-0.47}$   &          $2.23^{+0.04}_{-0.06}$	 &  $2.07^{+0.02}_{-0.01}$   &  $-1.37^{+0.05}_{-0.04}$  &  $-2.63^{+0.33}_{-0.25}$  &   $-4.31^{+0.41}_{-0.49}$  &  $1.20e53$  &    
$--$    & $--$ &   $0.82$  &  $2.19$  &  $1.75$ \\
\object{GRB 170202A$(R)^{(28)}$}    & 3.645  & $55.50^{+0.01}_{-0.01}$  &  $-2.77^{+0.08}_{-0.08}$  &          $2.90^{+0.01}_{-0.01}$	 &  $2.22^{+0.01}_{-0.01}$   &  $-1.68^{+0.05}_{-0.04}$  &  $-2.23^{+0.01}_{-0.01}$  &   $-3.66^{+0.06}_{-0.05}$  &  $1.70e53$  &    
$17.40^{+18.00}_{-14.40}$  &1147.32&  $0.54$  &  $0.19$  &  $19.16$ \\
\object{GRB 170519A$(R)^{(29)}$}    &  3.52  & $53.74^{+0.03}_{-0.04}$  &  $2.93^{+0.05}_{-0.10}$   &          $1.52^{+0.01}_{-0.01}$	 &  $2.07^{+0.01}_{-0.01}$   &  $-0.68^{+0.01}_{-0.01}$  &  $-0.42^{+0.03}_{-0.04}$  &   $-4.96^{+0.11}_{-0.08}$  &   $--$    &    
$--$  & $--$ & $--$    &  $1.24$  &  $0.52$ \\
\object{GRB 181110A$(R)^{(30)}$}    &  3.52  & $55.02^{+0.06}_{-0.04}$  &  $2.99^{+0.01}_{-0.02}$   &          $1.81^{+0.01}_{-0.01}$	 &  $2.06^{+0.01}_{-0.01}$   &  $--$  &  $-1.04^{+0.05}_{-0.06}$  &   $-4.95^{+0.06}_{-0.06}$  &   $1.34e53$    &    
$0.48^{+0.21}_{-0.61}$  & $132.44$ &  $1.26$  &  $2.63$  &  $0.85$ \\
\enddata
\tablerefs{
(1) \cite{Cenko_2006_Kasliwal_ApJ...652..490C, Willingale_2007_OBrien_arXiv0710.3727W} 
(2) \cite{Golenetskii_2005_Aptekar_GCN..4030....1G, Panaitescu_2006_Meszaros_MNRAS.369.2059P, Ukwatta_2010_Stamatikos_ApJ...711.1073U}
(3) \cite{Golenetskii_2005_Aptekar_GCN..4238....1G, Yost_2007_Swan_ApJ...657..925Y, Willingale_2007_OBrien_arXiv0710.3727W}
(4) \cite{Butler_2006_Li_ApJ...652.1390B, Yost_2007_Swan_ApJ...657..925Y, Krimm_2009_Yamaoka_ApJ...704.1405K}  
(5) \cite{Golenetskii_2006_Aptekar_GCN..4989....1G, Molinari_2007_Vergani_AA...469L..13M, Ukwatta_2010_Stamatikos_ApJ...711.1073U, Zitouni_2020_Guessoum_ApSS.365..177Z}
(6) \cite{Willingale_2007_OBrien_arXiv0710.3727W, Thone_2010_Kann_AA...523A..70T, Beskin_2015_Oganesyan_AstBu..70..400B}
(7)  \cite{Willingale_2007_OBrien_arXiv0710.3727W, Krimm_2009_Yamaoka_ApJ...704.1405K}
(8)  \cite{Butler_2006_Li_ApJ...652.1390B, Willingale_2007_OBrien_arXiv0710.3727W, Ferrero_2008_Kann_AIPC.1000..257F, Beskin_2015_Oganesyan_AstBu..70..400B}
(9)  \cite{Covino_2008_DAvanzo_MNRAS.388..347C, Beskin_2015_Oganesyan_AstBu..70..400B}
(10)  \cite{Kruhler_2009_Greiner_ApJ...697..758K, Beskin_2015_Oganesyan_AstBu..70..400B}
(11) \cite{Kruehler_2008_Schrey_GCN..8075....1K, Kann_2010_Klose_ApJ...720.1513K, Zitouni_2020_Guessoum_ApSS.365..177Z}
(12) \cite{Kann_2010_Klose_ApJ...720.1513K, Littlejohns_2012_Willingale_MNRAS.421.2692L, Beskin_2015_Oganesyan_AstBu..70..400B}
(13) \cite{Guidorzi_2011_Kobayashi_MNRAS.417.2124G}
(14) \cite{Kruhler_2009_Greiner_AA...508..593K, Beskin_2015_Oganesyan_AstBu..70..400B}
(15) \cite{Page_2009_Willingale_MNRAS.400..134P, Melandri_2010_Kobayashi_ApJ...723.1331M}
(16) \cite{Yuan_2010_Schady_ApJ...711..870Y, Kann_2010_Klose_ApJ...720.1513K}
(17) \cite{Jin_2009_Xu_MNRAS.400.1829J}
(18) \cite{Golenetskii_2008_Aptekar_GCN..8611....1G, Kuin_2009_Landsman_MNRAS.395L..21K, Nava_2012_Salvaterra_MNRAS.421.1256N}
(19)\cite{Melandri_2010_Kobayashi_ApJ...723.1331M, Beskin_2015_Oganesyan_AstBu..70..400B}
(20) \cite{McBreen_2009_GCN..9535....1M, Golenetskii_2009_Aptekar_GCN..9553....1G, Cano_2011_Bersier_MNRAS.413..669C}
(21) \cite{Golenetskii_2010_Aptekar_GCN.11251....1G, Gorbovskoy_2012_Lipunova_MNRAS.421.1874G, Beskin_2015_Oganesyan_AstBu..70..400B}
(22) \cite{Cano_2014_de_Ugarte_Postigo_AA...568A..19C, Huang_2018_Wang_ApJ...859..163H}
(23) \cite{Kruhler_2013_Ledoux_AA...557A..18K, Beskin_2015_Oganesyan_AstBu..70..400B}
(24) \cite{Trotter_2013_Haislip_GCN.14844....1T, Fitzpatrick_2013_Pelassa_GCN.14858....1F}
(25) \cite{Golenetskii_2014_Aptekar_GCN.16495....1G, Xin_2018_Zhong_ApJ...860....8X, Zitouni_2020_Guessoum_ApSS.365..177Z}
(26) \cite{Trotter_2014_Haislip_GCN.17210....1T, Atteia_2017_Heussaff_ApJ...837..119A, Wang_2018_Aimuratov_MmSAI..89..293W, Zitouni_2020_Guessoum_ApSS.365..177Z}
(27) \cite{Zitouni_2020_Guessoum_ApSS.365..177Z, Crisp_2021_Gendre_MNRAS.504..716C}
(28) \cite{Xin_2017_Wei_GCN.20576....1X, Zitouni_2020_Guessoum_ApSS.365..177Z, Gendre_2022_Orange_ApJ...929...16G}
(29) \cite{Hentunen_2017_Nissinen_GCN.21113....1H, Mazaeva_2017_Antonyuk_GCN.21169....1M, Mazaeva_2017_Pozanenko_GCN.21206....1M, Mazaeva_2017_Pozanenko_GCN.21208....1M}
(30) \cite{Zitouni_2020_Guessoum_ApSS.365..177Z, Han_2022_Li_Univ....8..248H}}
\label{Tab3}
\end{deluxetable*}

\clearpage

\begin{deluxetable*}{lccccc} 
\tabletypesize{\small}
\tablewidth{0pt}
\tablecaption{The distributions and our log-normal/bimodal fits of the jet properties for the bursts in our sample.}
\tablehead{
\colhead{Parameter} & \colhead{Range} & \colhead{Median} &\colhead{Average} & \colhead{Normality/Bimodality$^{a}$} & \colhead{Adj. $R^2$} } 
\startdata
 $\log E_{\rm k, iso}$/erg  &  $[52.5, 56.5]$ & $55.22$ &  $54.70$ &$N_1(53.81, 0.28)$  &  $0.90$ \\
 &   &   &   & $N_2(55.34, 0.45)$ &  \\ 
\hline
 $\log E_{\rm \gamma, iso}$/erg  &  $[52, 54]$ & $52.91$ & $53.07$ & $N(52.96, 0.37)$ & $0.92$ \\
\hline
$\log \theta_{\rm j}$/rad &	$[-2, 0.5]$ & $-1.47$ & $-1.24$ & $N(-1.51, 0.27)$ & $0.95$ \\
\hline
 $\log E_{\rm k, j}$/erg &	$[50, 53]$ & $51.71$ & $51.33$ & $N(51.78, 0.54)$ & $0.98$ \\
\hline
 $\log E_{\rm \gamma, j}$/erg & $[48, 51.5]$ & $49.57$ & $49.25$ & $N(49.47, 0.87)$ & $0.99$ \\
 \hline
 $\log \eta_{\gamma}/\%$  &	$[-1.4, 1]$ & $-0.27$ & $-0.10$ & $N(-0.38, 0.76)$ & $0.93$ \\
\hline
 $\log \Gamma_{0}$ &	$[1, 3]$ & $2.58$ & $2.57$ & $N(2.63, 0.27)$ & $0.83$ \\
 \hline
 $\log (M_{\rm fb}/M_{\odot})$ & $[-6.5, -3.5]$ & $-5.08$ & $-5.50$ & $N(-5.00, 0.75)$ & $0.95$\\
 \hline
 $\log n$/cm$^{-3}$ & $[-3, 3]$ & $-0.35$ & $-0.08$ & $N_1(-2.51, 0.50)$ & $0.88$ \\
 &  &  &  & $N_2(1.00, 1.34)$ &  \\
 \hline
 $\log R_{\rm dec}/{\rm 10^{17}cm}$ &	$[-1, 2]$ & $0.52$ & $0.45$ & $N(0.59, 0.56)$ & $0.99$ \\
 \hline
 $\log \epsilon_{B}$ & $[-7, -1.5]$ & $-3.91$ & $-4.40$ & $N(-4.09, 1.90)$ & $0.91$ \\
\hline
 $\log \epsilon_{e}$ & $[-3.5, -0.3]$ & $-2.11$ & $-2.20$ & $N_1(-2.39, 0.36)$ &  $0.88$ \\
 &  &  &  & $N_2(-0.95, 0.39)$ & \\
 \hline
 $\log \sigma_{B}$ &	$[-5.5, 0.5]$ & $-1.84$ & $-2.17$ & $N_1(-4.2, 1.06)$ & $0.89$ \\
 &   &   &   & $N_2(-1.30, 0.68)$ &  \\
 \hline
 $\log B_{0}$/Gs & $[-1.6, 1.2]$ & $-0.15$ & $-0.06$ & $N(-0.17, 0.54)$ & $0.99$
\enddata
$^a$ The results of our fits with a normal function $N(x_c,\sigma)$ or bimodal function  $N_1(x_{c,1},\sigma_{1})+N_2(x_{c,2},\sigma_{2})$. 
\label{Tab4}
\end{deluxetable*}

\clearpage

\begin{figure*}[htbp]
\centering
\includegraphics[angle=0,width=0.30\textwidth]{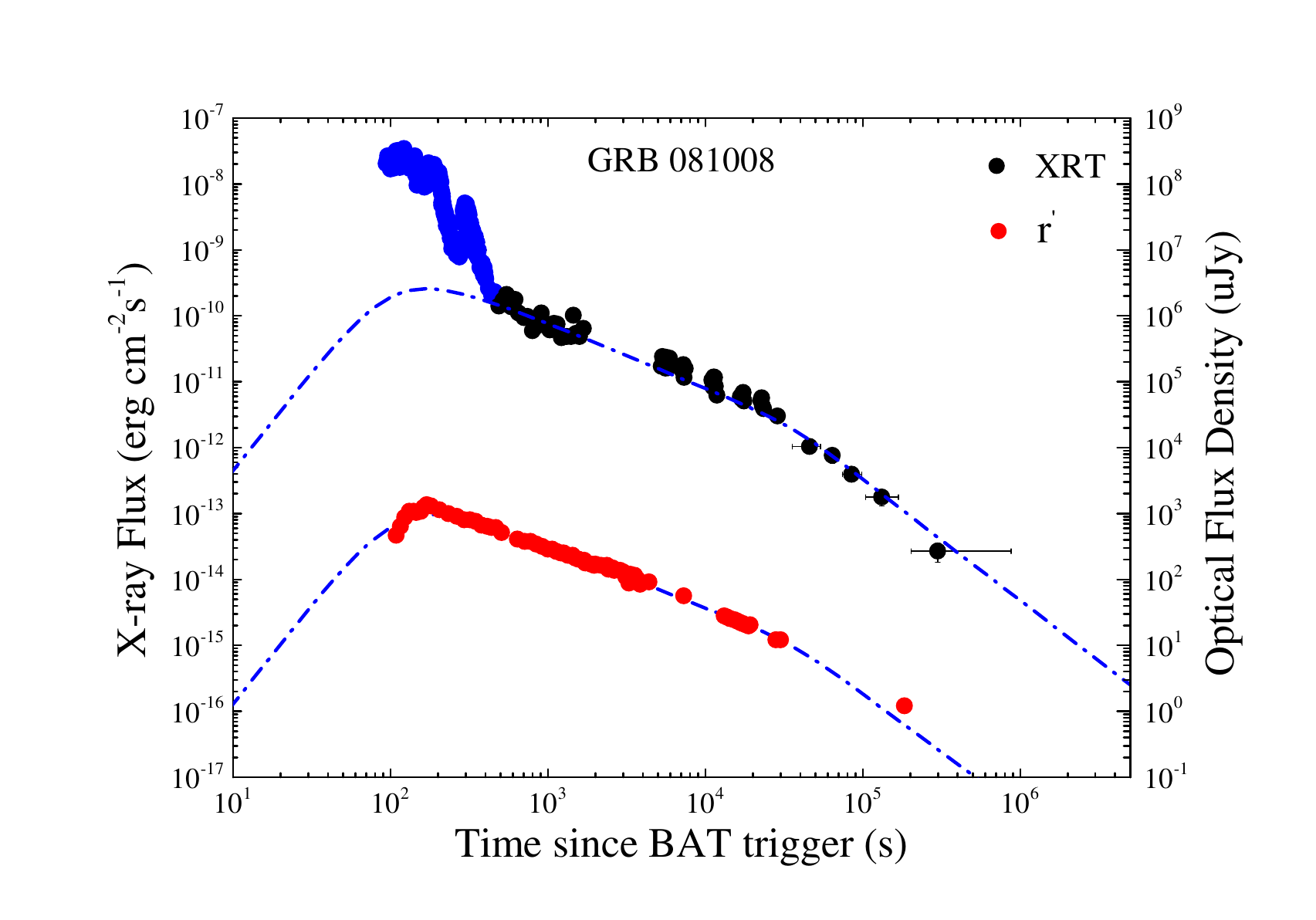}
\includegraphics[angle=0,width=0.30\textwidth]{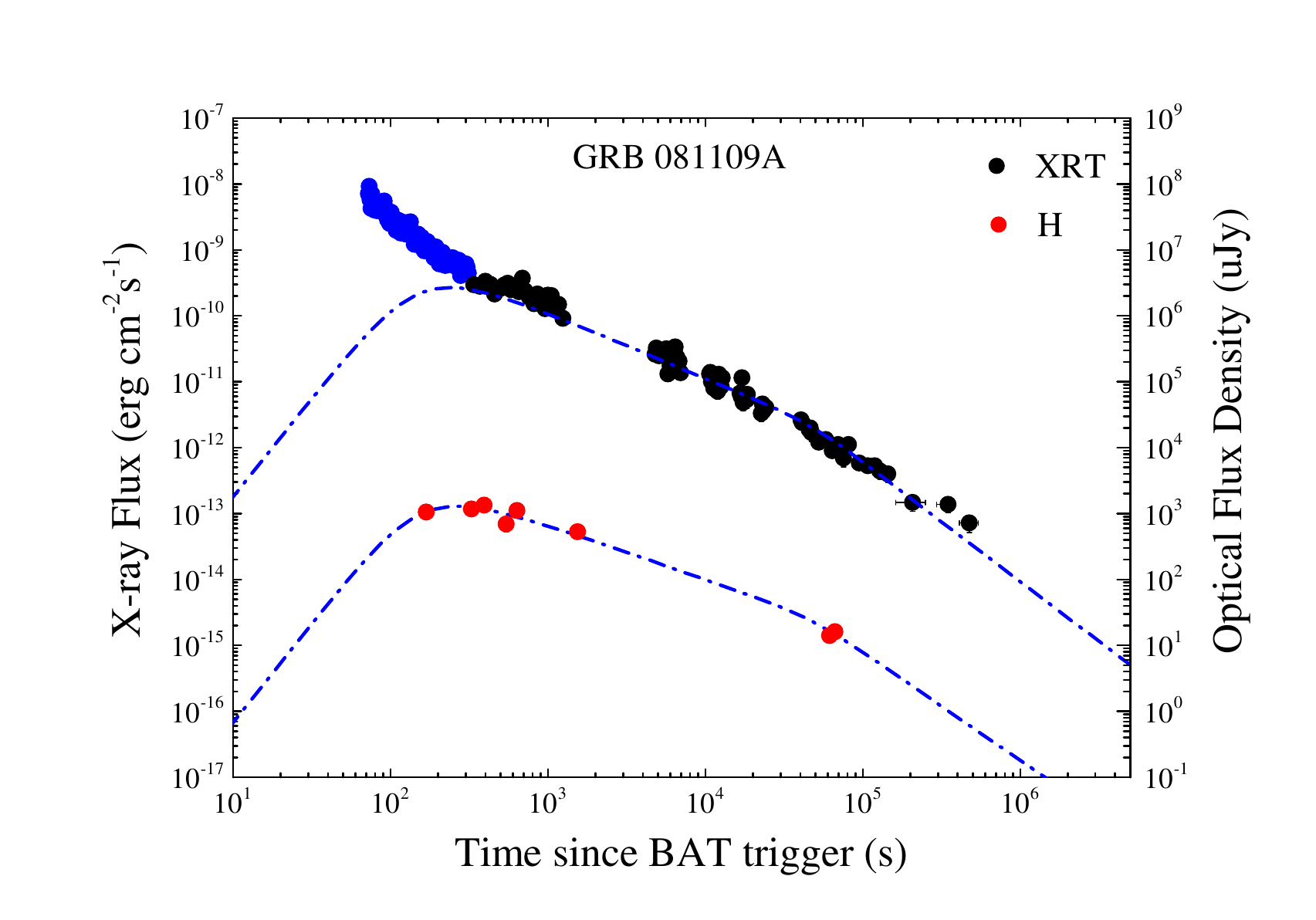}
\includegraphics[angle=0,width=0.30\textwidth]{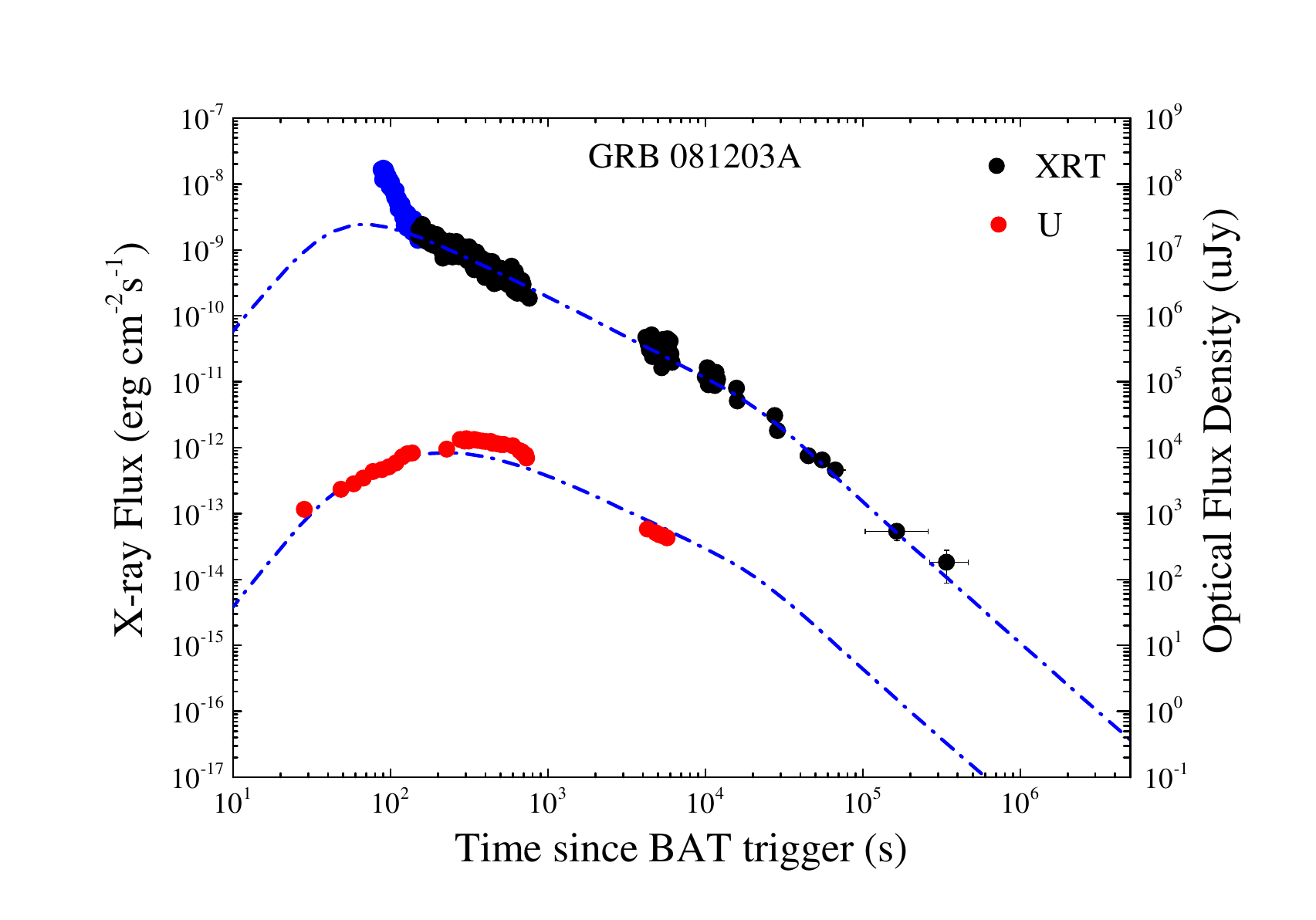}
\includegraphics[angle=0,width=0.30\textwidth]{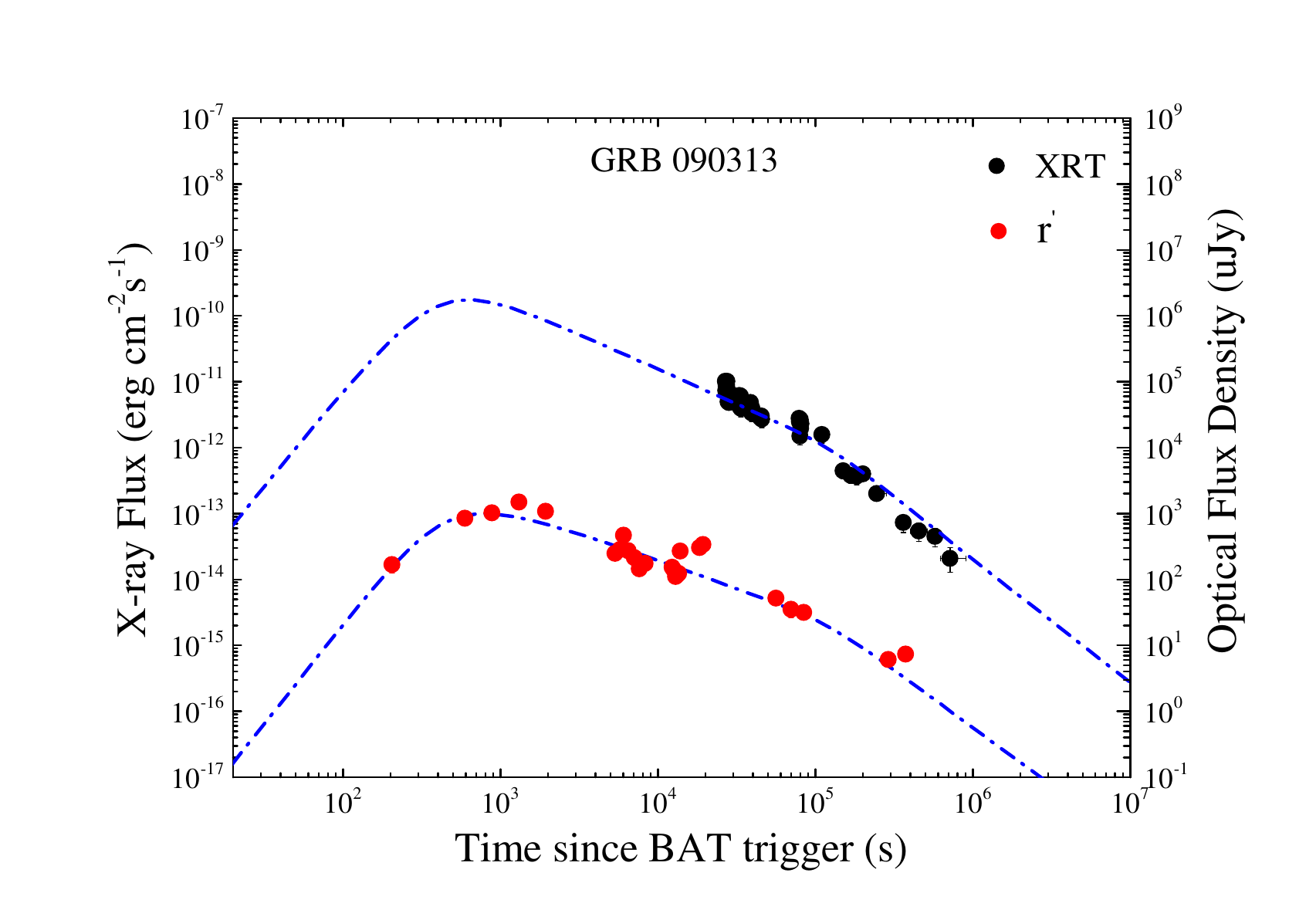}
\includegraphics[angle=0,width=0.30\textwidth]{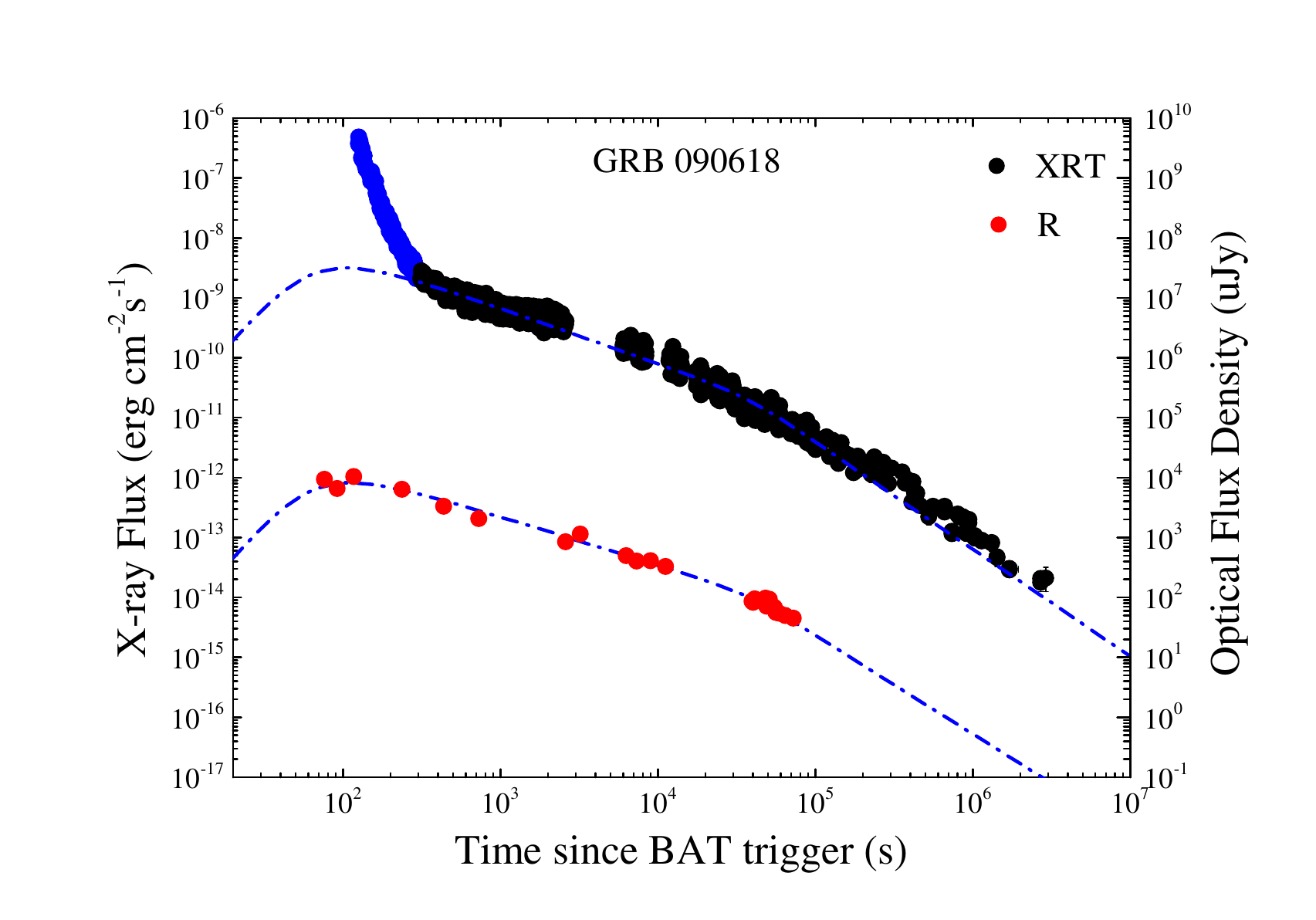}
\includegraphics[angle=0,width=0.30\textwidth]{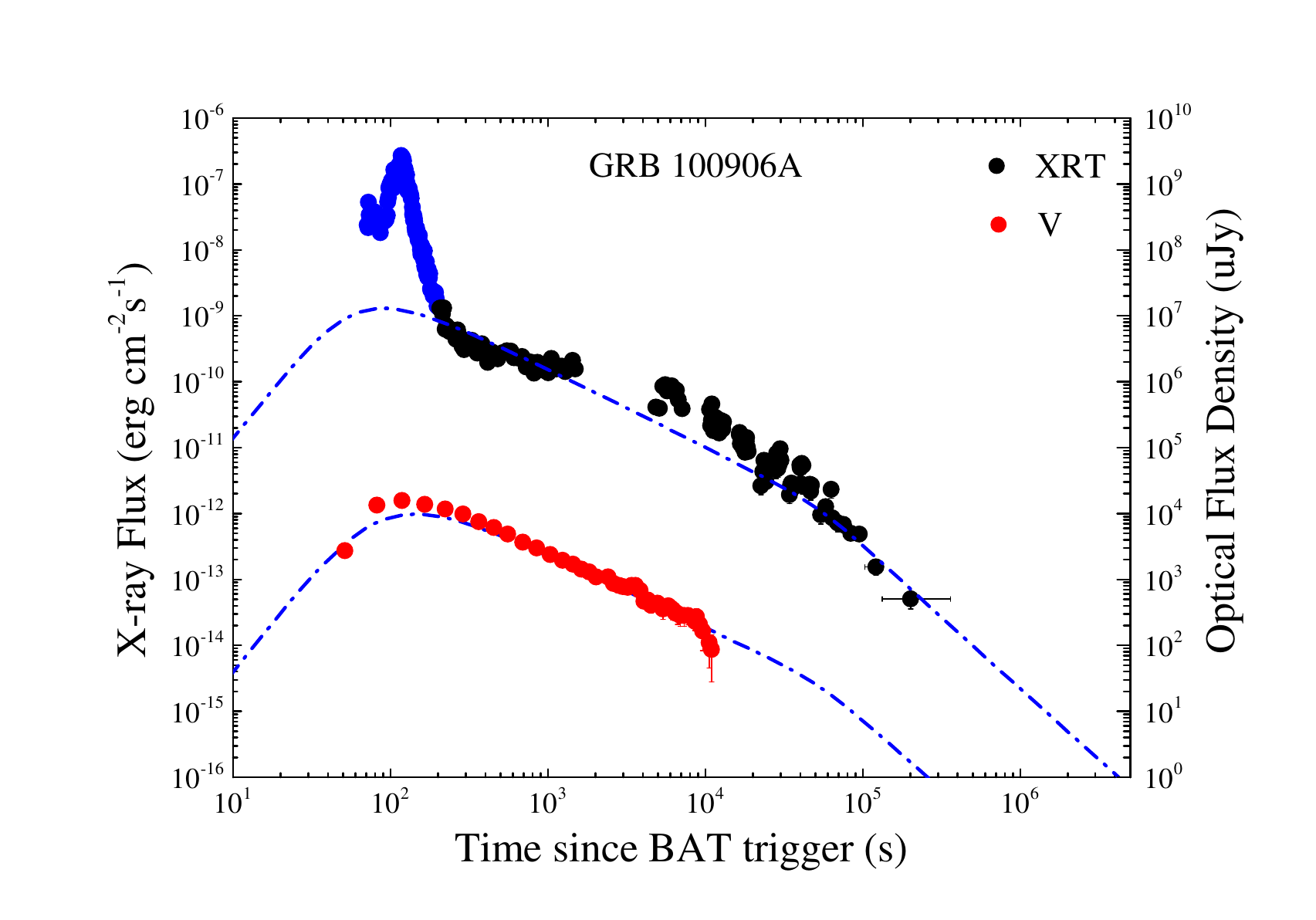}
\includegraphics[angle=0,width=0.30\textwidth]{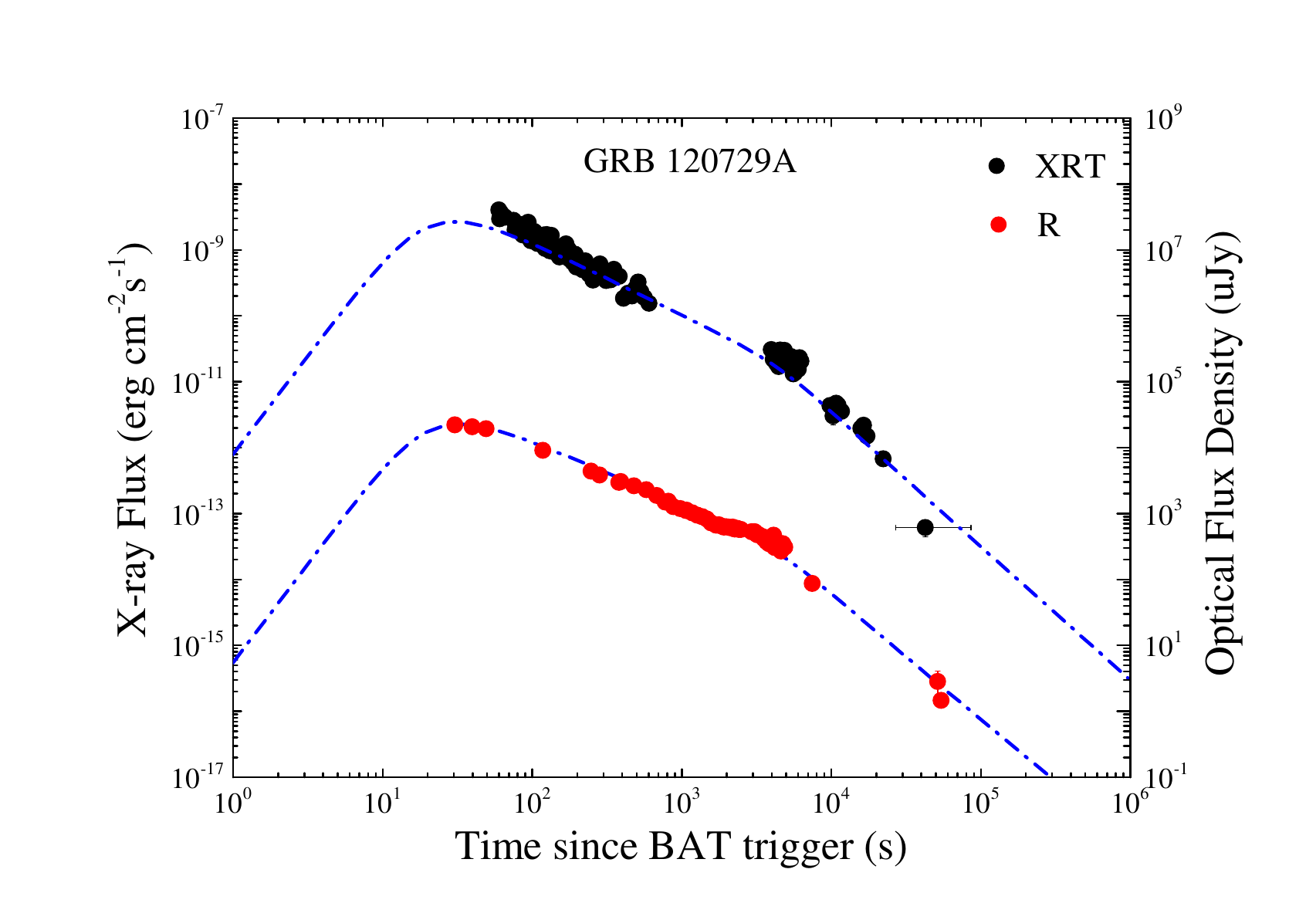}
\includegraphics[angle=0,width=0.30\textwidth]{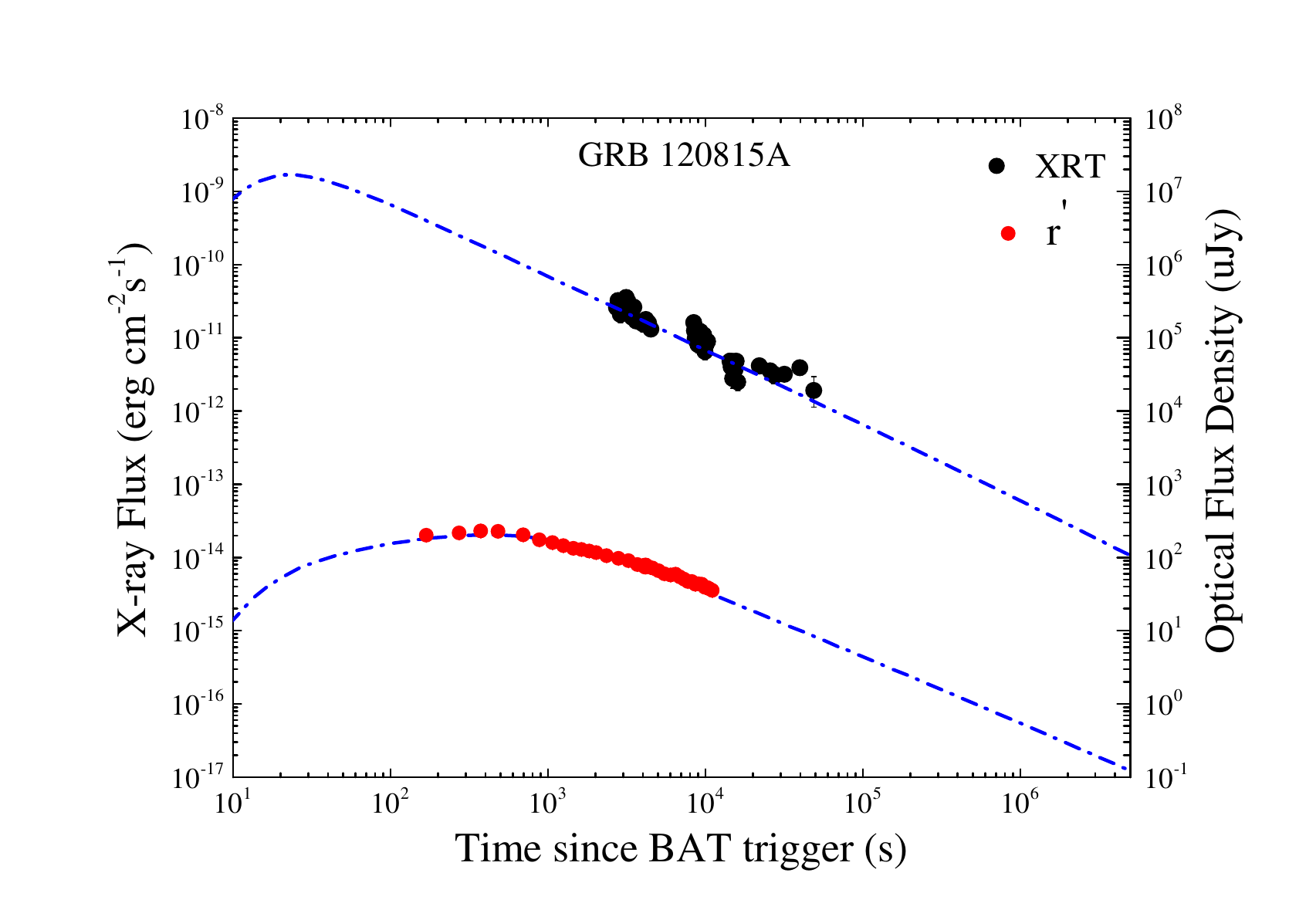}
\includegraphics[angle=0,width=0.30\textwidth]{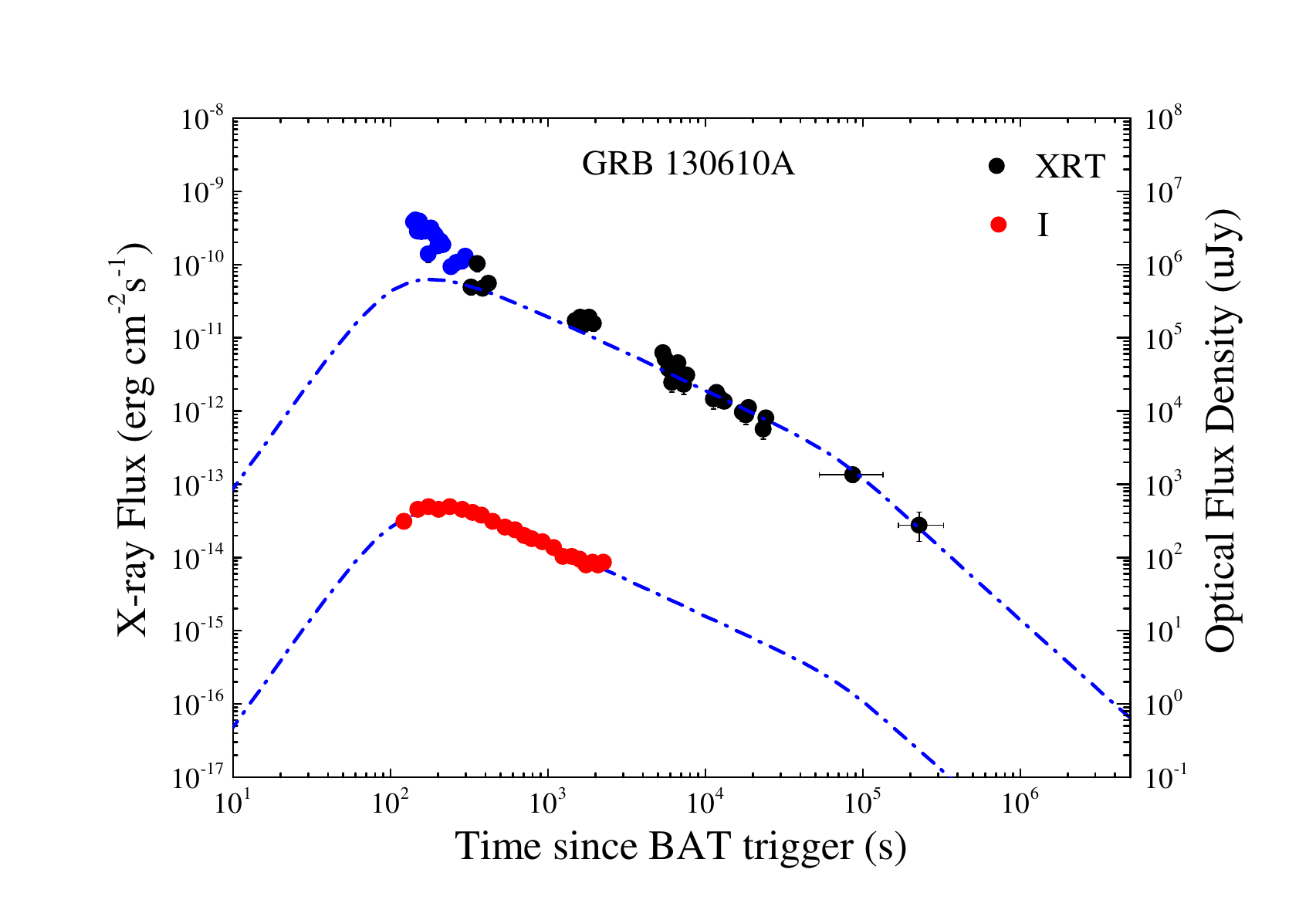}
\includegraphics[angle=0,width=0.30\textwidth]{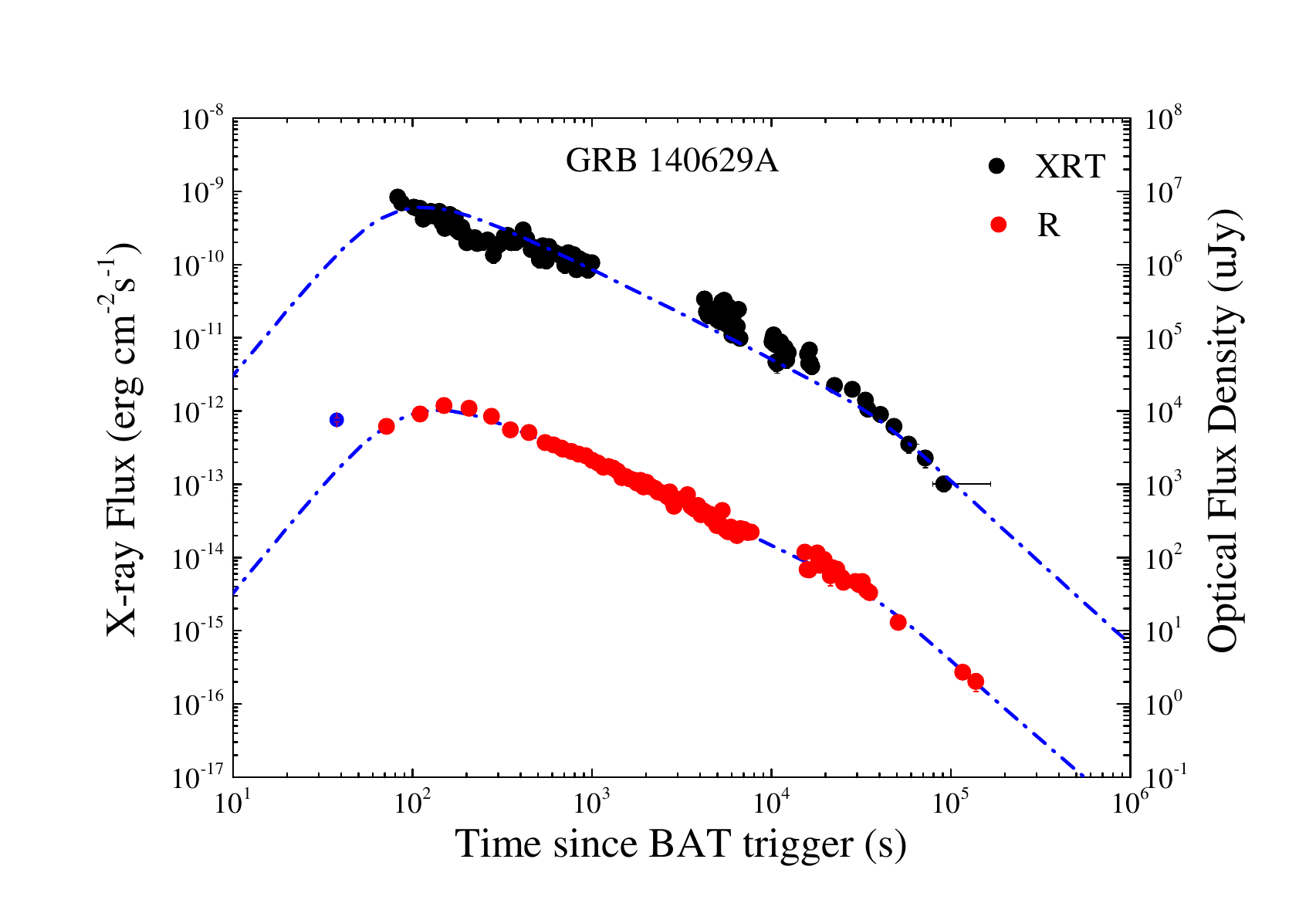}
\includegraphics[angle=0,width=0.30\textwidth]{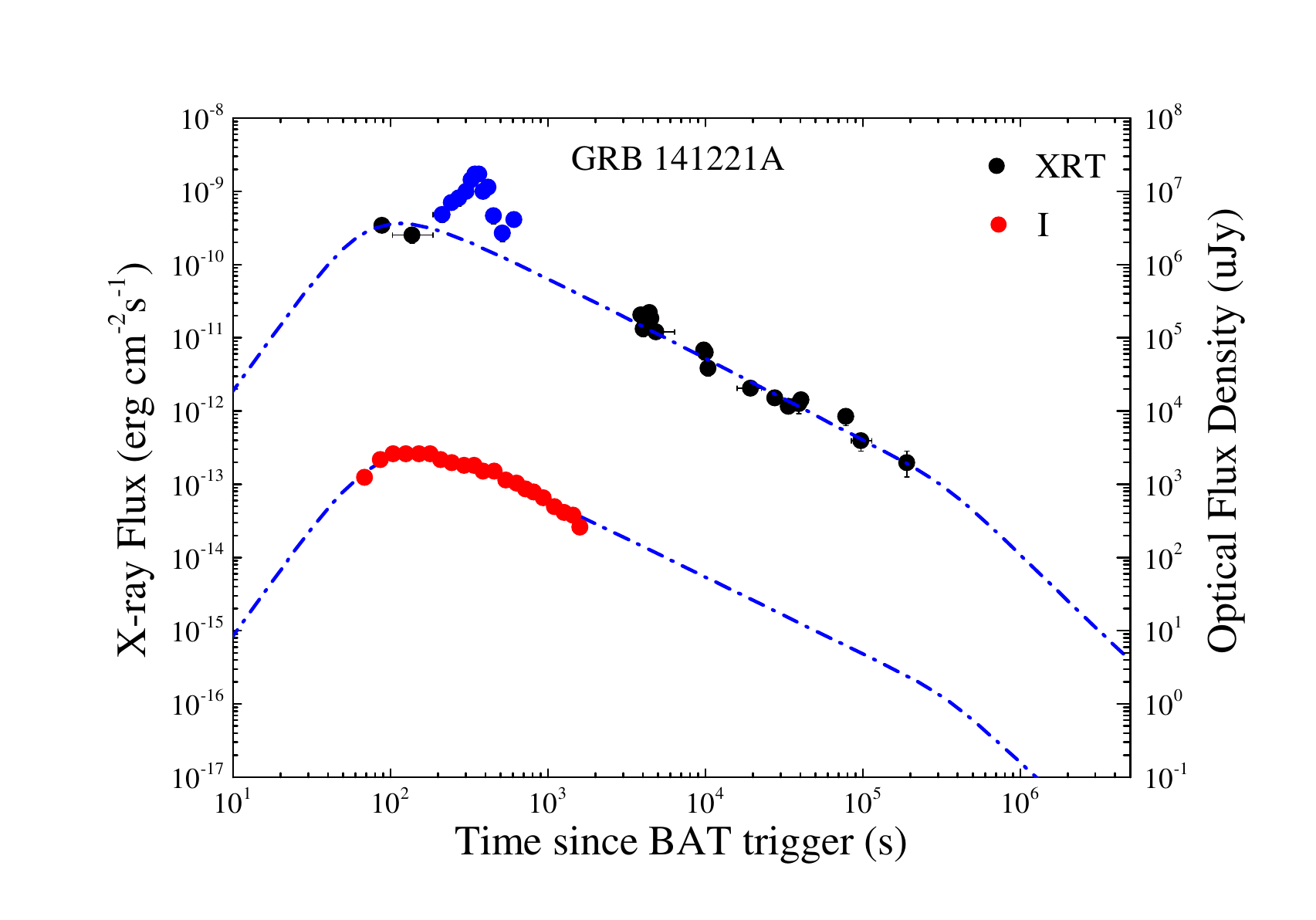}
\includegraphics[angle=0,width=0.30\textwidth]{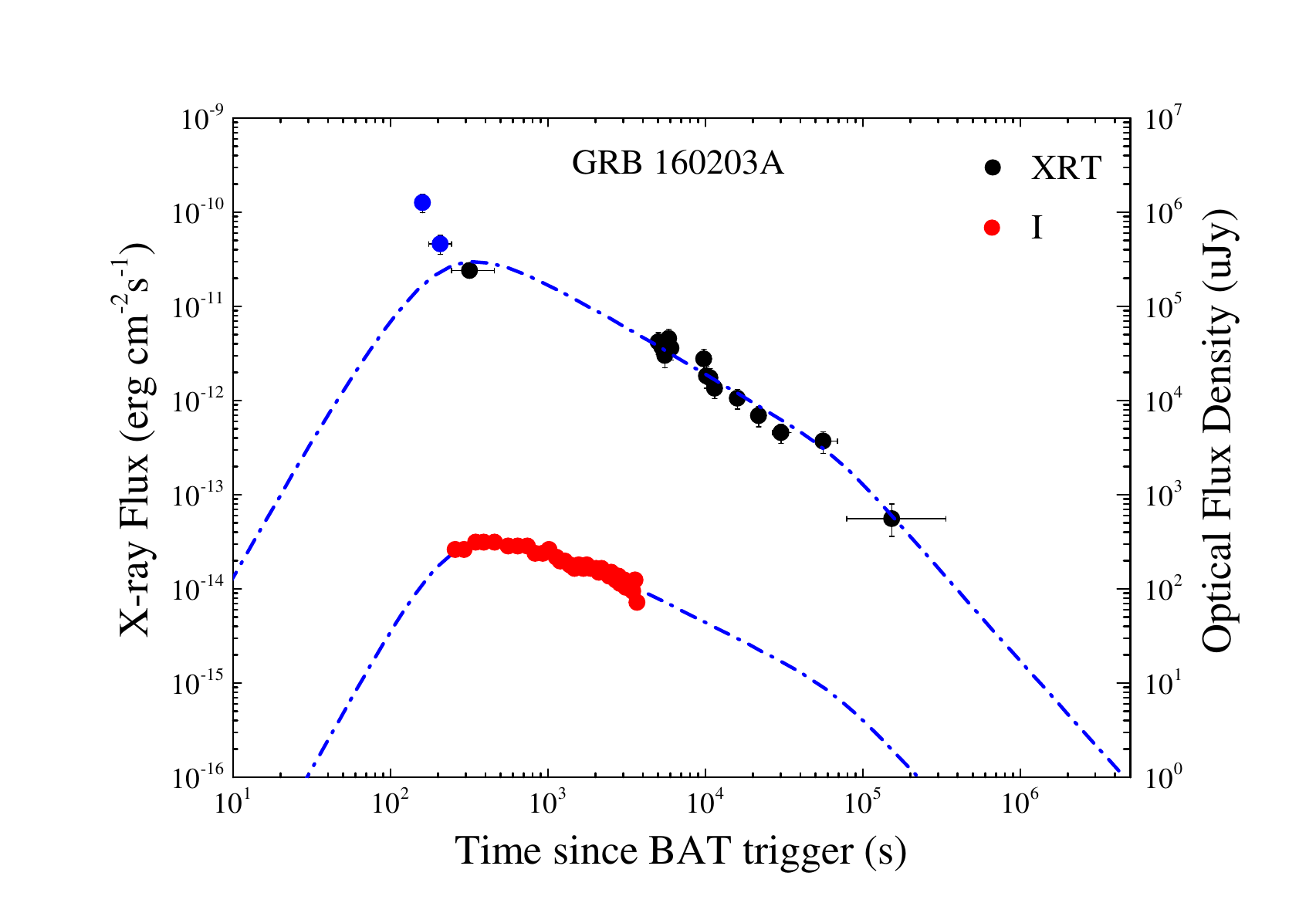}
\includegraphics[angle=0,width=0.30\textwidth]{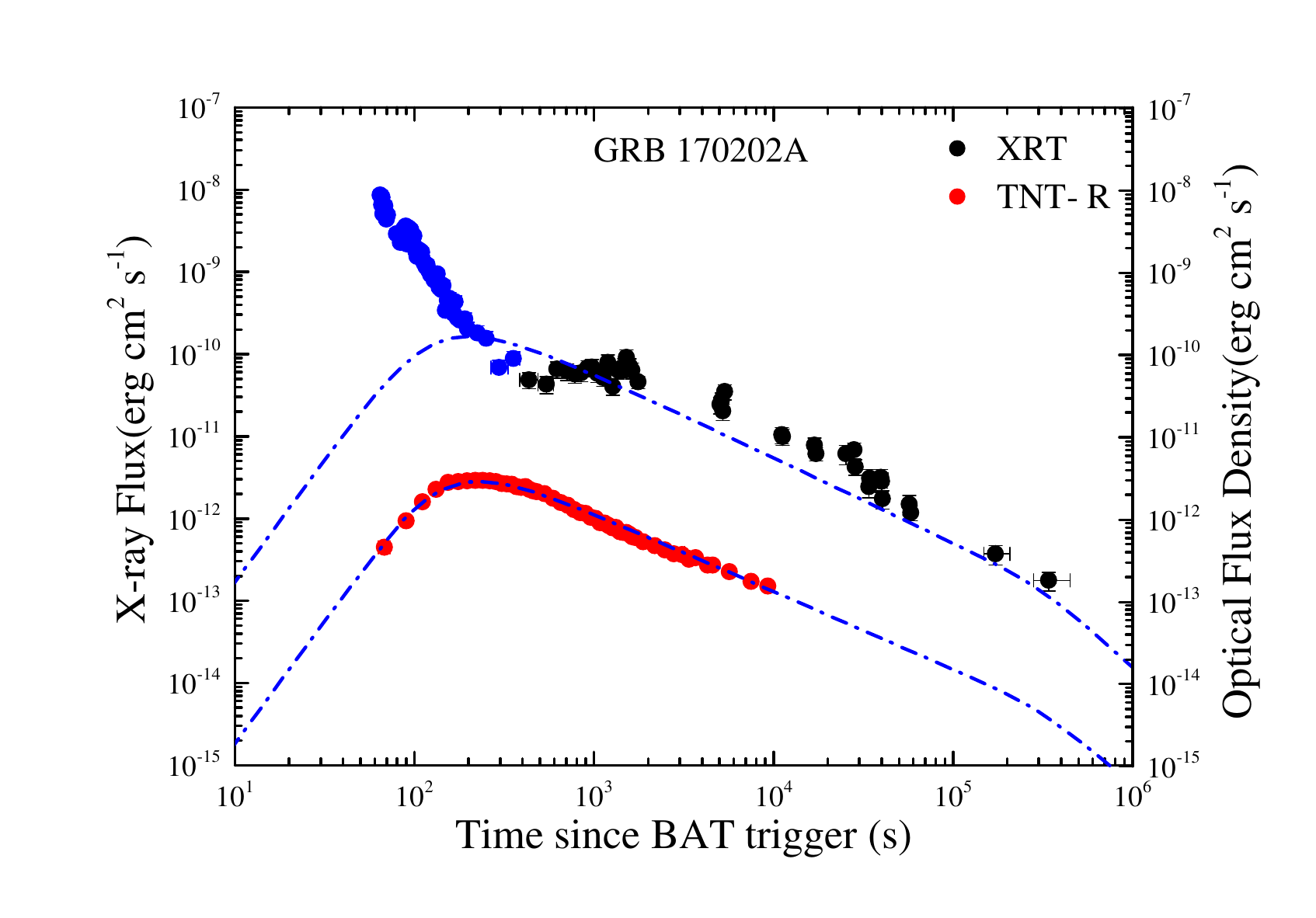}
\includegraphics[angle=0,width=0.30\textwidth]{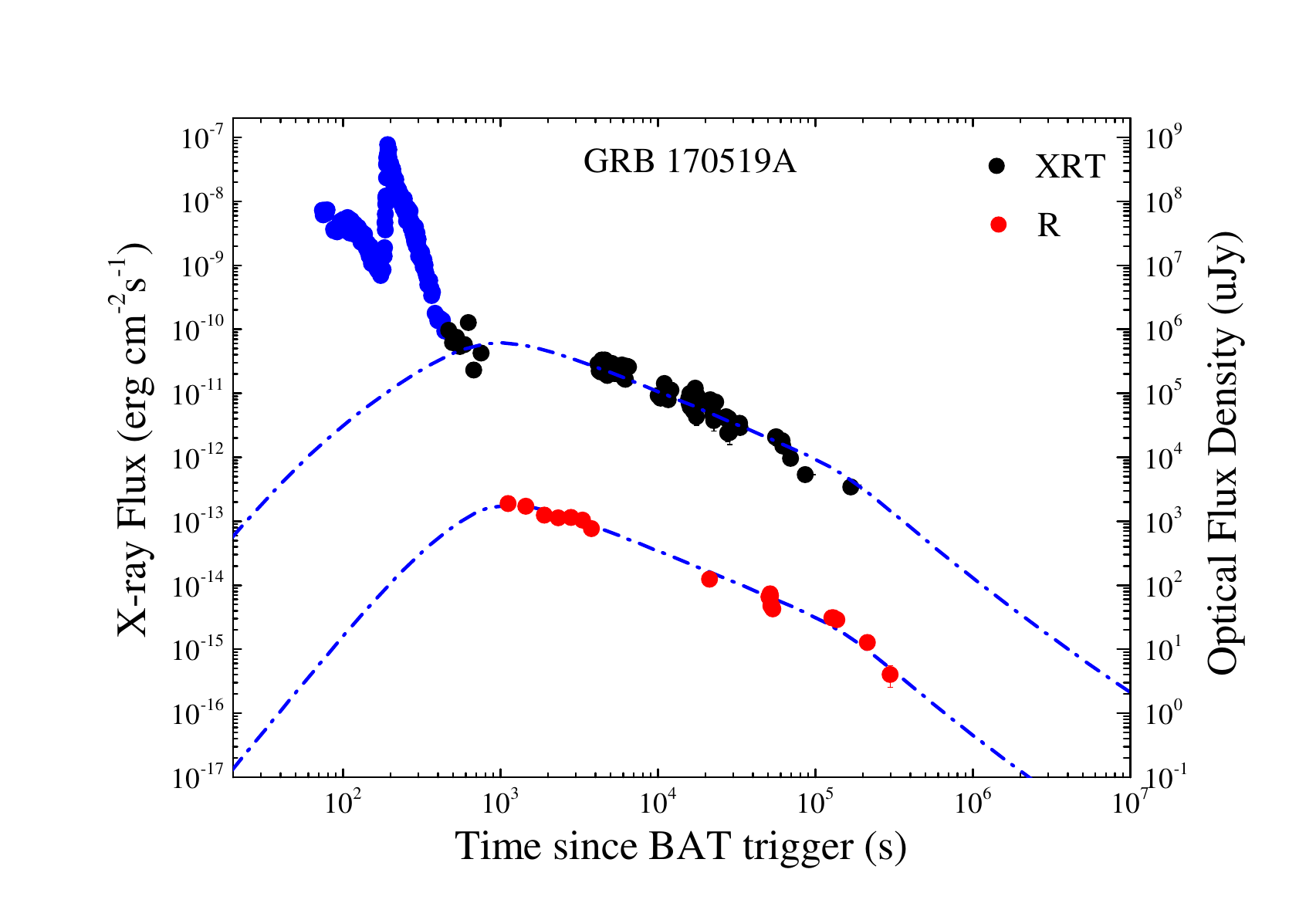}
\includegraphics[angle=0,width=0.30\textwidth]{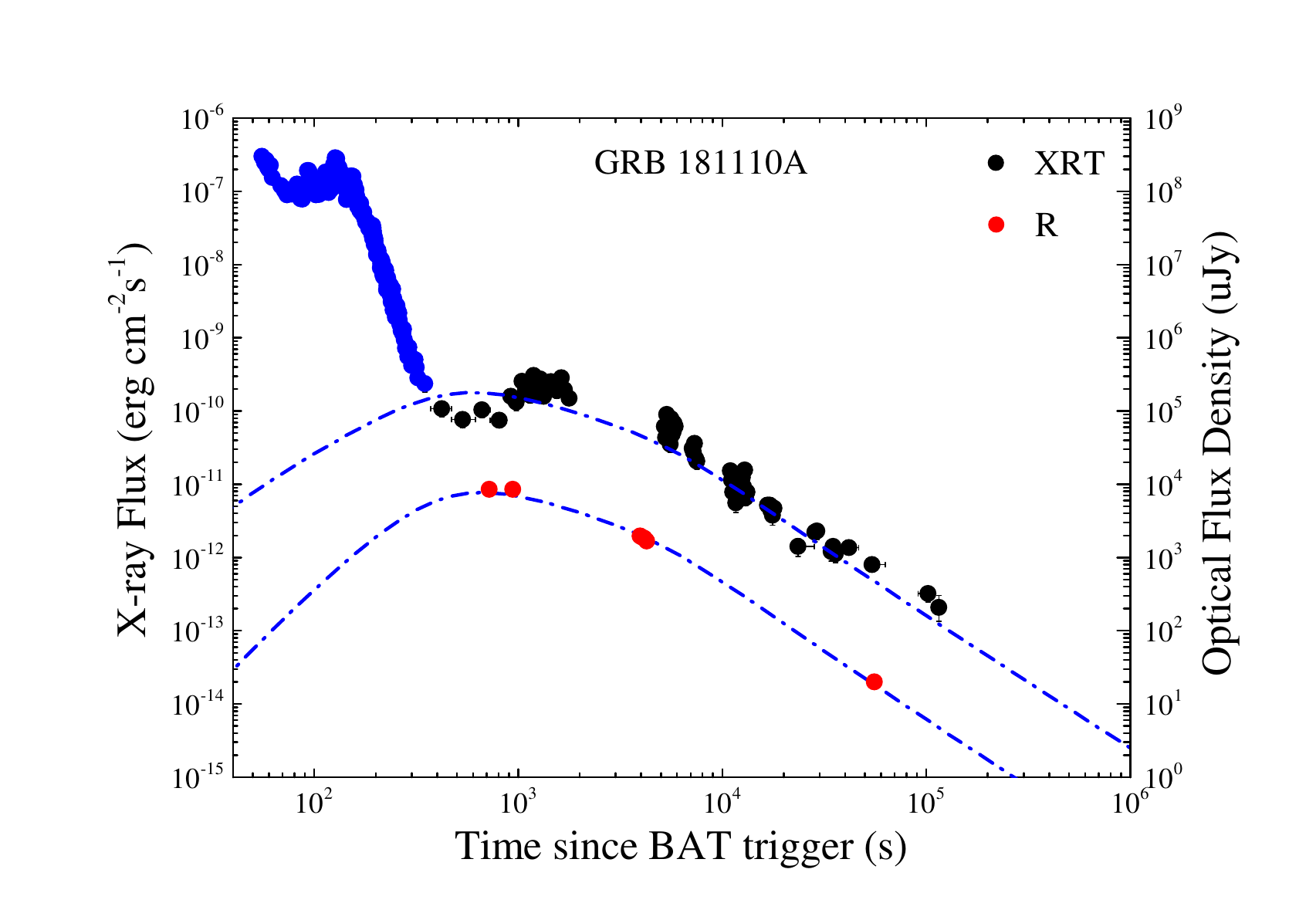}
\end{figure*}

\begin{figure*}[htbp]
\centering
\includegraphics[angle=0,width=1\textwidth]{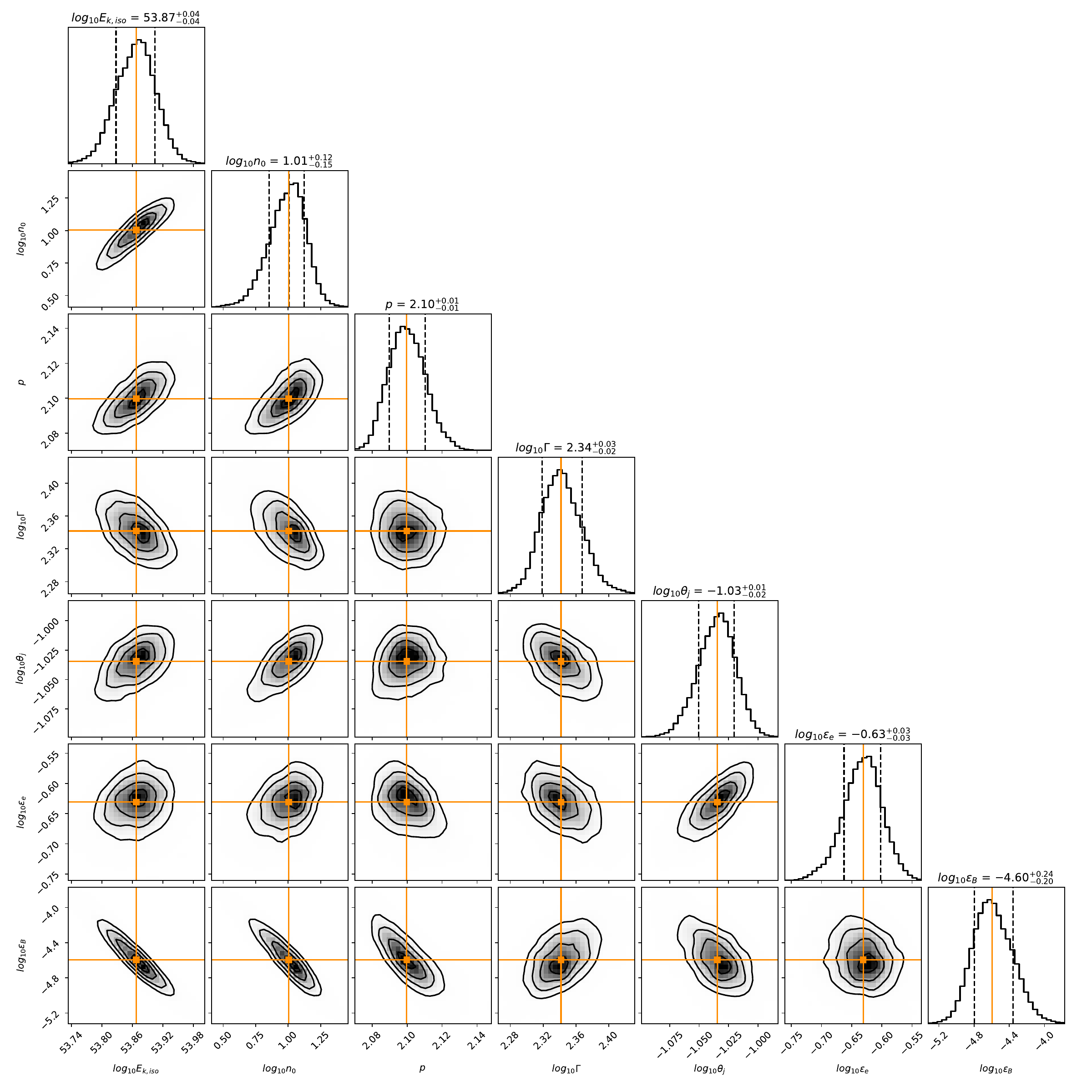}
\caption{Posterior probability distributions of the model parameters of GRB~080310 derived from our fit with the {\tt pymultinest} package. The vertical dashed lines mark the $1\sigma$~confidence level regions centering at the medians of the probability distributions.}
\label{fig1}
\end{figure*}

\begin{figure*}[htbp]
\centering
\subfigure{
\includegraphics[angle=0,width=0.45\textwidth]{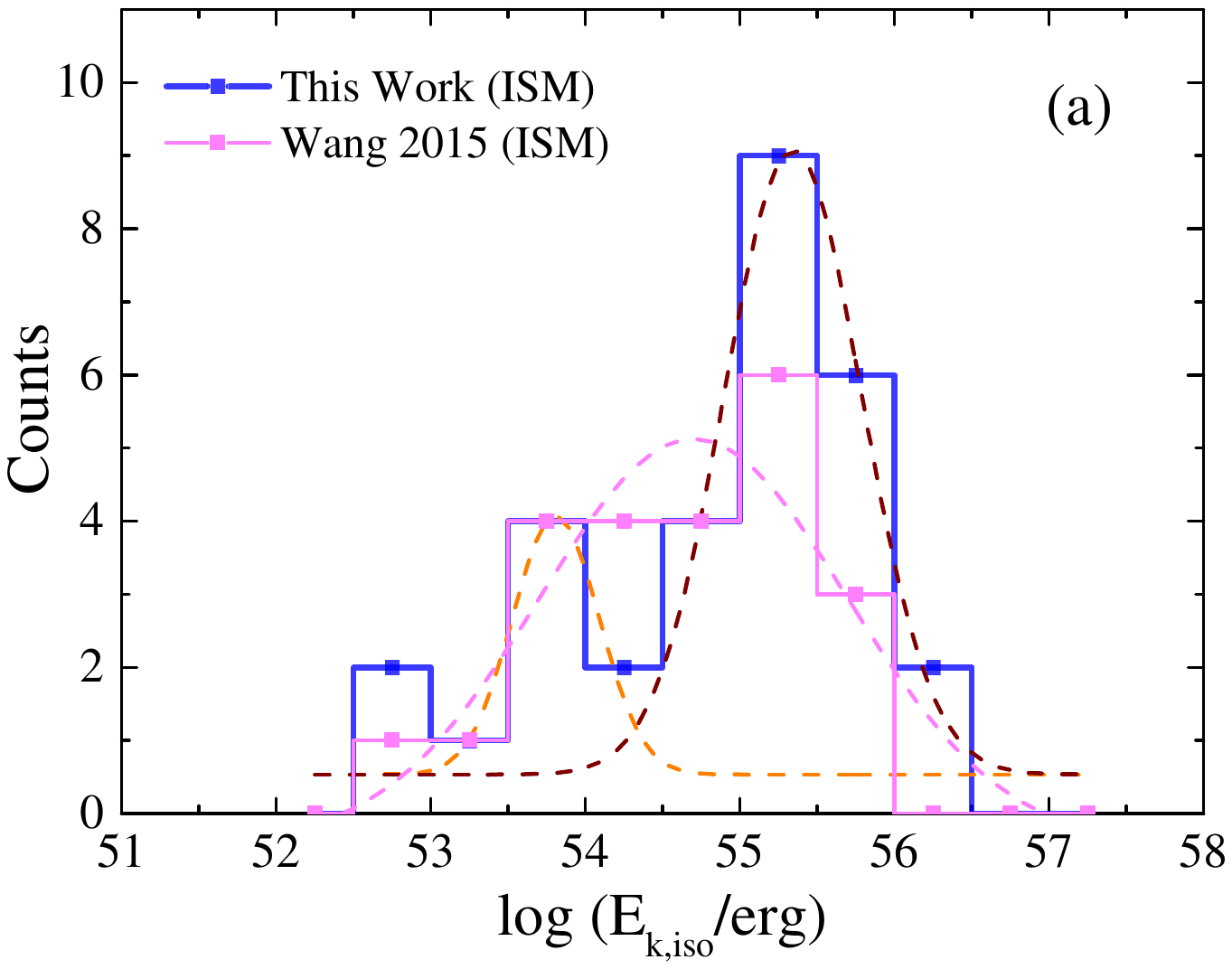}\label{Fig3a}}
\subfigure{
\includegraphics[angle=0,width=0.45\textwidth]{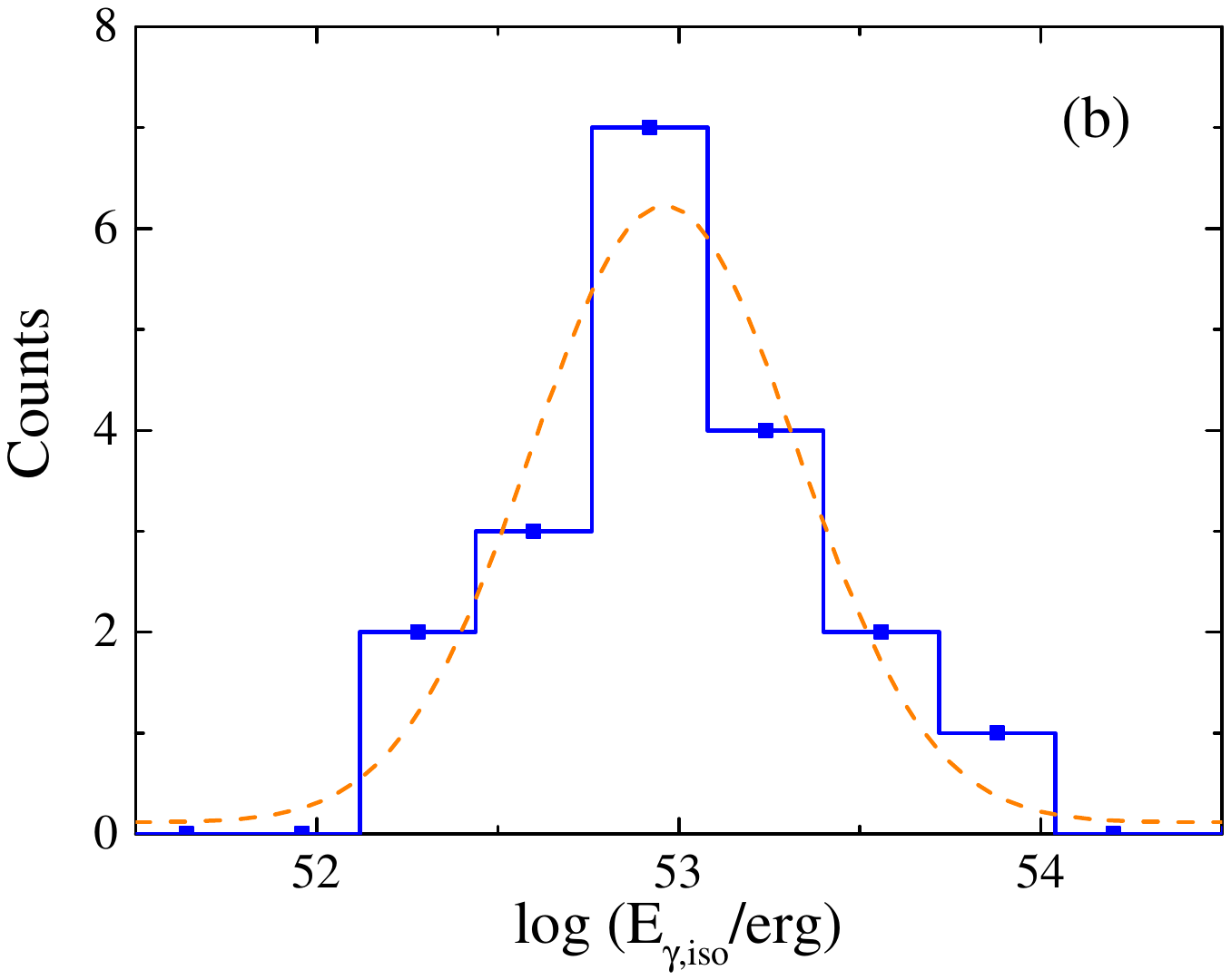}\label{Fig3b}}
\subfigure{
\includegraphics[angle=0,width=0.45\textwidth]{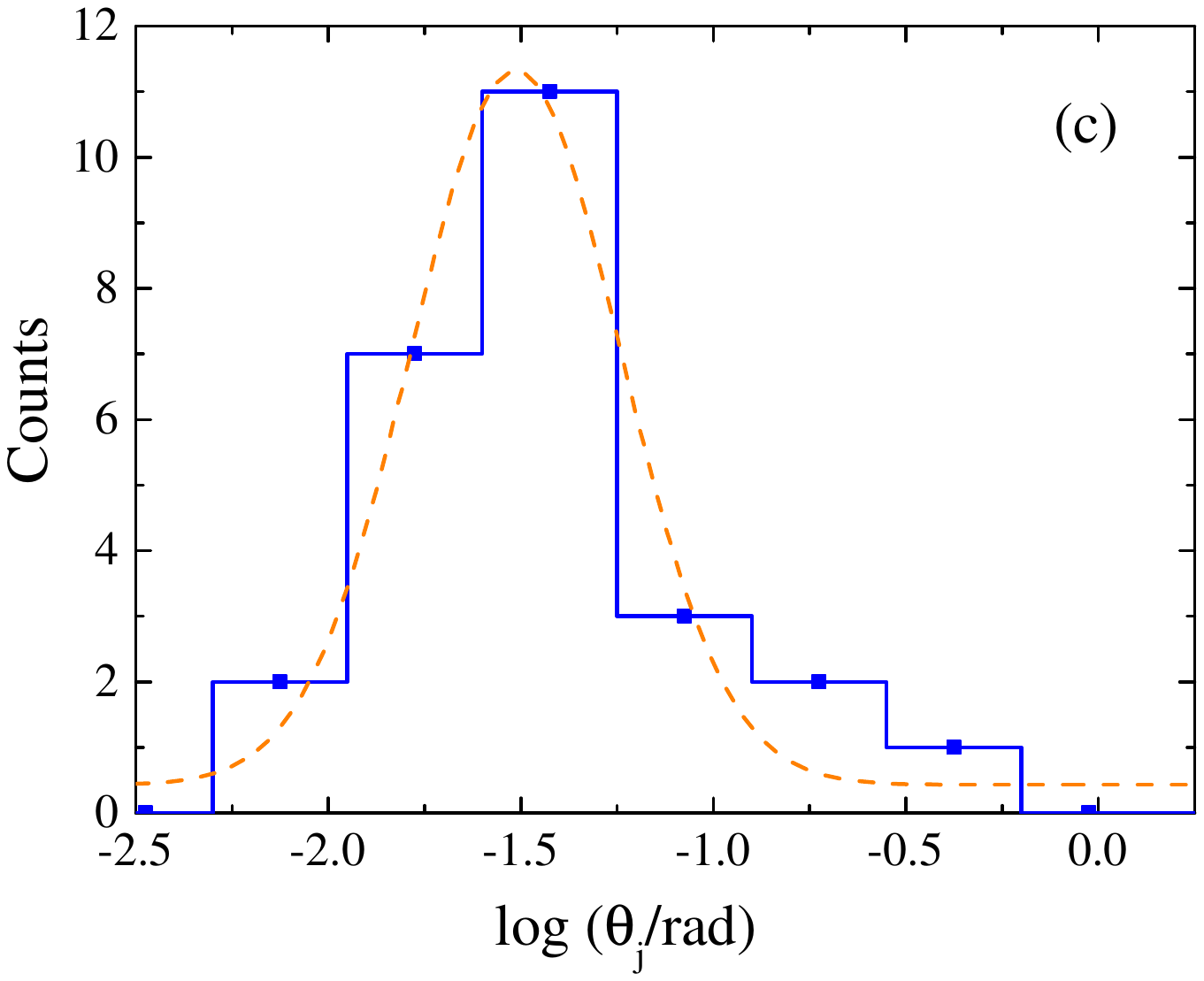}\label{Fig3c}}
\subfigure{
\includegraphics[angle=0,width=0.45\textwidth]{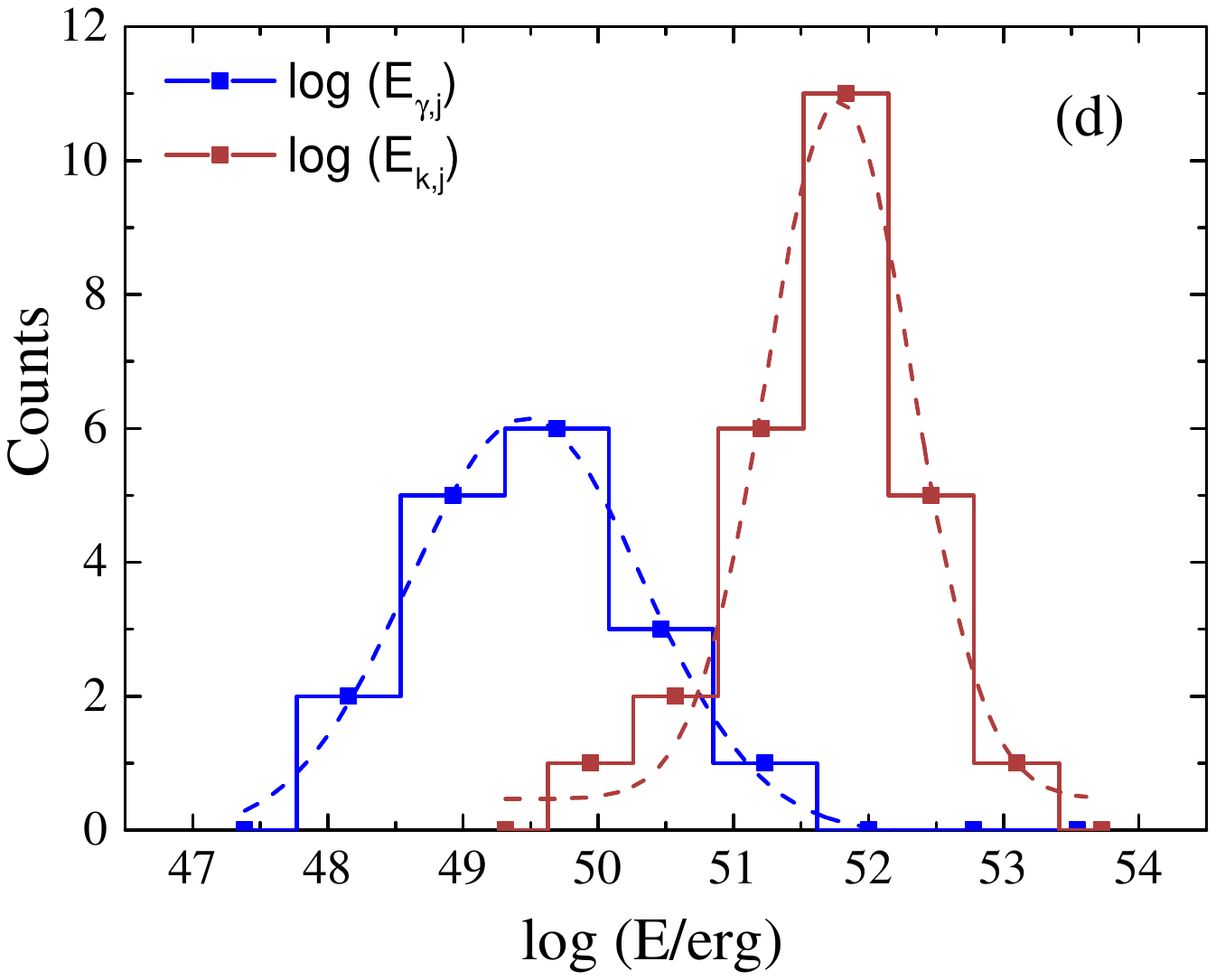}\label{Fig3d}}
\subfigure{
\includegraphics[angle=0,width=0.45\textwidth]{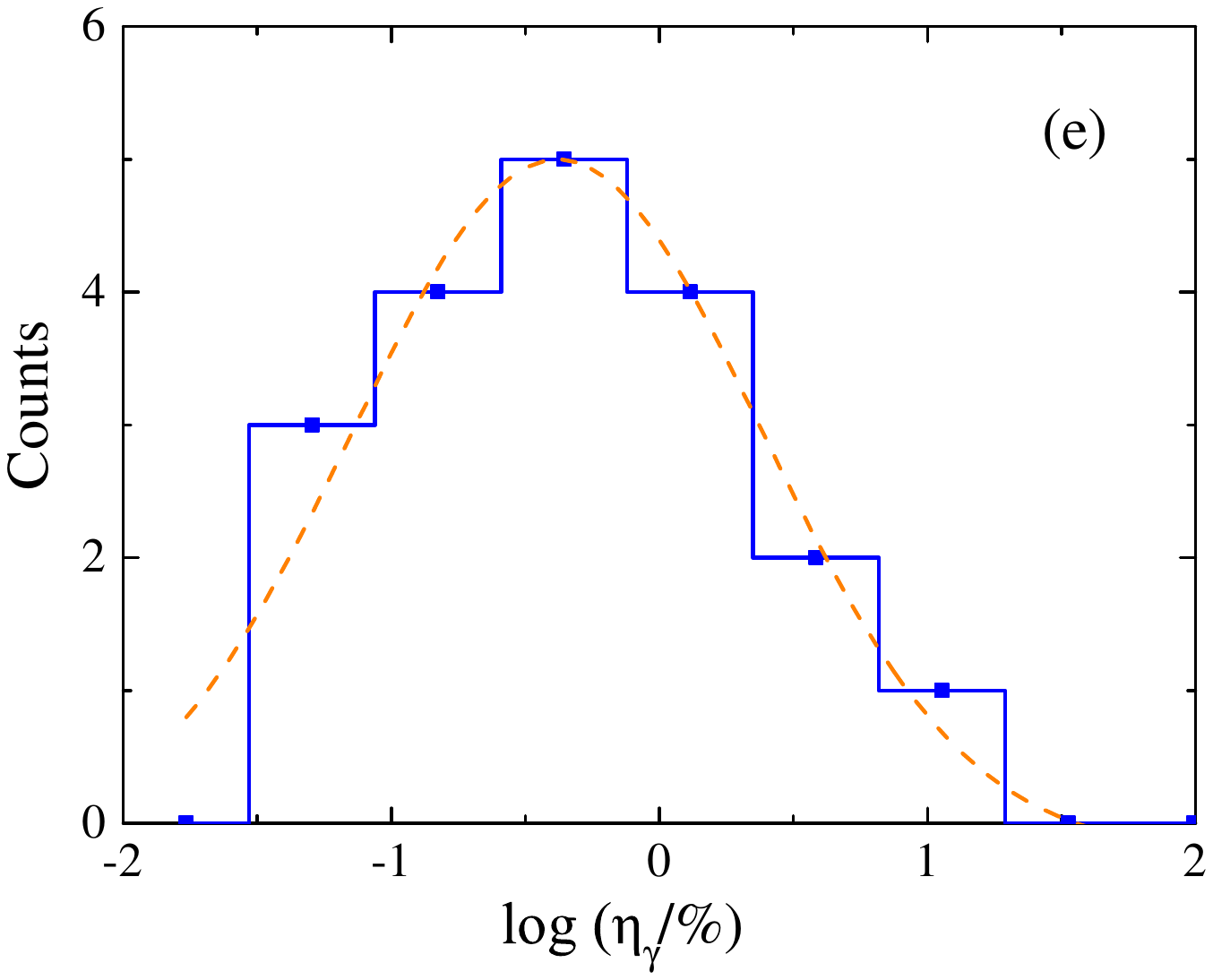}\label{Fig3e}}
\subfigure{
\includegraphics[angle=0,width=0.45\textwidth]{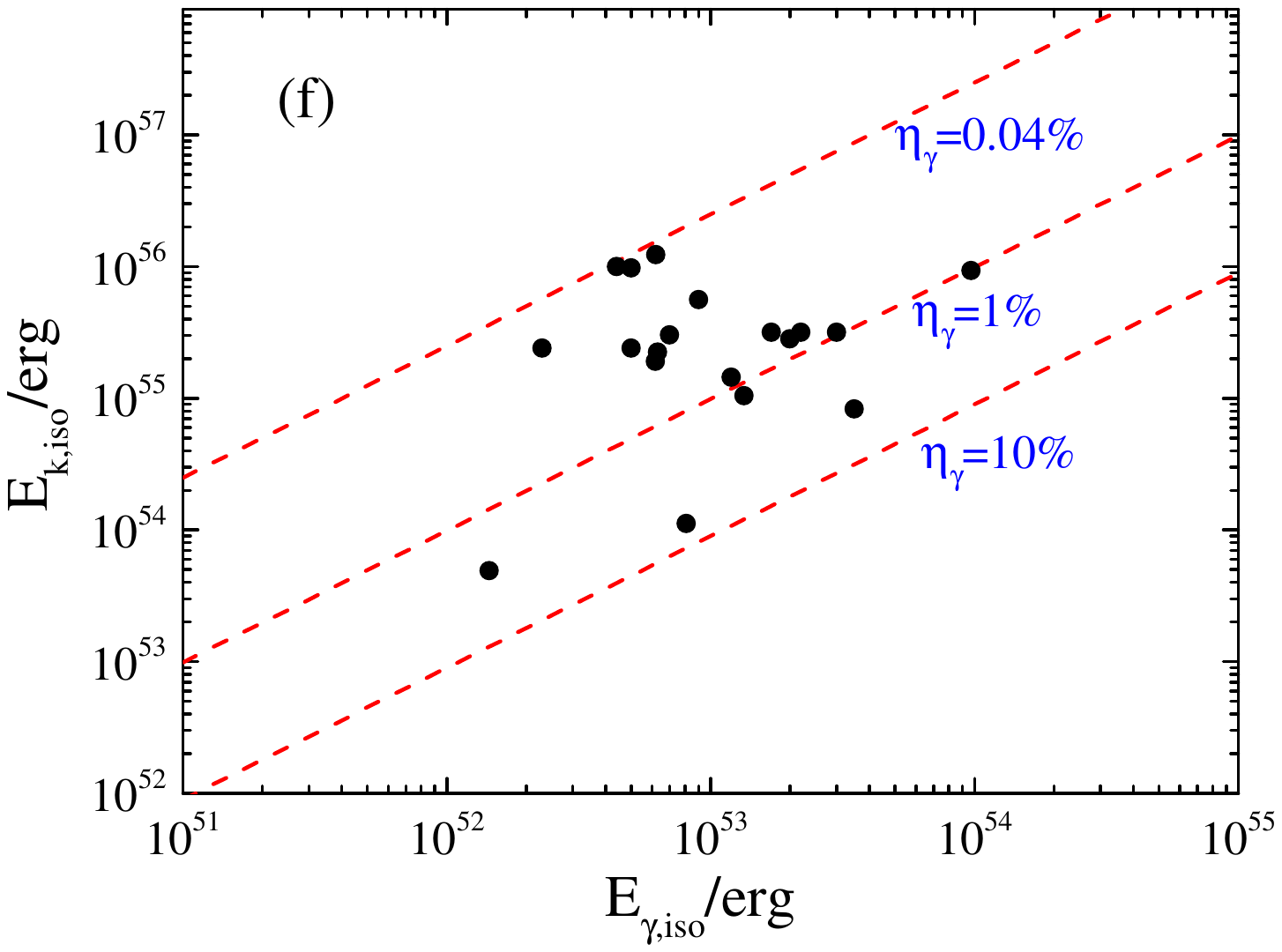}\label{Fig3f}}
\caption{Distributions of 
$\log (E_{\rm k,iso})$, $\log (E_{\rm \gamma,iso})$, 
$\log (\theta_{\rm j})$, $\log (E_{\rm k,j})$, 
$\log (E_{\rm \gamma,j})$, and $\log (\eta_{\rm \gamma})$ as well as $E_{\rm k,iso}$ as a function of $E_{\rm \gamma,iso}$ for the bursts in our sample. 
The dashed curves are the best fit with a Gaussian function or a two-Gaussian function model in panels (a)-(e). The $\log E_{\rm k, iso}$ distribution taken from \cite{Wang_2015_Zhang_ApJS..219....9W} is also shown with a magenta curve in panel (a) for comparison. The red dashed lines mark the line for $\eta_{\rm \gamma}=0.04\%, 0.1\%$, and $10\%$ in panel (f).}
\label{Fig3}
\end{figure*}
\newpage

\begin{figure*}[htbp]
\centering
\subfigure{
\includegraphics[angle=0,width=0.45\textwidth]{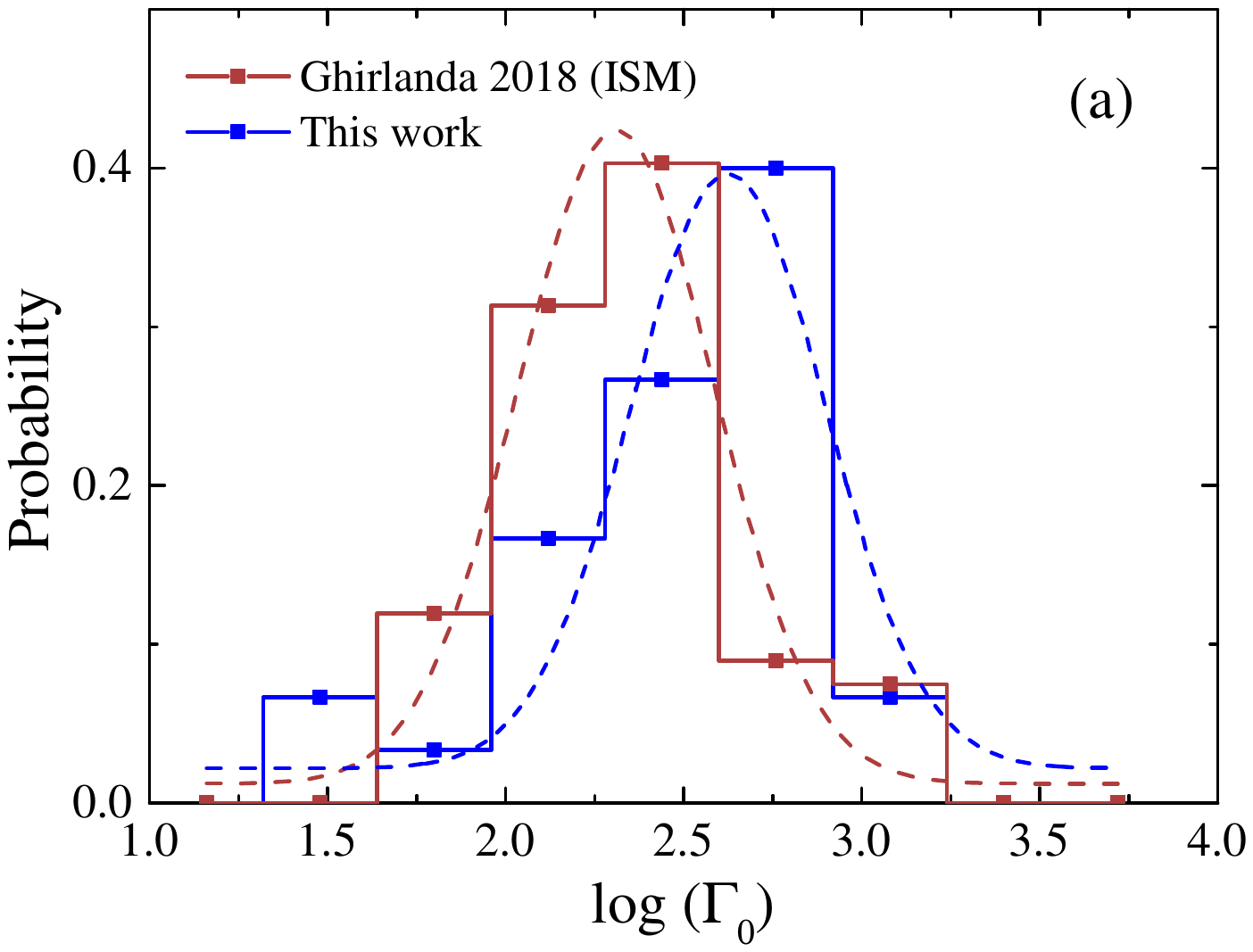}\label{Fig4a}}
\subfigure{
\includegraphics[angle=0,width=0.45\textwidth]{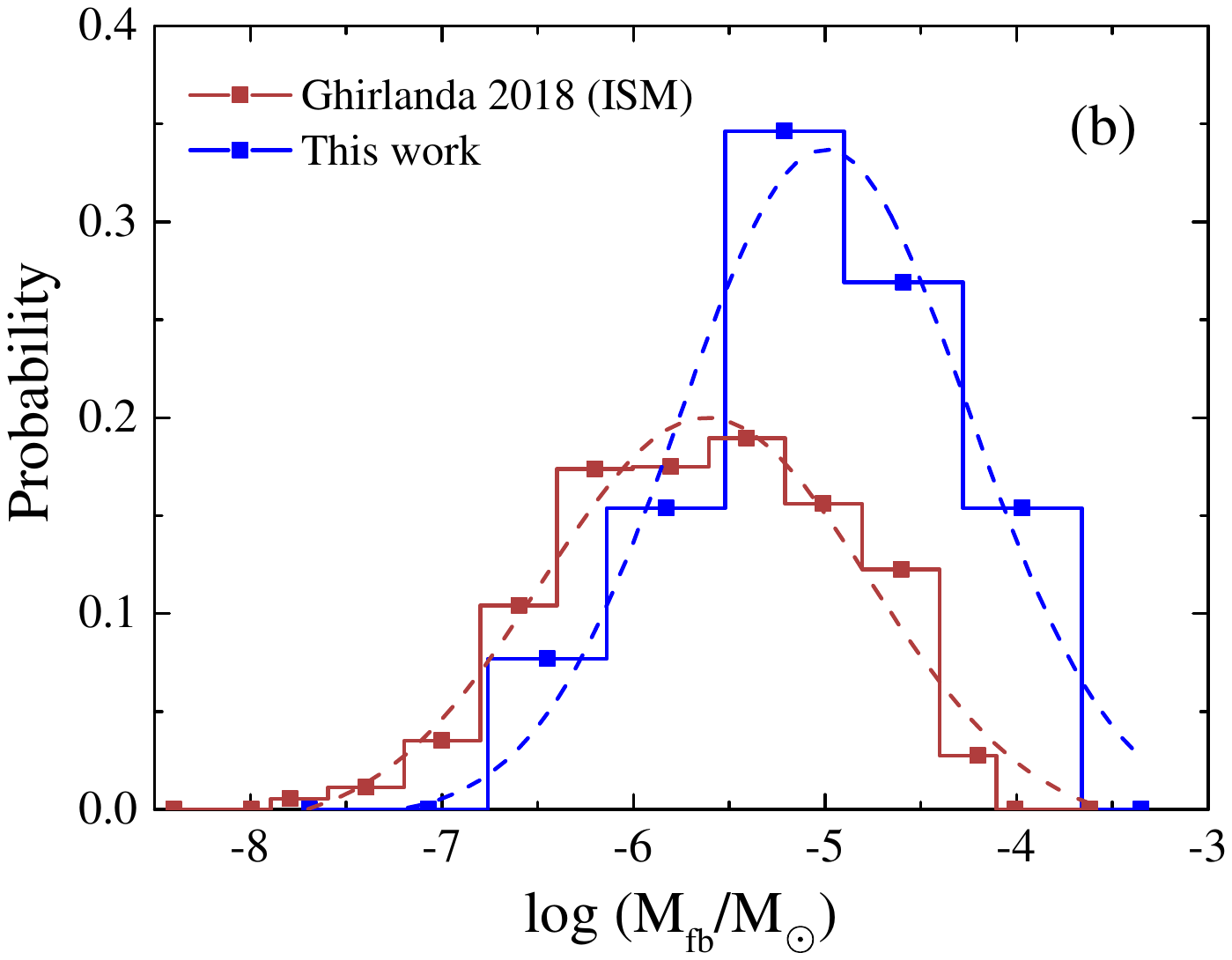}\label{Fig4b}}
\subfigure{
\includegraphics[angle=0,width=0.45\textwidth]{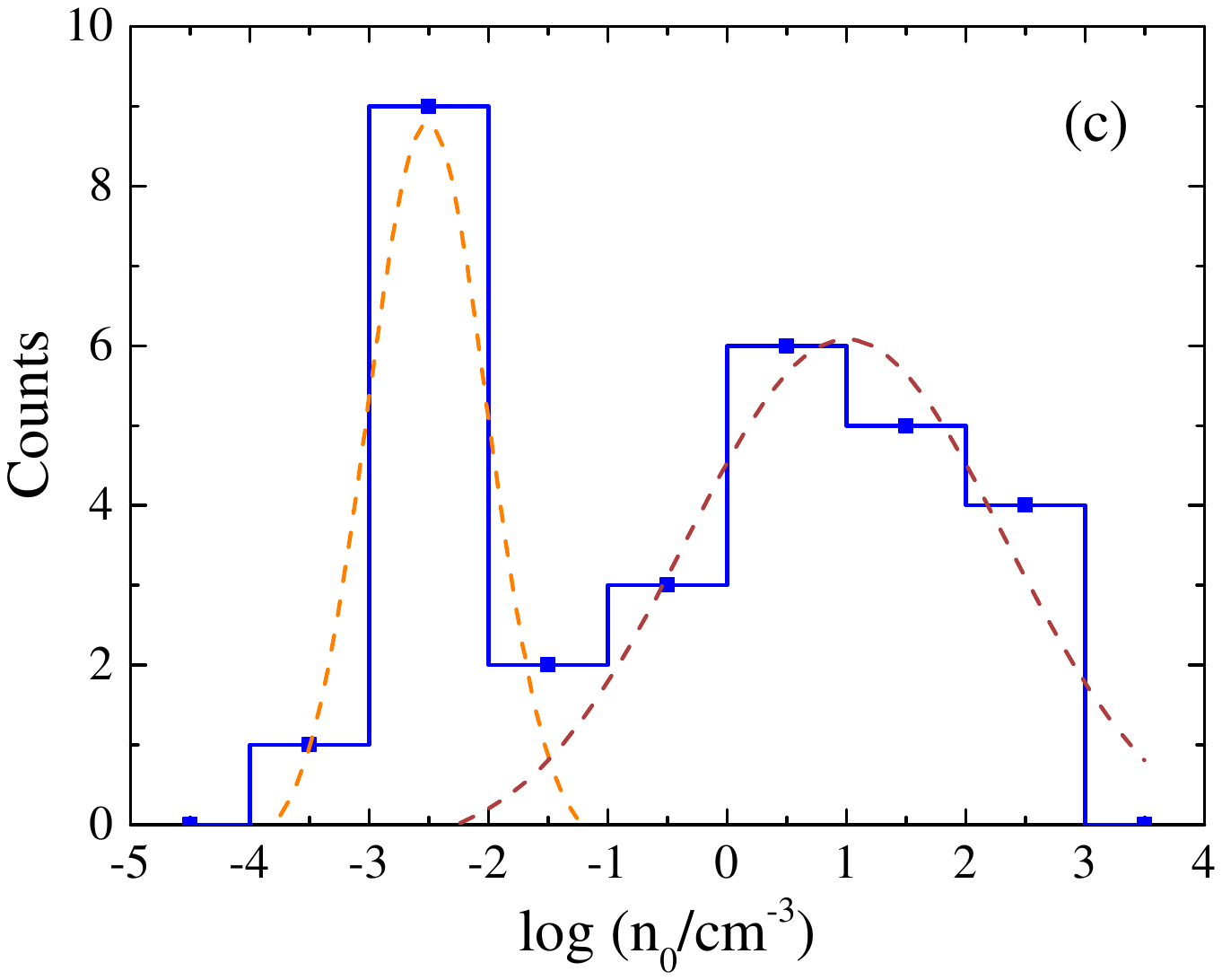}\label{Fig4c}}
\subfigure{
\includegraphics[angle=0,width=0.45\textwidth]{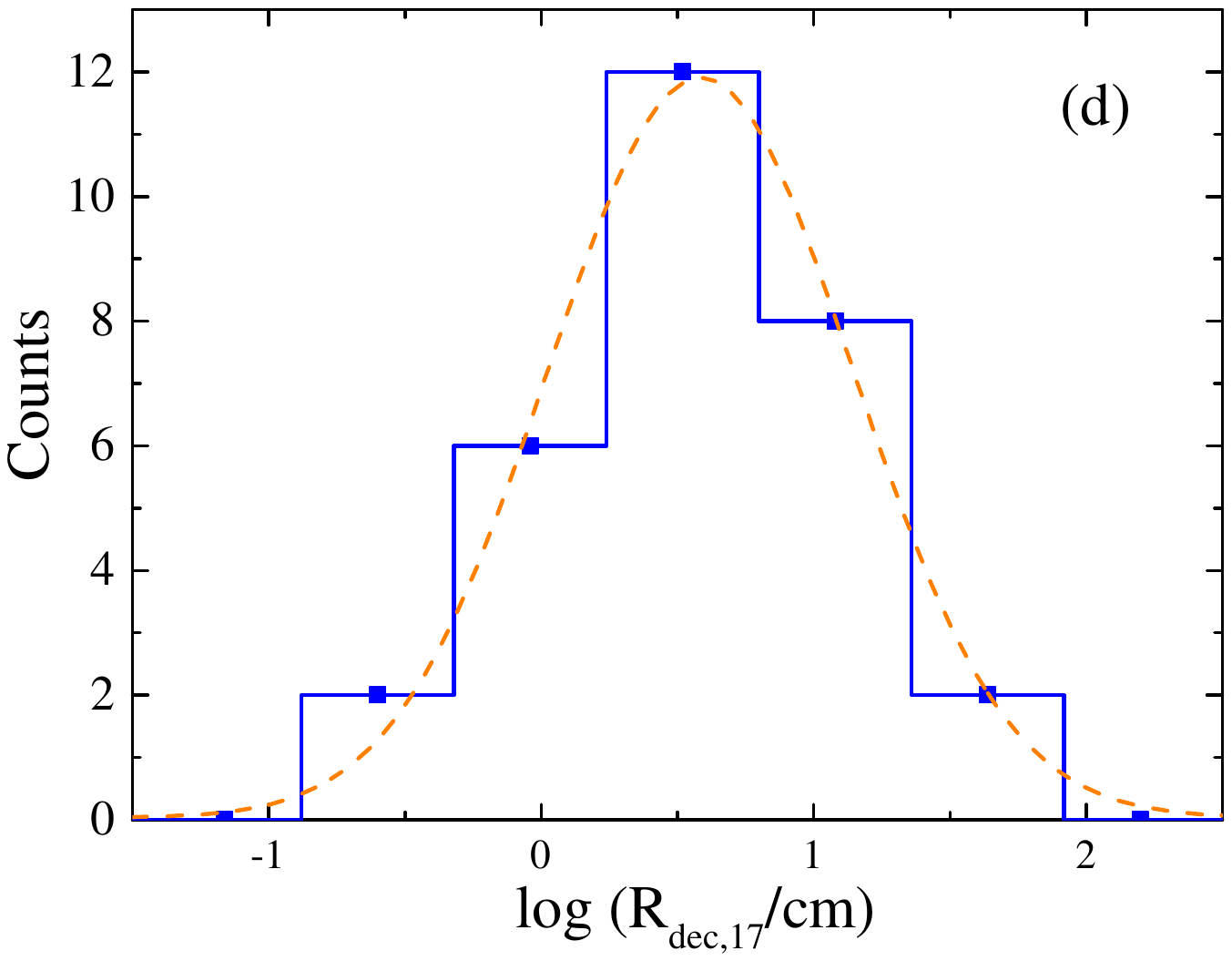}\label{Fig4d}}
\caption
{Distributions of 
$\log \Gamma_{0}$, $\log (M_{\rm fb}/M_{\rm \odot})$, $\log n_{0}$ and $\log R_{\rm dec,17}$ 
for the bursts in our sample. 
The $\log \Gamma_0$ and $\log (M_{\rm fb}/M_{\rm \odot})$ distributions taken from \cite{Ghirlanda_2018_Nappo_A_A...609A.112G} are also shown with red curves in panels (a) and (b) for comparison. The dashed lines in each panel represent the best fitting of the log-normal/bimodal Gaussian function model.}
\label{Fig4}
\end{figure*}

\begin{figure*}[htbp]
\centering
\subfigure{
\includegraphics[angle=0,width=0.45\textwidth]{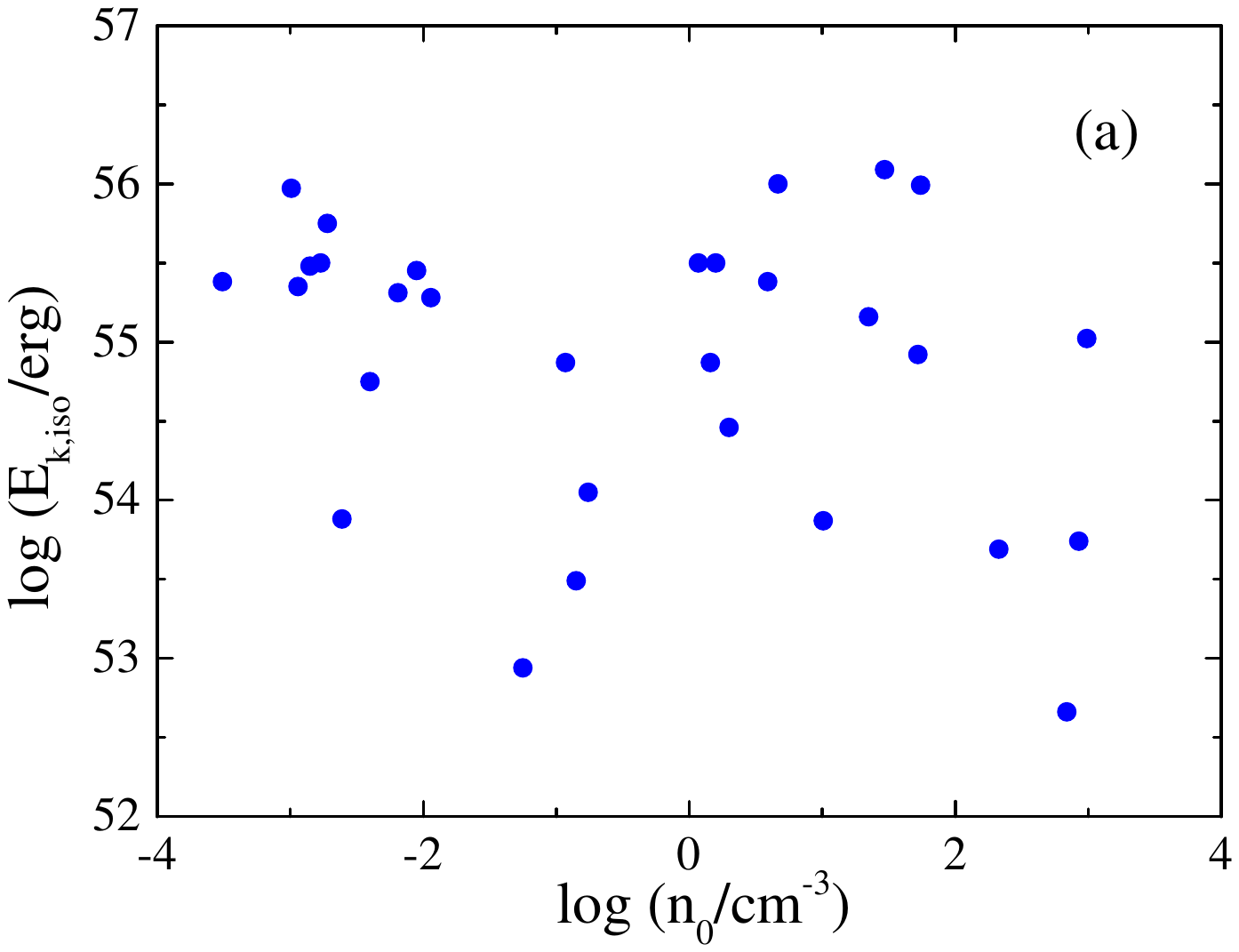}\label{Fig7a}}
\subfigure{
\includegraphics[angle=0,width=0.45\textwidth]{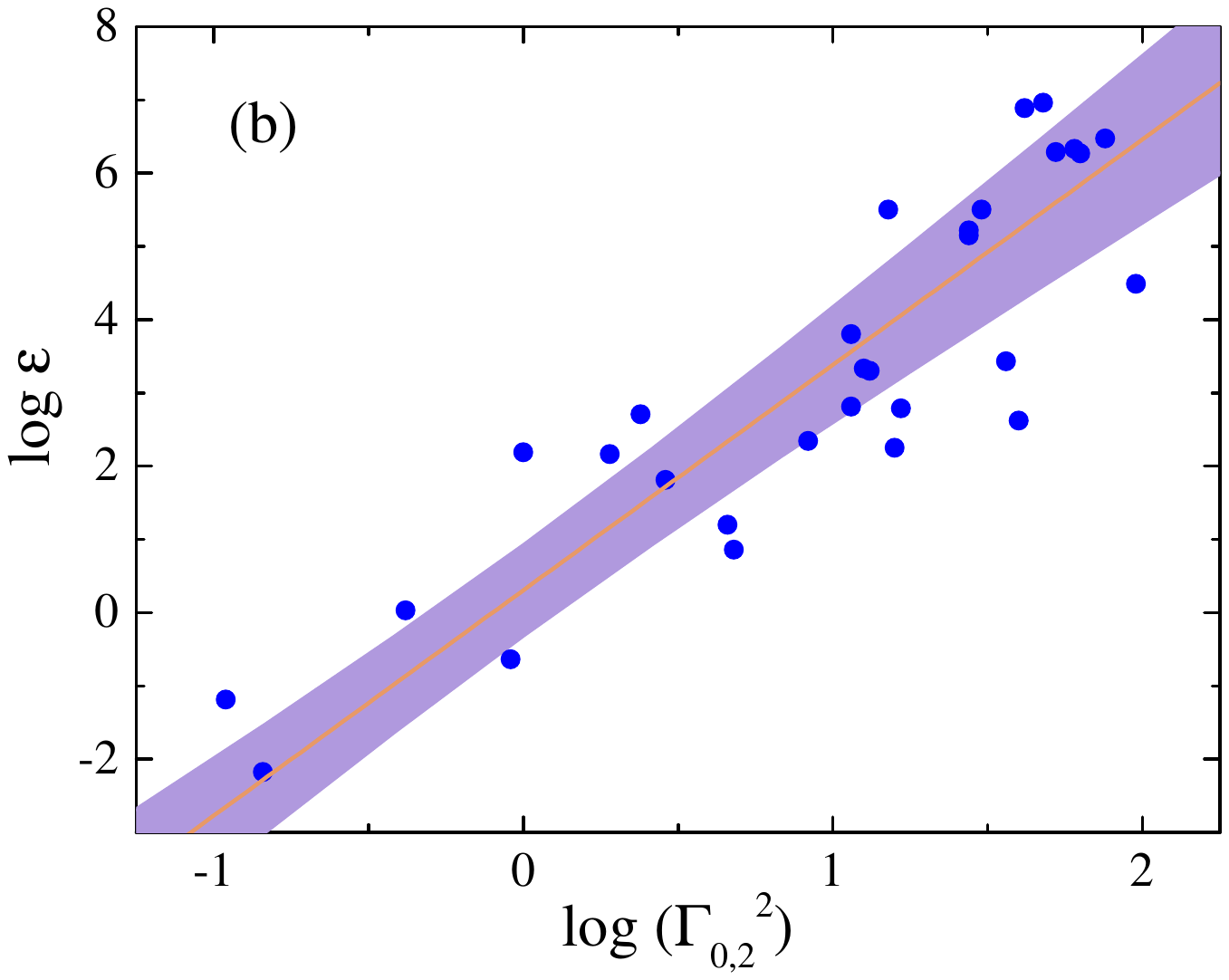}\label{Fig7b}}
\caption{Panel (a)--GRB distribution in the $\log n_{0}-\log E_{\rm k,iso}$ panel for the bursts in our sample. Panel (b)-- $\varepsilon$ as a function of $\log \Gamma_{0,2}^{2}$. Their relation derived from the best fit with the OLS bisector algorithm is marked with an orange solid line and its confidence level of $2\sigma$ is marked with a purple band.}
\label{Fig7}
\end{figure*}

\begin{figure*}[htbp]
\centering
\subfigure{
\includegraphics[angle=0,width=0.45\textwidth]{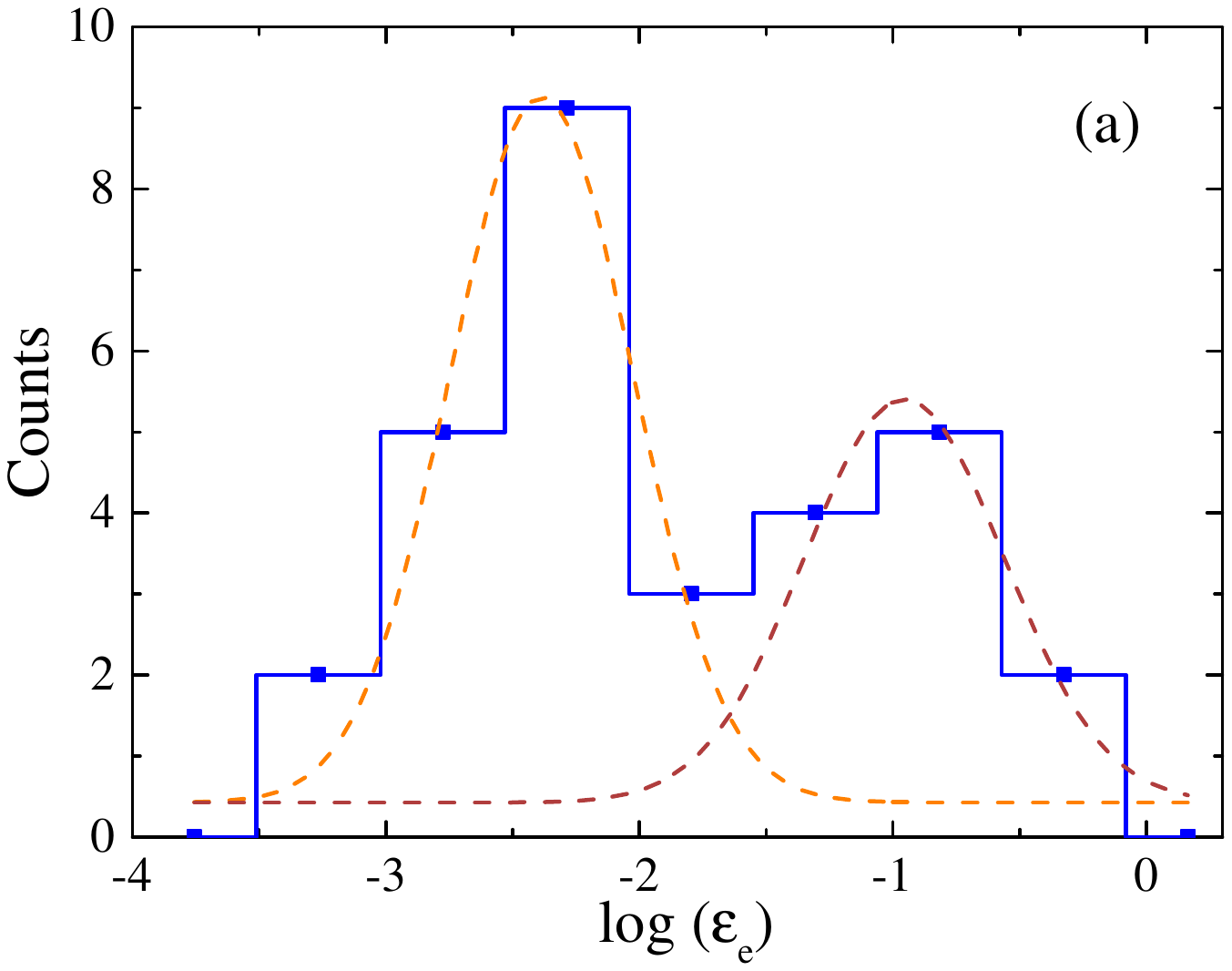}\label{Fig5a}}
\subfigure{
\includegraphics[angle=0,width=0.45\textwidth]{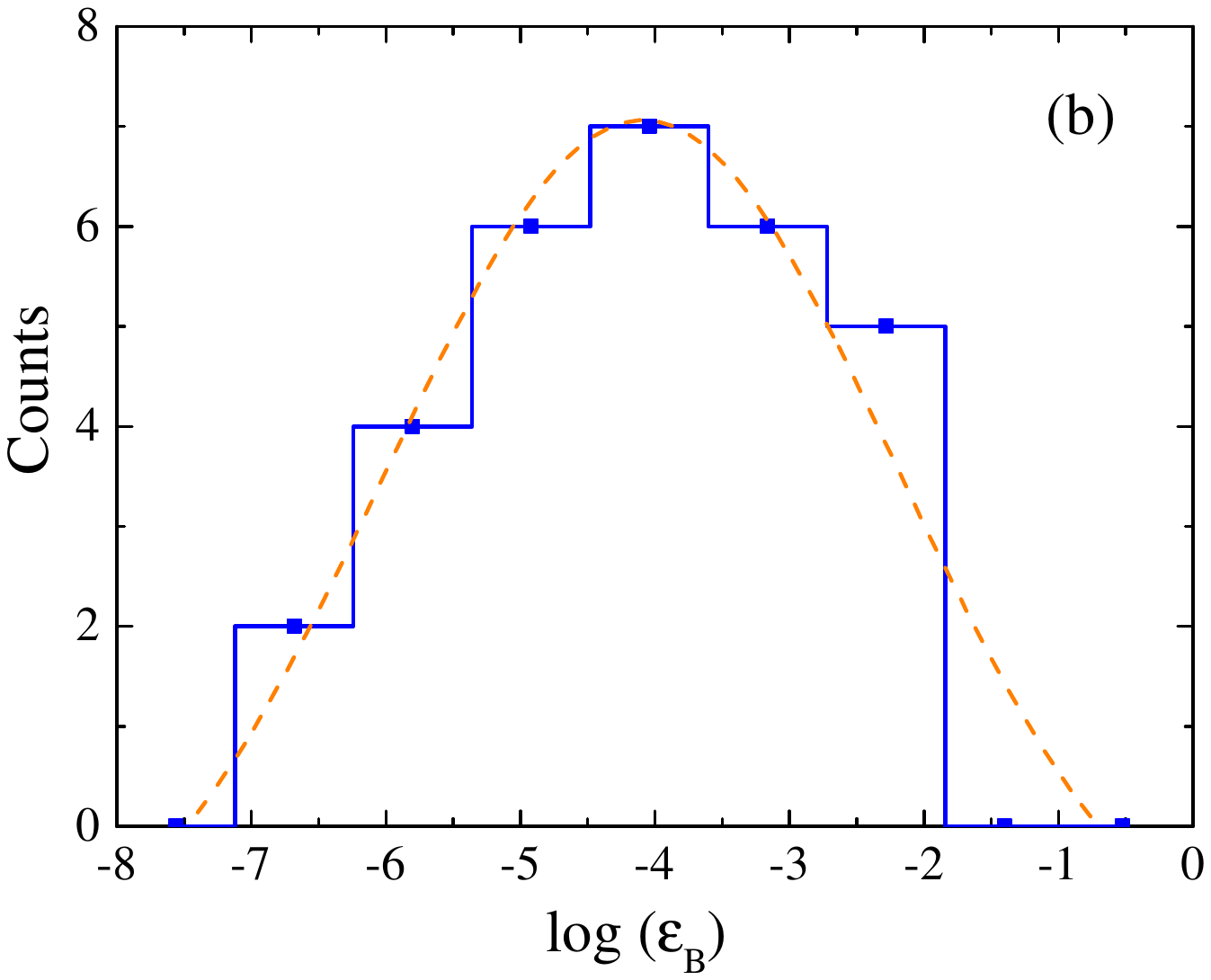}\label{Fig5b}}
\subfigure{
\includegraphics[angle=0,width=0.45\textwidth]{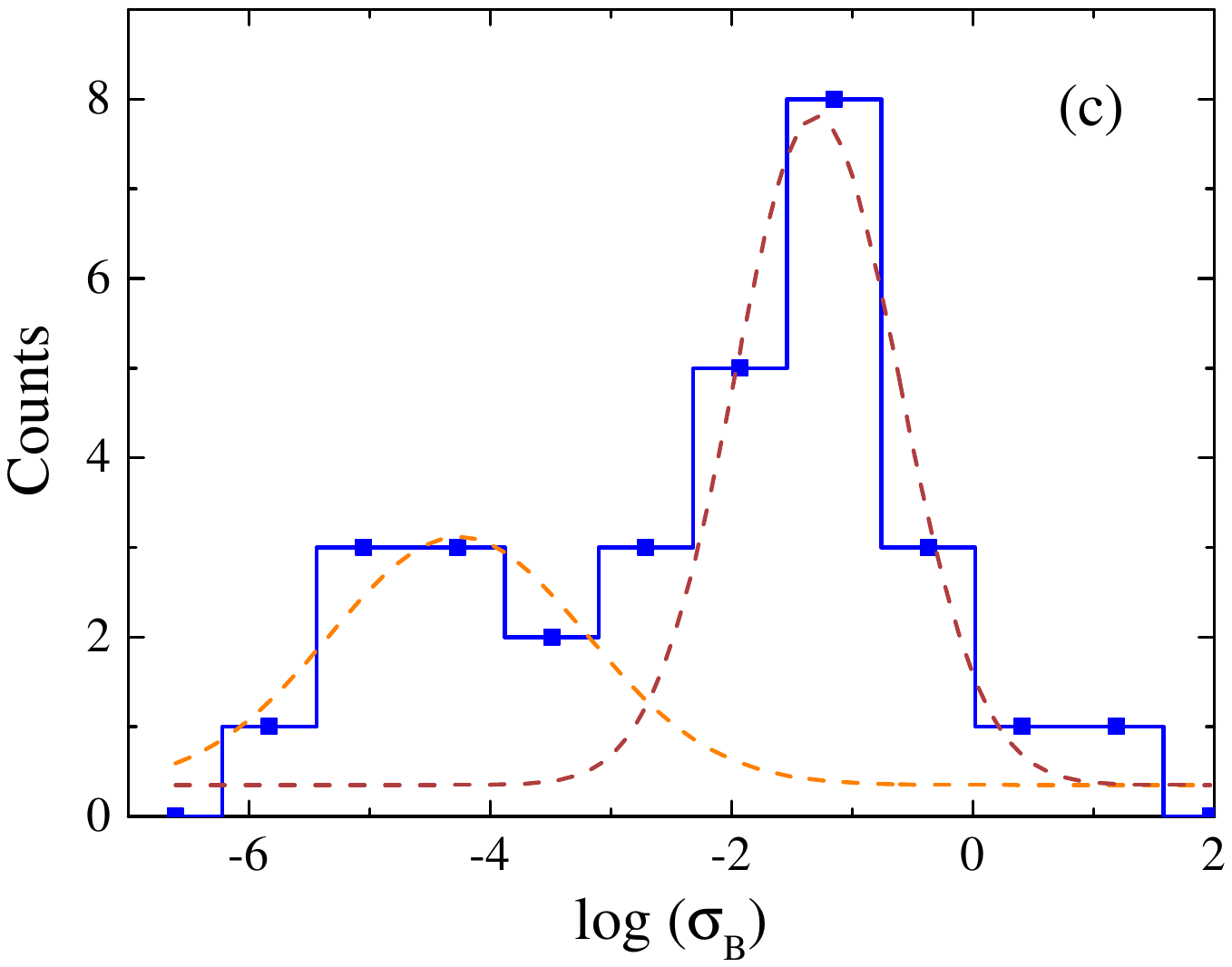}\label{Fig5c}}
\subfigure{
\includegraphics[angle=0,width=0.45\textwidth]{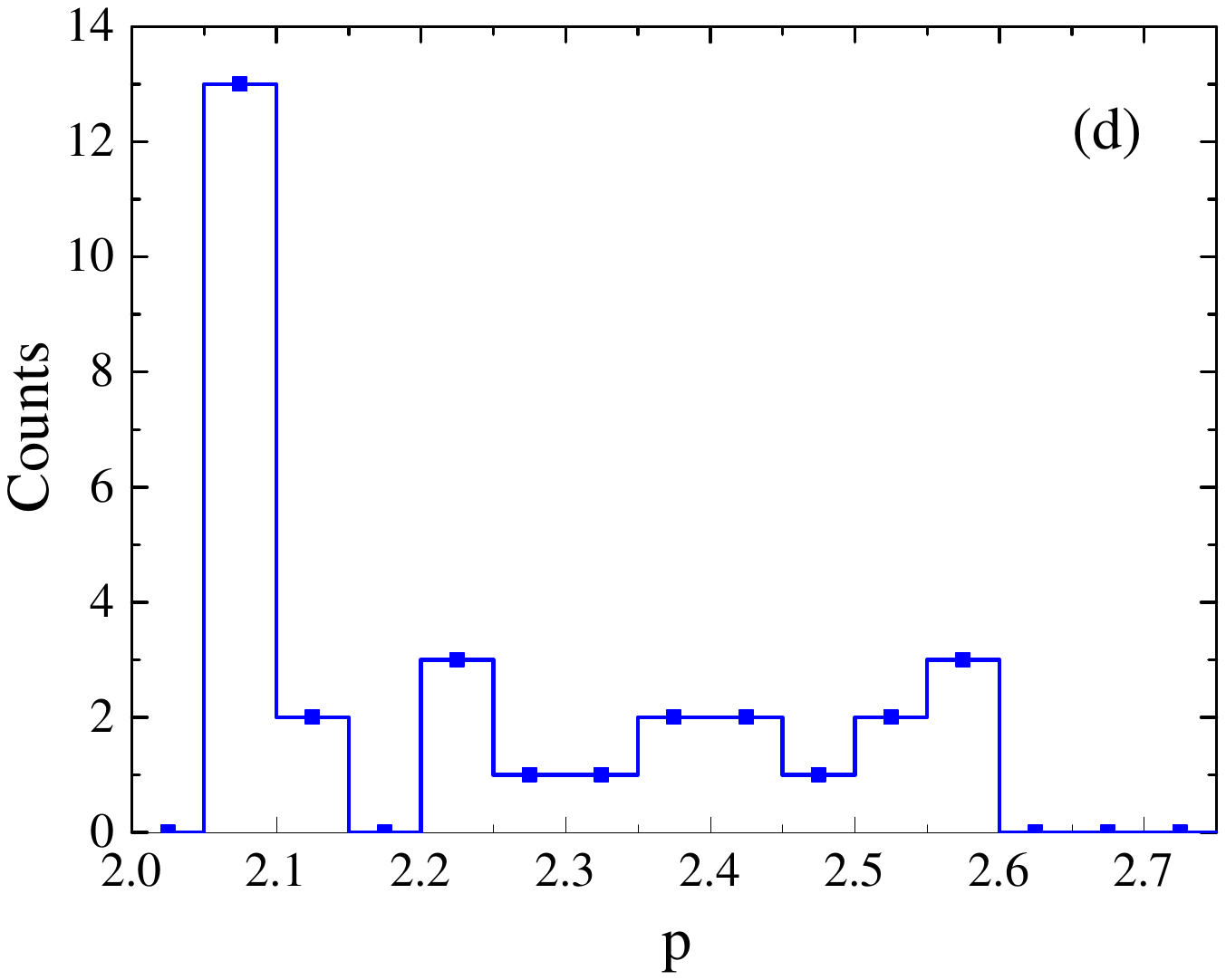}\label{Fig5d}}
\subfigure{
\includegraphics[angle=0,width=0.45\textwidth]{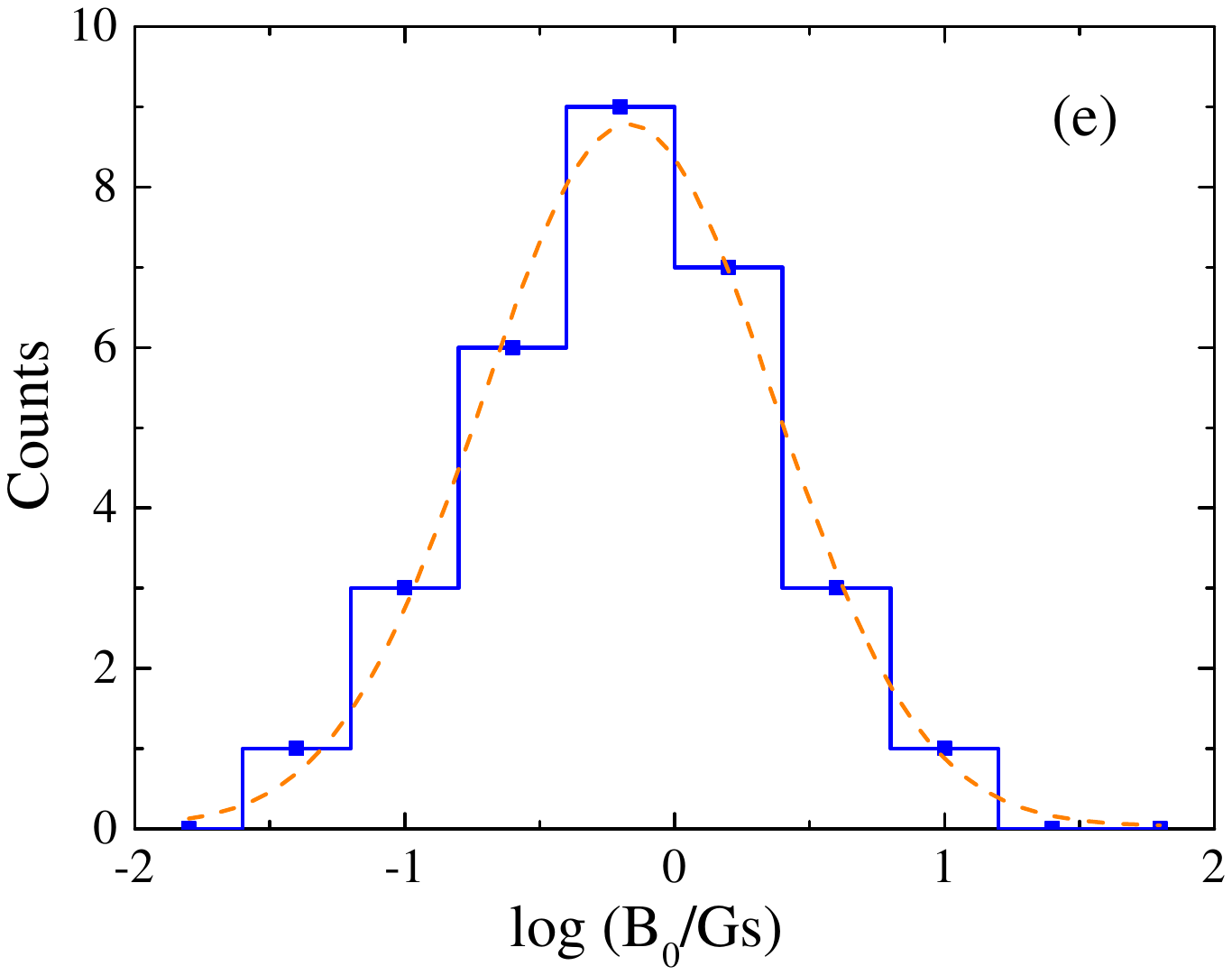}\label{Fig5e}}
\caption{Panel (a)-(e)-- Distributions of $\log (\epsilon_{e})$, $\log (\epsilon_{B})$, 
$\log (\sigma)$, $p$ and $\log (B_{0})$ for the bursts in our sample, respectively. The dashed lines represent the best fitting with a Gaussian function or a two-Gaussian function model.}
\label{Fig5}
\end{figure*}

\begin{figure*}[htbp]
\centering
\subfigure{
\includegraphics[angle=0,width=0.45\textwidth]{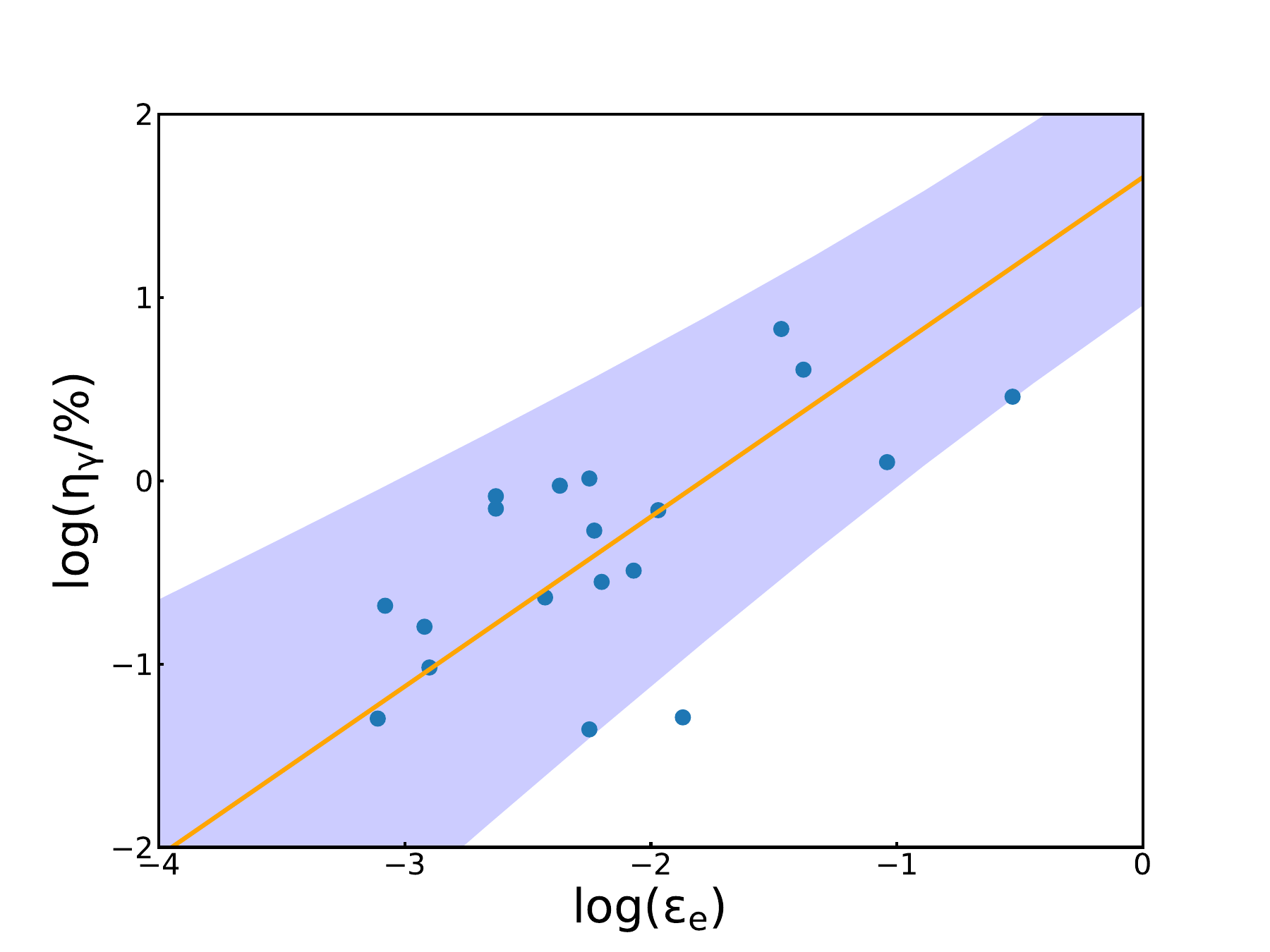}\label{Fig5f}}
\caption{Pair correlations between $\log (\epsilon_{e})$ and $\log (\eta_{\gamma})$ for the bursts in our sample. The solid line represents the best-fitting result with the OLS bisector algorithm model, and the purple area is the confidence level of $2\sigma$.}
\label{Fig5_6}
\end{figure*}

\begin{figure*}[htbp]
\centering
\subfigure{
\includegraphics[angle=0,width=0.45\textwidth]{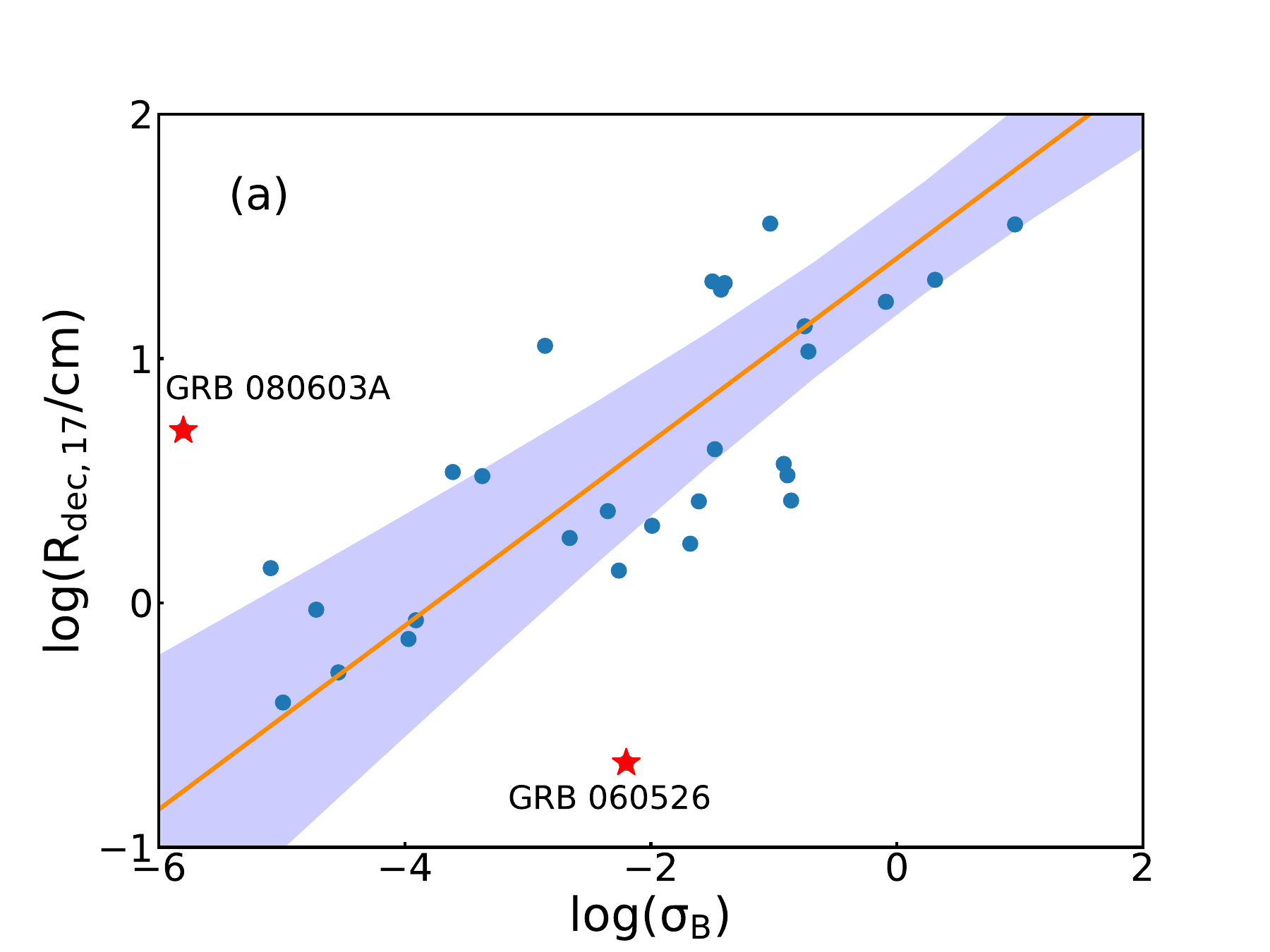}\label{Fig6a}}
\subfigure{
\includegraphics[angle=0,width=0.45\textwidth]{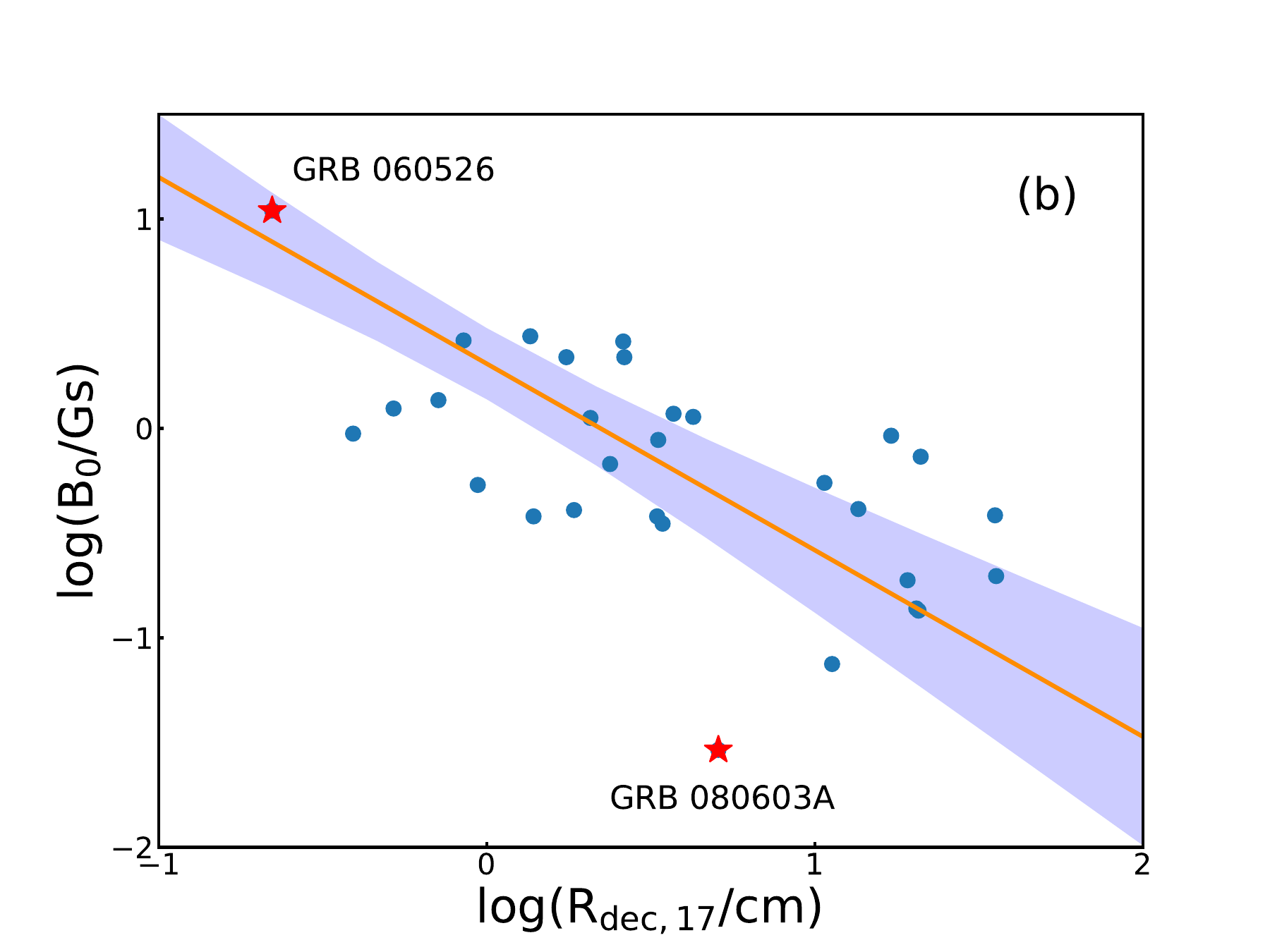}\label{Fig6b}}
\subfigure{
\includegraphics[angle=0,width=0.45\textwidth]{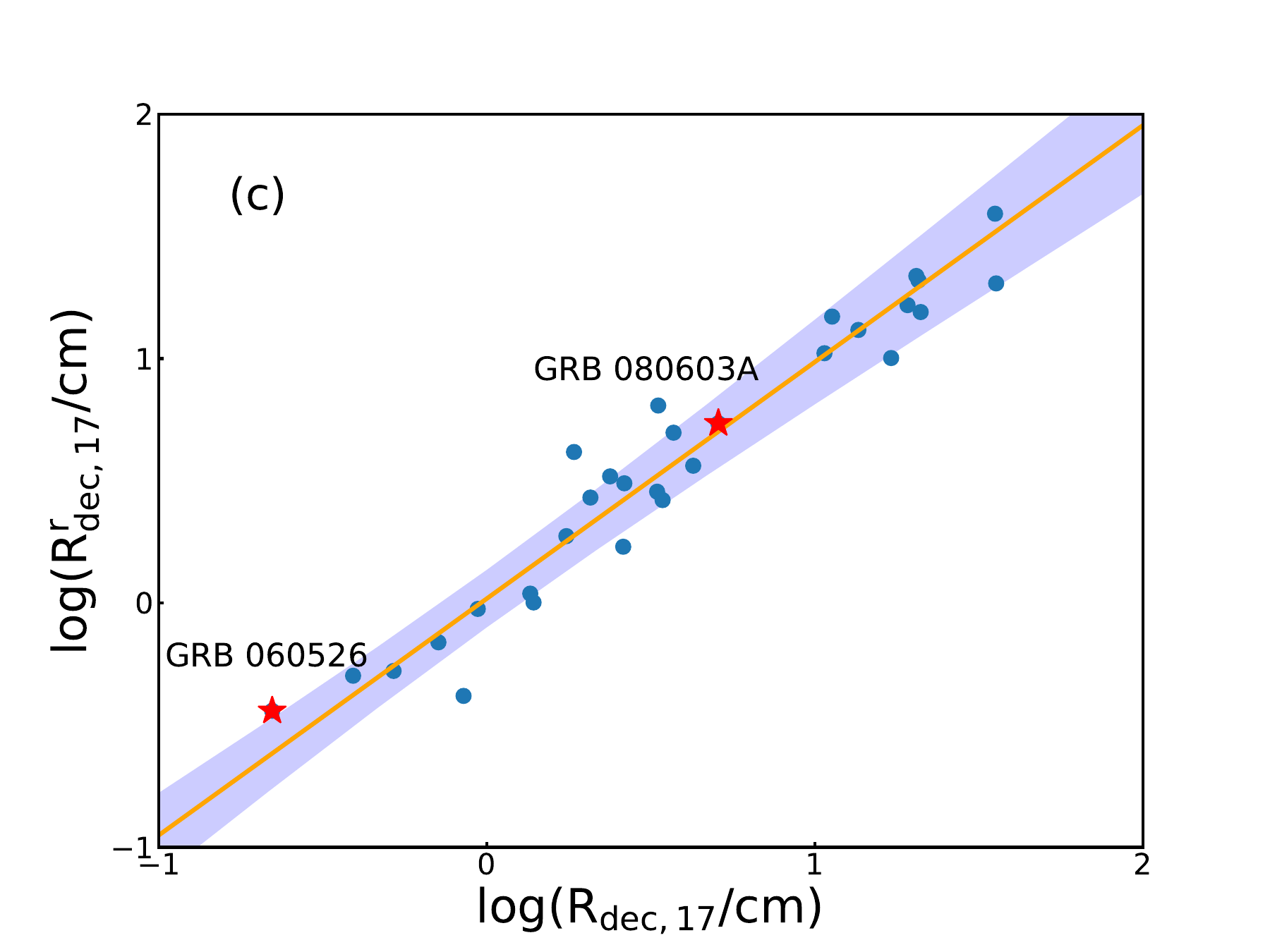}\label{Fig6c}}
\caption{Pair correlations among $R_{\rm dec,17}$, $\sigma_{B}$, $B_{0}$, and the correlation between $R^{r}_{\rm dec,17}(B_{0},\sigma_{B})$ and $R_{\rm dec,17}$, where $R^{r}_{\rm dec,17}(B_{0},\sigma_{B})$ is derived from our multiple variable stepwise regression analysis. The red star represents GRBs~060526 and 080603A.
The light purple area represents the $2\sigma$ confidence level. The orange lines are the best-fitting results by utilizing the OLS bisector.}
\label{Fig6}
\end{figure*}

\begin{figure}[htbp]
\centering
\includegraphics[angle=0,width=0.9\textwidth]{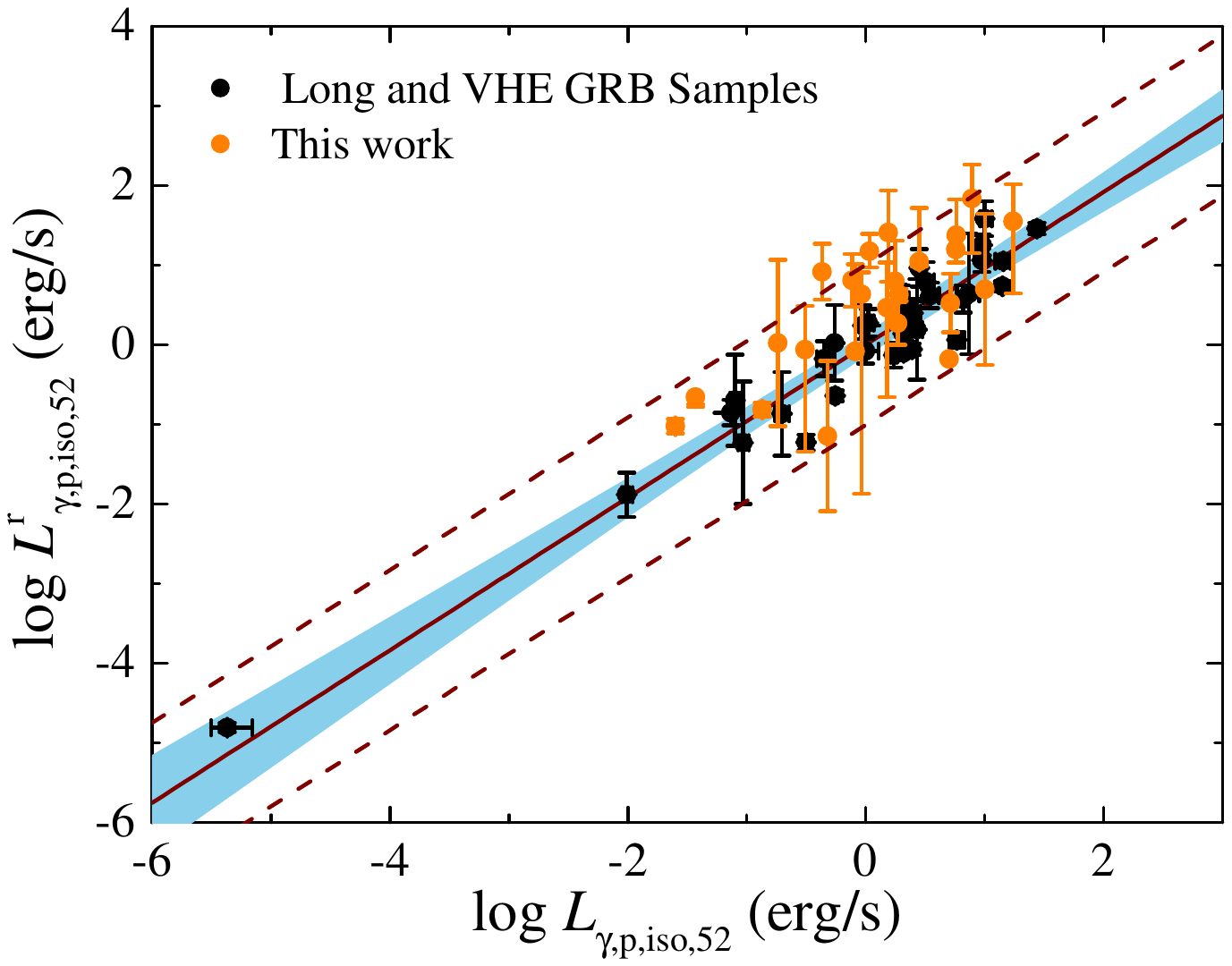}
\caption{Distribution of the GRBs in our sample (orange dots) in the $\log L^{\rm r}_{\gamma,p,\rm iso}-\log L_{\gamma,p,\rm iso}$ plane, where $\log L^{\rm r}_{\gamma,p,\rm iso}$ is calculated with the $L_{\rm \gamma, iso}-\Gamma_0-E_{\rm p,z}$ relation reported by \cite{Liang_2015_Lin_ApJ...813..116L}. The typical GRBs, low-luminosity GRBs, high-luminosity GRBs and VHE GRBs from  \cite{Liang_2015_Lin_ApJ...813..116L}, \cite{Huang_2020_Liang_ApJ...903L..26H}, and \cite{Zhang_2023_Ren_ApJ...952..127Z} are also shown for comparison (black dots).  
The solid and dashed lines are the best fit and its dispersion in the $3\sigma$ confidence level for the relation between $\log L^{\rm r}_{\gamma,p,\rm iso}$ and $\log L_{\gamma,p,\rm iso}$ reported by \cite{Liang_2015_Lin_ApJ...813..116L}.
}
\label{three parameters}
\end{figure}

\end{document}